\documentclass[twocolumn,a4paper,10pt,aps,prd,amsmath,amssymb,preprintnumbers,longbibliography]{revtex4-1}%\documentclass[twocolumn,aps,prd,amsmath,amssymb,preprintnumbers,longbibliography]{revtex4-1}
\usepackage{amsmath} 
\usepackage{amsfonts} 
\usepackage{amssymb}
\usepackage{bbm}
\usepackage{graphics}
\usepackage{graphicx}
\usepackage{titlesec}		%to adjust appendix title size
\usepackage{mathtools}
\usepackage{environ}
\usepackage{dsfont}
\usepackage{tensor}
\usepackage{xcolor}		%use colors
\usepackage{paralist}	%enumerate in line
\usepackage{todonotes}
\usepackage{braket}
\usepackage[inline]{enumitem} % to use enumertate*
	\usepackage[colorlinks=true,hyperindex,breaklinks]{hyperref}
	\hypersetup{urlcolor=black,linkcolor=black,citecolor=black}
	\usepackage[toc,page,title]{appendix}
	\usepackage{makecell,tabularx}
	\setcellgapes{3pt}
	\usepackage{natbib} % umgestellt auf biblatex
	\setcitestyle{sort&compress,cite}
\makeatletter
\renewcommand*\env@matrix[1][c]{\hskip -\arraycolsep
 \let\@ifnextchar\new@ifnextchar
 \array{*\c@MaxMatrixCols #1}}
\makeatother
\newcommand{\be}{\begin{equation}}
\newcommand{\ee}{\end{equation}}
\newcommand{\ba}{\begin{eqnarray}}
\newcommand{\ea}{\end{eqnarray}}
\newcommand{\nn}{\nonumber}

\newcommand{\tr}{\mathrm{tr}}

\titleformat{\subsection}[block]{\normalfont\bfseries}{\thesubsection.}{1ex}{}
\titlespacing{\subsection}{0pt}{10pt}{1pt}[0pt]
\titleformat*{\section}{\large\bfseries}
\renewcommand{\thesubsection}{\arabic{subsection}}
\titlespacing{\paragraph}{0pt}{15pt}{5pt}[0pt]
\titleformat{\paragraph}[block]{\normalsize\bfseries}{\theparagraph}{1ex}{}

	\definecolor{refkey}{rgb}{0,0,1}
	\definecolor{labelkey}{rgb}{0,1,0}
	\renewcommand{\thesection}{\arabic{section}}
	\renewcommand{\thesubsection}{\thesection.\arabic{subsection}}
	
	\makeatletter %um @ in name zu verwenden
	\renewcommand{\p@subsection}{}
	\renewcommand{\p@subsubsection}{}
	\makeatother % macht einstellung rückgängig

\DeclareMathOperator{\diag}{diag}
\DeclareMathOperator{\sign}{sign}
\let\Re\relax
\DeclareMathOperator{\Re}{Re}
\let\Im\relax
\DeclareMathOperator{\Im}{Im}
\renewcommand{\d}{\mathop{}\!\mathrm{d}}
\newcommand{\D}{\mathop{}\!{D}}
\renewcommand{\epsilon}{\varepsilon}
\begin{document}

\title[ ]{Pregeometry and spontaneous time-space asymmetry}

\author{C. Wetterich}
\affiliation{Institut  f\"ur Theoretische Physik\\
Universit\"at Heidelberg\\
Philosophenweg 16, D-69120 Heidelberg}

% change width of abstract, to make the rest fit better %%%%%%%%%%%%%%%%%%%%%%%%%%%%%%%%%%%%%%%%
%\makeatletter
%\renewcommand\frontmatter@abstractwidth{\dimexpr\textwidth-3cm\relax}
%\makeatother
%%%%%%%%%%%%%%%%%%%%%%%%%%%%%%%%%%%%%%%%%%%%%%%%%%%%%%%%%%%%%%%%%%%%%%%%%%%%%%%%%%%%%%%%%%%%%%%%
\begin{abstract}
In pregeometry a metric arises as a composite object at large distances. We investigate if its signature, which distinguishes between time and space, could be a result of the dynamics rather than being built in already in the formulation of a model. For short distances we formulate our model as a Yang-Mills theory with fermions and vector fields. For the local gauge symmetry we take the non-compact group SO(4,\,$\mathbb{C}$). The particular representation of the vector field permits us to implement diffeomorphism invariant kinetic terms. Geometry and general relativity emerge at large distances due to a spontaneous breaking of the gauge symmetry which induces masses for the gauge bosons. 
The difference between time and space arises directly from this spontaneous symmetry breaking. For a euclidean metric all fields have a standard propagator at high momenta. Analytic continuation to a Minkowski-metric is achieved by a change of field values. We conjecture that this type of model could be consistent with unitarity and well behaved in the short distance limit.
\end{abstract}

\maketitle

\section{Introduction}

%Absatz
Within special and general relativity time and space are integrated into a common spacetime. Still, time and space have different characteristics. Timelike correlation functions often show an oscillatory behavior, while spacelike correlations typically decay exponentially or with a power law. This time-space asymmetry results from the particular signature ($-,\, +\,,+\,, +$) of the metric.

%Absatz
In a unified treatment of time and space one may ask what causes this asymmetry. We will consider a setting without a difference between time and space on a fundamental level. Both are just coordinates which label families of observables. In four dimensions this coordinate manifold can be taken as $\mathbb{R}^{4}$. Geometry, as given the metric and its signature, arises only in terms of the expectation value of a suitable metric field $g_{\mu\nu}(x)$, $ x^{\mu}=(t,\vec{x})$, $\mu = 0 \dots 3$, which can be considered as a collective degree of freedom.

%Absatz
One possibility for the emergence of a time-space asymmetry could be simply the selection of observables. Different types of observables are selected for the description of the evolution in time and the space-dependence of expectation values and correlations. The selection proceeds by defining a family of hypersurfaces along which the ``evolution" is investigated - these hypersurfaces are labeled by the time coordinate~$t$~\cite{CWPW}.
The direction of the sequence of hypersurfaces in $\mathbb{R}^{4}$, or in a corresponding four-dimensional hypercubic lattice of space-time points for a discrete formulation, is arbitrary. Observables are selected such that their dependence on $t$ can catch a unitary evolution along the $t$-hypersurfaces. An example for this setting is given by a simple probabilistic cellular automaton that describes a discrete Thirring type quantum field theory for interacting fermions~\cite{CWCA,CWCAF}. A setting on a two-dimensional square lattice results in the continuum limit in a 1+1-dimensional fermionic theory with Lorentz-symmetry. Lorentz-symmetry entails time-space asymmetry. The selection of the $t$-axis and the $x$-axis on the two-dimensional square lattice is arbitrary, however. 

%Absatz
One may ask if an arbitrariness in the selection of a time-axis can be encoded in terms of spontaneous symmetry breaking~\cite{CWTS}. Consider metric fields $g_{\mu\nu}(x)$ that are independent of $x$ and diagonal. For an expectation value $g_{\mu\nu}=\diag (-1,1,1,1)$ the $x^{0}$-axis is the time direction, while for $g_{\mu\nu}= \diag (1,-1, 1,1)$ the time direction is associated with $x^{1}$. The selection of the time axis is related in this case to a particular expectation value of a field.  Once expectation values $\diag (-1,1,1,1)$ and $\diag (1,-1,1,1)$ are admitted there seems to be no reason why to exclude the value $\diag (1,1,1,1)$. This corresponds, however, to a different signature of the metric. 

%Absatz
If the dynamics makes no difference between values $\diag (-1,1,1,1)$ and $\diag (1, -1, 1, 1)$, these two values may be related by a symmetry which is respected by the dynamics. The selection of the time axis is then formulated as a result of a spontaneous breaking of this symmetry. If dynamically selected ``ground states" with metrics $\diag (-1,1,1,1) $ or $\diag (1,-1,1,1) $ are preferred as compared to a state with $\diag (1,1,1,1)$, this can yield a dynamical explanation of the time-space asymmetry. In contrast, if $\diag (1,1,1,1) $ is also related to $(-1,1,1,1)$ by a symmetry, not only the direction of time is an effect of spontaneous symmetry breaking, but also the time-space asymmetry itself. In this paper we are open to these different possibilities. What is central, however, is that no signature constraint is imposed on the metric from the beginning. 

%Absatz
In a theory with fermions a pseudo-metric can be formulated as a bilinear in the complex generalized vierbein $e_{\mu}{}^{m}$,
\begin{equation}
G_{\mu\nu}=e_{\mu}{}^{m}e_{\nu}{}^{n}\delta_{mn}\;.
\label{I1}
\end{equation}
Since we do not want to introduce a time-space bias from the beginning, the contraction of the ``Lorentz indices" $m,n=0\dots 3$ proceeds by $\delta_{mn}$. A negative sign of a real $G_{00}$ obtains in this case by an imaginary value of  $e_{0}{}^{0}$, e.g. for $k,l=1,2,3$
\begin{equation}
e_{0}{}^{0}=i\,,\quad e_{k}{}^{l}=\delta_{kl}\,,\quad e_{0}{}^{k}=0\,,\quad e_{k}{}^{0}=0\;.
\label{I2}
\end{equation}
A formulation without time-space bias is therefore based on complex values of the vierbein. The ground state and spontaneous symmetry breaking select an expectation value of the vierbein, which in turn determines the signature of the metric by eq.~\eqref{I1}.

%Absatz
In this formulation a state with euclidean signature can be obtained by replacing in eq.~\eqref{I2} the value $e_{0}{}^{0}=i$ by $e_{0}{}^{0}=1$. A powerful and complete formulation of analytic continuation from euclidean to Minkowski space can be realized by investigating a family of complex vierbeins $e_{0}{}^{0}=e^{i\varphi}$ and varying $\varphi$~\cite{CWES}. In the formulation of the present paper or ref.~\citep{CWTS} this formal property is directly related to a trajectory in field space. If we admit a complex vierbein, the analytic continuation proceeds by evaluating observables for different values of the vierbein. 

%Absatz
Arbitrary values of a complex vierbein translate to complex $G_{\mu\nu}$ by eq.~\eqref{I1}. The pseudo-metric retains a simple geometric interpretation as a metric only for the particular cases where it is real. This challenges the fundamental microscopic formulation of a quantum theory for gravity based only on geometric concepts.

%Absatz
So far, the short distance behavior of quantum gravity remains an important open problem. If one formulates quantum gravity in terms of a fundamental metric degree of freedom, one can obtain a renormalizable and asymptotically free model involving up to four derivatives\,\cite{STE,FRATSE,AB}. It is not unitary, however, since it contains ghost degrees of freedom. A consistent perturbative quantum field theory on this basis remains elusive, despite interesting recent developments\,\cite{AGRA1,AGRA2,SAREP}. If an ultraviolet fixed point of quantum gravity exists, asymptotic safety\,\cite{WEIN,MR,DPER,SOUM,LAUR,RSAU} should lead to a consistent unitary short distance behavior. In particular, it should establish a consistent short distance behavior of the graviton propagator. There has been progress in this direction\,\cite{GRP1,GRP2,GRP3}, but the issue is rather complex and it seems fair to say that up to now no convincing understanding of the short distance graviton propagator has emerged in this approach.

%Absatz
These difficulties of the pure metric formulation make it tempting to ask if different degrees of freedom at short distances could lead to a simpler description. Superstring theories are a rather maximal approach in this direction, involving ``infinitely many particles`` at short distances. At tree level the short distance behavior of the graviton propagator and graviton scattering find a consistent and satisfactory picture. Unfortunately, no non-perturbative approach to superstrings is available for an inclusion of the effects of fluctuations of the graviton. Since graviton fluctuations appear to play a crucial role for all other approaches to quantum gravity, this shortcoming turns the string solution for the short distance behavior of gravity to at best a partial answer.

%Absatz
Our approach is based on the functional integral formulation of quantum field theories. In this setting the fluctuations, which are a central ingredient of any quantum theory, are directly incorporated and can be computed, in principle, by various perturbative or non-perturbative methods. For the basic degrees of freedom used for the formulation of the functional integral we use degrees of freedom different from the metric. This defines some type of ``pregeometry". The metric has to emerge as the expectation value of some composite field or collective degree of freedom. The present paper explores pregeometries with a complex vierbein, such that the quest for a short-distance formulation of the quantum theory of gravity can be combined with an investigation of the origin of time-space asymmetry.

%Absatz
A rather minimalistic approach to ``pregeometry'' tries to formulate the short distance behavior of the gravitational theories in terms of fermionic degrees of freedom\,\cite{AK,AV,DS}. In spinor gravity\,\cite{CWSG1,CWSG2,CWHEB,CWSGA} the vierbein, metric and spin connection are all described as composite objects built from fermions. These theories possess beyond diffeomorphism invariance a local gauge symmetry that includes the local Lorentz transformations\,\cite{CWSG1,CWSG2}. In this context the time-space asymmetry induced by spontaneous symmetry breaking has been investigated in ref.~\cite{CWTS}. Practical computations in this approach to quantum gravity are not sufficiently advanced, however, in order to assess the consequences for the short distance behavior of the graviton propagator and similar issues.

%Absatz
In the present work we follow an intermediate approach to pregeometry.  We formulate pregeometry as a diffeomorphism invariant Yang-Mills gauge theory based on the non-compact gauge group SO(4,\,$\mathbb{C}$). Besides the gauge bosons $A_\mu^z$ it contains fermions $\psi_\alpha$ and vectors $e_\mu^a$ in suitable representations of the gauge group. They are coupled to the gauge bosons by standard covariant derivatives in their respective kinetic terms. These kinetic terms involve up to two derivatives and may lead at short distances to a standard form of the propagators.
For the gauge symmetry we choose the complexified group SO(4,\,$\mathbb{C}$) since it contains both the Lorentz group SO(1,\,3) and the euclidean rotations SO(4) as subgroups and therefore does not induce a bias for the signature. Non-compact groups could, however, lead to inconsistencies for a quantum field theory.

%Absatz
A quantum field theory based on this type of pregeometry has to obey several requirements. First, a functional integral for the fields of a given model should be well defined. Second, the associated time evolution should be unitary. Third, flat Minkowski space should be a stable approximate solution. The present paper constitutes a first exploration if these requirements lead to objections against the use of a non-compact gauge group. While the first two points concern the possible form of the microscopic or classical action, the third point concerns the properties of the quantum effective action, see sect.~\ref{sec:2}.

%Absatz
We explore a microscopic action for which the inverse propagator for all physical bosonic modes grows for large momenta $\sim q^2=q_\mu q^\mu$. This is the standard behavior for massless or massive particles. Avoidance of ghosts or tachyonic instabilities is therefore easier than for higher-derivative theories as Stelle's gravity \cite{STE}.
Nevertheless, if we would construct the kinetic term for the gauge fields in the usual form with the metric, we would immediately find tachyonic modes. We will show that the use of a complex vierbein can avoid this problem. We present a classical action that has neither ghost nor tachyon modes. For flat euclidean space the inverse propagator for all physical excitations grows for high momenta proportional to $q^2$ with a positive coefficient.
If a renormalizable theory can be built on this type of pregeometry, it would constitute a good candidate for an ultraviolet completion of quantum gravity. 

%Absatz
Quantum fluctuations induce additional terms in the quantum effective action. They could lead to new instabilities, connected to the form of the full propagator for physical excitations. Ghosts, tachyons or other problems with the analytic structure of the propagator could arise at this level. We investigate such additional terms, assuming that they do not alter qualitatively the high momentum limit $q^{2}\to 0$. 
These terms are needed for a realistic form of low-energy gravity. A well known issue is the form of the full graviton propagator. Expanding the inverse propagator in powers of $q^2$ and truncating at any finite order beyond $q^2$ necessarily leads to ghosts or tachyons.

%Absatz
We will argue that the full form of the graviton propagator can avoid this potential disease.
For an expansion around flat space we find a range of parameters of a possible quantum effective action for which no instability in the form of tachyons or ghosts occurs for the graviton. The graviton propagator is given by a momentum dependent function $G_\mathrm{grav}(q^2)$ multiplying an appropriate projector on the transversal traceless metric fluctuations that involves the index structure. We obtain for the inverse propagator function
\begin{align}
\label{eq:IN1}
G^{-1}_\mathrm{grav}&(q^2) = \frac{m^2}{8} \left\{ (Z+1)q^2 +m^2 -M^2 \vphantom{\frac{0}{0}}\right. \\
\nonumber
	&- \left. \sqrt{\big[ (Z-1)q^2 +m^2 -M^2 \big]^2 +4\frac{q^2}{m^2} (m^2-M^2)^2 } \right\}.
\end{align}
Here $M^2$ is the (reduced) squared Planck mass and $m^2$ corresponds to the squared mass for the gauge bosons after spontaneous symmetry breaking. These free parameters of our model have to be in the range $0 < M^2 < m^2$. A third parameter $Z$ of our model multiplies the kinetic term for the gauge bosons. It has to obey the restriction
\begin{equation}
0 < Z < \frac{M^2}{m^2} \left( 1 - \frac{M^2}{m^2} \right)^{-1}.
\label{eq:IN3}
\end{equation}
This graviton propagator seems to be well behaved in the entire complex plane for $q^{2}$. It can serve as an example for a consistent graviton propagator in quantum gravity.

%Absatz
The squared momentum involves the inverse vierbein~$e_{m}{}^{\mu}$,
\begin{equation}
q^{2}=q_{\mu}q_{\nu} e_{m}{}^{\mu}e_{n}{}^{\nu}\delta^{mn}\;.
\label{I3}
\end{equation}
The complex $q^{2}$-plane can be spanned by complex values of the vierbein. For the appropriate range of parameters the only pole of $G_\mathrm{grav}(q^2)$ in the complex $q^2$-plane occurs for $q^2 = 0$. One can analytically continue this propagator from euclidean flat space with positive $q^2$ to Minkowski space with $q^2 = -q_0^2 + \vec{q}^2$ without any obstruction. We can define complex momenta by
\begin{equation}
q_{m}=e_{m}{}^{\mu}q_{\mu}\;.
\label{I3a}
\end{equation}
Translating to the complex $q_0$-plane in Minkowski space the poles occur for real $q_0 = \pm\sqrt{\vec{q}^2}$, typically accompanied by branch cuts. For a suitable parameter range they occur on the real axis for $q_0^2 > |q_c^2| + \vec{q}^2$, with $|q_c^2|$ a positive value of the order $m^2$ different from zero.

%Absatz
For low momenta the inverse graviton propagator admits a polynomial expansion in $q^2$,
\begin{equation}
G^{-1}_\mathrm{grav}(q^2) = \frac{M^2 q^2}{4} + \frac{1}{4} \left( Z - \frac{M^2}{m^2} \right) q^4 + ...
\label{eq:IN2}
\end{equation}
For a truncation at the order $q^4$ the graviton propagator \eqref{eq:IN2} shows a ghost instability. This ghost is not present for the full expression \eqref{eq:IN1}. It therefore is an artifact of the truncation. This is a direct example how pregeometry can cure shortcomings of metric gravity.

%Absatz
For the momentum range $|q^2|\ll m^2$ the metric appears as a composite object which dominates the dynamics. The low momentum effective theory is described by general relativity with the Einstein-Hilbert action based on the curvature scalar. Expanding the low momentum effective theory in powers of derivatives or momenta one recovers Stelle's gravity \cite{STE} in a truncation in fourth order in derivatives. This is reflected in the expansion \eqref{eq:IN2}. The ghost observed in this truncation turns out to be an artifact of the approximation \cite{PLCW}. It is not present in our model of pregeometry if the full momentum dependence of the graviton propagator is taken into account.

%Absatz
Besides the graviton our model of geometry involves other scalar, vector and tensor fluctuations. We will discuss the corresponding propagators and stability properties for parts of these modes. In this sense our investigation does not yield a complete answer to the question if flat space is a stable ground state for the proposed effective action. Nevertheless, many qualitative and partly quantitative features emerge from this investigation.

%Absatz
Even if a satisfactory form of the quantum effective action can be found a central question remains: Can such an effective action describe the effects of quantum fluctuations for the functional integral based on the proposed classical action? This question concerns the renormalizability of the model of pregeometry. It may be attacked by the functional renormalization flow from the classical action to the quantum effective action as quantum fluctuations are included in a stepwise manner. The present paper does not address this key issue for a consistent model containing quantum gravity. We remain on the preliminary level of exploring possible consistent forms of the classical and quantum effective action.

%Absatz
This paper is organized as follows. The first part of sects.~\ref{sec:2}-\ref{sec:5} addresses the question of consistency of the classical action. We discuss in sect.~\ref{sec:2} the basic setting of our proposal for pregeometry. Sect.~\ref{sec:3} formulates pregeometry as a SO(4,\,$\mathbb{C}$)-gauge theory with diffeomorphism symmetry. We introduce the diffeomorphism invariant kinetic terms for the complex gauge fields and vierbeins. Sect.~\ref{sec:4} investigates the mode expansion around flat space. Flat space solutions are stable, and the short-distance behavior of the inverse propagator for most physical modes is proportional to $q^2$, without any tachyons or ghosts. At this stage there are still a few physical modes that do not have a kinetic term, and therefore admit no well defined propagator. In sect.~\ref{sec:5} we introduce a complex pseudo-metric as a bilinear of the vierbeins. Additional invariants for the vierbein can be formulated via the use of this pseudo-metric. In the presence of these additional invariants all physical degrees of freedom have a well behaved propagator.
At this stage we have candidates for valid microscopic or classical actions. 

%Absatz
In the second part comprising sects.~\ref{sec:6}-\ref{sec:8} we deal with possible forms of the quantum effective action. We use an (incomplete) set of invariants with no more than two derivatives for an ansatz for the quantum effective action. Sect.~\ref{sec:6} demonstrates that this includes a type of potential for the pseudometric. Solutions with a real pseudo-metric are singled out, allowing for a definition of a real metric and an understanding of its particular significance. 
In sect.~\ref{sec:7} we turn to the low momentum effective theory. We describe the emergence of general relativity. The stability of the propagators is investigated in sect.~\ref{sec:8}. In particular, we find the well behaved graviton propagator \eqref{eq:IN1}. The ghosts in four-derivative gravity are understood as truncation artifacts. 

%Absatz
The discussion in sect.~\ref{sec:9} points to the tasks that need to be accomplished for establishing the proposed pregeometry as a consistent quantum field theory. An extended appendix \ref{sec: Ap.A} performs the field decomposition for fluctuations around flat space. It establishes the propagators for all modes and defines the low-momentum effective theory as the theory for the massless modes. In particular we compute the graviton propagator and find a consistent form for a suitable range of parameters.   This appendix partly recapitulates material of ref.~\cite{Wetterich:2021ywr} for convenience of the reader.

\section{Pregeometry as a Yang-Mills theory\label{sec:2}}

%Absatz
We propose that a pregeometry based on a Yang-Mills gauge theory could serve as a starting point for a quantum theory of gravity. For the particular case of a gauge group SO(1,\,3)~- local Lorentz symmetry -- or the euclidean counterpart SO(4), the real vector field $\tensor{e}{_\mu^m}$ plays the role of a generalized vierbein, $m=0 ...3$, for which the covariant derivative does not vanish. For this case our approach follows the discussion in refs. \cite{CWGG},\cite{CWFSI} see also \cite{RP84}. We discuss the euclidean setting with gauge group SO(4) in detail in ref.~\cite{Wetterich:2021ywr} and appendix \ref{sec: Ap.A}. Our approach in the main text will be more general. We consider a complex generalized vierbein and concentrate on the complexified orthogonal group SO(4,\,$\mathbb{C}$). This setting does not favor a particular signature a priori.

\paragraph*{Classical and quantum effective action\label{par: Classical and quantum}}

%Absatz
It is our aim to find a possible starting point for a consistent quantum field theory that includes gravity. A non-perturbative formulation is based on a functional integral. A well-defined functional integral requires certain properties of the classical action such that the weight factor $\exp (-S)$ is well behaved, not diverging in an unacceptable way. (We take here the ``Euclidean action"  $S$, related to the ``Minkowski action" $S_{M}$ by a factor $i$.)
Since the weight factor can be complex the precise conditions on $S$ are not simple. We require here positive eigenvalues of the inverse propagator matrix (no tachyons or ghosts) for vierbein configurations corresponding to flat euclidean space and vanishing gauge fields. This is often considered as a rather minimal requirement, even though strictly speaking it is neither sufficient nor necessary. Analytic continuation to complex vierbeins may then preserve a well defined functional integral.

%Absatz
A consistent quantum theory should have a unitary time evolution. From the functional integral point of view one introduces a slicing of space by some type of time-hypersurfaces. The evolution of the probabilistic information from one time slice to a neighboring one defines a generalized Schrödinger equation and the associated operator formalism. The Hamiltonian of the generalized Schrödinger equation should be hermitian (self adjoint) in order to guarantee an unitary evolution where information is not lost. For a general complex $S$ the conditions for a unitary evolution are not well known. It is usually thought that this requires the absence of ghosts or tachyons. We will restrict our discussion to this issue.

%Absatz
A realistic theory of quantum gravity needs a consistent phenomenology. For low momenta it has to reproduce general relativity or mild modifications thereof, possibly with additional massless or light fields. In particular, for length scales probed by present observations which are substantially smaller than the horizon of our present Universe, flat Minkowski space should be a stable solution. Otherwise our local environment would quickly deviate from what is observed. In a theory of quantum gravity one would like to extend this stability requirement to a range of length scales down to the Planck length.

%Absatz
The field equations needed for the analysis of stability of flat space have to include the effects of quantum fluctuations. The relevant object for this purpose is the quantum effective action which includes all effects of quantum fluctuations. The field equations derived by variation of the quantum effective action are exact. Computing the quantum effective action amounts to solving the quantum field theory. We are obviously far from realizing this task.

%Absatz
What we can exploit, nevertheless, are general features of the quantum effective action, as symmetries and certain locality (or smoothness) properties resulting in the validity of an expansion in the number of derivatives for sufficiently large length scales. We will focus on the question if a quantum effective action for pregeometry can be found that leads effectively to general relativity and stable Minkowski space for low orders in a derivative expansion. Stability can be linked to the absence of tachyons for physical excitations, but now concerning the quantum effective action instead of the classical action. The issues are different since quantum fluctuations typically produce additional terms in the effective action that do not need to be present in the classical action. We will see that such terms are required for a consistent phenomenology. If an acceptable quantum effective action can be found, the big task remains to relate it to a classical action which is used to formulate the functional integral.

\paragraph*{Covariant derivative of generalized vierbein\label{par: Covariant derivative}}

%Absatz
Before presenting the detailed definition of the classical action for our model of pregeometry in the next section, it may be useful to put this formulation into a more general geometric context. This should help to see the relations with previous work on related or somewhat similar concepts.

%Absatz
For our model of pregeometry an important role is played by the non-vanishing covariant derivative of the generalized vierbein, $D_\mu \tensor{e}{_\nu^m}$. The square of this tensor can be used for the construction of a diffeomorphism invariant kinetic term for the vierbein. In the limit where this covariant derivative vanishes and the composite metric is real one recovers general relativity. Our approach employs two different connections. One is the gauge connection and given by the gauge fields, the other is the geometric connection given by the Levi-Civita connection expressed in term of derivatives of the vierbein. The geometric connection is torsion free and there is no non-metricity. This is the standard setting for Yang-Mills theories coupled to gravity. The two connections correspond to different fiber bundles -- this was historically the reason why for this version of generalized gravity chiral fermions could be obtained from dimensional reduction of higher dimensional theories with compact internal space \cite{CWGG}.

%Absatz
In this language there seems at first sight only little overlap with formulations of metric affine gravity with torsion and/or non-metricity, for which a dynamical spin connection plays the role of the gauge bosons of the Lorentz-group \cite{CAR,UTI,KIB,SCI,HH,SHA,HCM,HEI,RP,PS,SSTZ,MMOY}. We will see, however, that it is possible to choose a different geometric connection with torsion and a vanishing covariant derivative of the vierbein that brings this particular version of pregeometry much closer to versions of metric affine gravity. Furthermore, for a subclass of field configurations the bundle structure of the gauge sector can be identified with the one relevant for geometry. 
The difference between two connections with the same bundle structure is a tensor. We can then express the covariant derivative of the vierbein as a linear combination of the gauge field and the torsion tensor, thereby establishing a close connection to the formulation of Poincaré gravity in refs.~\cite{HS1,HS2,HS3,SVN,NARR,NIRRU}. 
Some aspects are also similar to the gauge theory in ref. \cite{KR1,KR2}.

%Absatz
Within a more general setting a model is specified by the choice of the gauge group and the representation for the fermions $\psi$ and vector fields $\tensor{e}{_\mu}$. One possible choice for the gauge group could be the Lorentz group SO(1,\,3), with $\psi$ in a spinor representation and $e_\mu$ in a vector representation. For euclidean gravity, the gauge group is instead SO(4), and we discuss this choice in detail in appendix \ref{sec: Ap.A}. For these rather minimal settings the time-space asymmetry of SO(1,\,3), which corresponds to the signature of the invariant tensor $\eta_{mn}=diag(-1,+1,+1,+1)$, has to be postulated a priori. It will determine the signature of the composite metric. We pursue here an extended setting that admits both Minkowski space and euclidean flat space as possible field configurations. The observed time-space asymmetry arises from spontaneous symmetry breaking that selects Minkowski space, or a corresponding cosmological solution, dynamically \cite{CWTS,CWSG1,CWSG2}. For alternative ideas in this direction see ref.~\cite{MKZU,ZK}.

\paragraph*{Non-compact gauge group and ghosts\label{par: Non-compact gauge group}}

%Absatz
For the gauge group we choose SO(4,\,$\mathbb{C}$). This extends the orthogonal transformations in four dimensions to complex infinitesimal transformation parameters, thereby doubling the number of generators and gauge fields. Both SO(4) and SO(1,\,3) are subgroups of SO(4,\,$\mathbb{C}$). The vector field $\tensor{e}{_\mu^m}$ belongs to the complex four component vector representation, $m=0 ...3$, corresponding for each $\mu$ to eight real components. We may call it a ''complex vierbein``. The fermions are complex four-component Dirac spinors or two component Weyl spinors. The six complex (or twelve real) gauge fields $\tensor{A}{_{\mu mn}}=-\tensor{A}{_{\mu nm}}$ belong to the adjoint representation. One could restrict the discussion to real gauge fields and real vierbeins by considering the gauge groups SO(1,\,3) or SO(4). Many points of our discussion can directly be taken over to these simpler cases. In particular, a discussion of euclidean quantum gravity for the gauge group SO(4) and real $e_{\mu}{}^{m}$ can be found in ref.~\cite{Wetterich:2021ywr}. Our setting will be more general and involve additional degrees of freedom. Larger gauge groups with an SO(1,\,3)-subgroup, or both SO(1,\,3) and SO(4)-subgroups, are also possible.

%Absatz
The gauge group SO(4,\,$\mathbb{C}$) is non-compact. This brings new potential problems since the kinetic terms for the gauge bosons may not all have the same sign. As a result, ghost instabilities could be present. On the level of the microscopic or classical action the issue of ghosts concerns the question if the functional integral is well defined. On this level we have found gauge-invariant terms which provide for euclidean flat space a positive kinetic term for all gauge bosons, despite the non-compact character of SO(4,\,$\mathbb{C}$). This demonstrates that non-compact gauge groups do not necessarily lead to ghosts.  Similarly, for suitable invariants the kinetic terms for the complex vierbein are all positive on euclidean flat space.

%Absatz
On the level of the quantum effective action possible problems from ghosts concern the stability of solutions of the field equations derived from it. More precisely, this involves the question if a supposed ground state or cosmological solution is stable with respect to perturbations. While tachyons indicate an instability on the level of linear fluctuations, the issue of ghosts is more complex. In the absence of tachyons problems can arise beyond the linear approximation, while stability of small fluctuations is maintained. Furthermore, the presence of ghosts could induce in the effective action additional terms which affect the complex structure of propagators or similar, for example by generating a decay width with the wrong sign due to quantum pair production of particles with negative energy. In our view the issue of ghosts is not fully settled at the present stage.

\paragraph*{Analytic continuation\label{par: Analytic continuation}}

% Absatz
The fact that excitations of Minkowski space and euclidean space can be related by analytic continuation in field space  may shed new light on the questions concerning ghosts and possible instabilities.  For stable classical propagators (absence of ghosts) in euclidean flat space analytic continuation to Minkowski signature may still permit a well defined functional integral. We will address several aspects of the issue of analytic continuation, without a complete answer about the conditions under which negative kinetic terms are not a cause of a physical instability of Minkowski space.

%Absatz
Analytic continuation is usually thought as a map between two different quantum field theories. In the present approach analytic continuation can be realized \textit{within} a given theory, by continuously changing the complex vierbein field \cite{CWES}. This offers new prospects of exploring the configuration space of the quantum effective action of our theory by analytic continuation in field space starting, for example, from flat euclidean space and reaching Minkowski space. We will establish stability of the microscopic theory in euclidean space. Analytic continuation in field space suggests that this carries over to Minkowski space. 

%Absatz
A second version of analytic continuation does not only change the value of the complex vierbein, but in addition the value of certain components of the complex gauge fields. For this type of analytic continuation the euclidean SO(4)-subgroup of SO(4,\,$\mathbb{C}$) is analytically continued to the SO(1,\,3)-subgroup.

\paragraph*{Spontaneous gauge symmetry breaking \label{par: Spontaneous gauge symmetry breaking}}

%Absatz
Any non-vanishing expectation value of the vierbein breaks spontaneously both the local gauge symmetry SO(4,\,$\mathbb{C}$) and the gauge symmetry of diffeomorphisms. For suitable constant vierbeins a global symmetry SO(1,\,3) or SO(4), together with translation symmetry, remains preserved, corresponding to the symmetry of Minkowski space or euclidean space. Analogously to the Higgs mechanism, the ''spontaneous breaking`` of the gauge symmetry provides mass terms for the gauge bosons -- for the simplest case an equal mass $m$ for all gauge bosons. The effective low momentum theory for momenta $|q^2|<<m^2$ turns out to be some version of general relativity.

%Absatz
A composite real metric emerges as a bilinear in the vierbeins, $g_{\mu\nu}=Re(\tensor{e}{_\mu^m}\tensor{e}{_\nu^n}\delta_{mn})$. In the presence of suitable invariants the solutions of the field equations lead to vierbein-configurations for which the complex vierbein bilinear $G_{\mu \nu}=\tensor{e}{_\mu^m}\tensor{e}{_\nu^n}\delta_{mn}$ assumes real values. This does not imply real vierbeins, however. Since $G_{\mu \nu}$ is SO(4,\,$\mathbb{C}$)-invariant, all vierbeins related by local SO(4,\,$\mathbb{C}$)-transformations lead to the same real $G_{\mu \nu}$. Fluctuations for which $G_{\mu \nu}$ becomes complex are suppressed, however, by mass terms. They are absent in the effective low-momentum theory. The bosonic physical degrees of freedom in the low-momentum effective theory are all contained in the real metric $g_{\mu \nu}$.

%Absatz
Local gauge symmetries are never spontaneously broken in a strict sense. This feature is manifest if we formulate the effective low-momentum theory in terms of vierbeins. By simple field-redefinitions the vierbein can be chosen to be real, with a metric given now by $g_{\mu \nu}=\tensor{e}{_\mu^m}\tensor{e}{_\nu^n}\eta_{mn}$, where $\eta_{mn}$ has the signature appropriate to SO(1,\,3) or SO(4), depending on the particular solution of the field equations. In the effective low-momentum theory the gauge fields $\tensor{A}{_{\mu mn}}$ equal the spin connection $\tensor{\omega}{_{\mu mn}}$ which is formed in the usual way from derivatives of the vierbein. They are no longer independent degrees of freedom. Local SO(1,\,3)-symmetry or SO(4)-symmetry is still manifest, acting on the vierbein and fermions. The degrees of freedom in the vierbein that lead to the same metric are gauge degrees of freedom. The effective low-momentum theory also remains diffeomorphism invariant. In the presence of suitable invariants within pregeometry, the low-momentum effective theory is general relativity, with the Einstein-Hilbert action involving the curvature scalar formed from $g_{\mu\nu}$ supplemented by higher derivative terms.

\paragraph*{Ultraviolet completion of general relativity\label{par: Ultraviolet completion}}

%Absatz
The full theory is actually much simpler than the low-momentum effective theory. The inverse propagators of all physical degrees of freedom grow $\sim q^2$ for large momenta. One may consider the analogy to strong interactions in particle physics. The metric corresponds to the pions, or more generally mesons, in the effective low-momentum theory (chiral perturbation theory). The degrees of freedom of pregeometry correspond to the gluons, for which the short distance behavior becomes simple due to asymptotic freedom. 

%Absatz
Embedding a pure metric formulation as an effective low-momentum theory in a more complete theory of pregeometry sheds light on the origin of the complexity that is apparent in a pure metric formulation of asymptotic safety. One can always formally integrate out all degrees of freedom except for the composite metric. As a result, a simple local action for pregeometry takes the form of a complicated non-local action for the metric in the range of momenta $q^2>>m^2$. This is due to the presence of additional degrees of freedom in the range of high momenta. In our analogy one could formally use mesons also for momenta of $100$ GeV or larger. The resulting effective action would be complicated, obscuring the simplicity of QCD with gluons. 

%Absatz
The present paper can be considered as a starting point for an investigation if this type of pregeometry can lead to a consistent quantum field theory. One has to find out if a simple action with parameters in the range appropriate for a stable theory can be considered as a good approximation to the quantum effective action. This will need functional renormalization group (FRG) investigations since not all couplings are dimensionless small parameters and a free short distance theory is not expected for gravitational interactions. First FRG - investigations for models with vierbeins and an independent spin connection are already available \cite{DR1,DR2,HR} and would need to be focused on our proposal. 

%Absatz
If our proposal leads to a consistent quantum field theory, this does not imply that it is the most fundamental formulation of pregeometry.
In spinor gravity the vierbein and the gauge fields arise as composite objects formed from fermions. One can reformulate a purely fermionic theory by adding composite fields, as well known from the Nambu-Jona-Lasinio model for strong interactions or from solid state models as the Hubbard model. If spinor gravity is a consistent quantum field theory, also its reformulation with composite fields, which takes the structure of the present model of pregeometry, must be a consistent quantum field theory.

\section{Local non-compact SO(4,\,$\mathbb{C}$)-gauge symmetry\label{sec:3}}

%Absatz
This section presents the key objects of our model of pregeometry. Fields depend on coordinates $x=x^\mu=(x^0,x^k)$.
There is no difference between the coordinates $x^0=t$ and $x^k$. These coordinates parametrize $\mathbb{R}^4$ and are kept fixed. A particular geometry will only arise with a composite metric $g_{\mu \nu}(x)$.  The signature, which can differentiate between time and space, is only fixed by the expectation value of this metric. 

\paragraph*{Complex vierbein and gauge transformations}

%Absatz
A basic object of our discussion is the complex generalized vierbein $\tensor{e}{_\mu^m}$. It can be considered as a complex $4\times 4$-matrix, with spacetime index $\mu = 0...3$, and Lorentz index $m=0..3$. With respect to general coordinate transformations (diffeomorphisms) it transforms as a covariant vector, and it is also a vector of SO(4,\,$\mathbb{C}$), with infinitesimal transformation
\begin{equation}
\delta \tensor{e}{_\mu^m} = -\tensor{e}{_\mu^m} \tensor{\epsilon}{_n^m} = \tensor{\epsilon}{^m_n} \tensor{e}{_\mu^n}.
\label{eq:1}
\end{equation}
We raise and lower Lorentz indices with $\delta^{mn}$ or $\delta_{mn}$, such there is actually no difference between upper and lower Lorentz-indices. The infinitesimal transformation parameters $\epsilon_{mn}$ are complex and antisymmetric,
\begin{equation}
\epsilon_{mn} = -\epsilon_{nm},\quad \tensor{\epsilon}{_m^n} = \epsilon_{mp} \delta^{pn}.
\label{eq:2}
\end{equation}
These are six independent complex or twelve independent real parameters, corresponding to the generators of SO(4,\,$\mathbb{C}$). The transformations of the vierbein field $e_{\mu}{}^{m}(x)$ are local, with $\varepsilon_{mn}(x)$ depending on $x$.

%Absatz
Real parameters $\epsilon_{mn}$ generate the euclidean subgroup SO(4). Under this subgroup the real and the imaginary parts of the vierbein transform separately, such that $\tensor{e}{_\mu^m}$ consists of two independent vector representations. For the subgroup SO(1,\,3) -- the Lorentz group -- the parameters $\epsilon_{0k} = -\epsilon_{k0}$ are purely imaginary, while $\epsilon_{kl}$ are real, where $k,l=1...3$. We may define a ``Minkowski vierbein'' by multiplication of the zero component with $-i$, 
\begin{equation}
e^{\mathrm{(M)}0}_\mu=-i \tensor{e}{_\mu^0} ,\quad  \tensor{e}{_\mu^{\mathrm{(M)}k}}=\tensor{e}{_\mu^k}.
\label{eq:5A}
\end{equation}
The ``Minkowski transformation parameters'',
\begin{equation}
\epsilon^\mathrm{(M)}_{0k} = -\epsilon^\mathrm{(M)}_{k0} = i\epsilon_{0k} = -i\epsilon_{k0},\quad
\epsilon^\mathrm{(M)}_{lk} = \epsilon_{lk}, 
\label{eq:4}
\end{equation}
are real and antisymmetric for the subgroup SO(1,\,3). We can write eq.\,\eqref{eq:1} equivalently as
\begin{equation}
\delta e^{\mathrm{(M)}\,m}_{~~\mu} = -e^{\mathrm{(M)}\,n}_{~~\mu} \;\epsilon^{\mathrm{(M)}\,m}_{~~n},
\label{eq:5}
\end{equation}
where for Minkowski parameters we employ $\eta^{mn}$ and $\eta_{mn}$ for raising and lowering indices,
\begin{equation}
\begin{split}
\eta_{mn} = \eta^{mn} = &\diag(-1,1,1,1), \\
\epsilon^{\mathrm{(M)}\,m}_{~~n} = \epsilon^{\mathrm{(M)}}_{~~np} \;\eta^{pm},\;\; &\epsilon^{\mathrm{(M)}\,k}_{~~0} = \epsilon^{\mathrm{(M)}\,0}_{~~k} = i\tensor{\epsilon}{_0^k} = -i\tensor{\epsilon}{_k^0}.
\end{split}
\label{eq:6}
\end{equation}
More generally, the Lorentz indices of all Minkowski quantities are raised and lowered with $\eta$, e.\,g.\ $e^{\mathrm{(M)}}_{\mu m} = e^{\mathrm{(M)}\,m}_{~~\mu} \eta_{nm}$. For the SO(1,\,3)- subgroup the real and imaginary parts of $e^{\mathrm{(M)}\,m}_{~~\mu}$ transform independently, in contrast to SO(4). 

%Absatz
By taking suitable combinations of purely imaginary and real $\epsilon_{mn}$ one obtains the other SO($n$,\,$4-n$)-subgroups of SO(4,\,$\mathbb{C}$). We will not discuss these possibilities here since they do not seem to play a role for observations.

%Absatz
In general, a SO(4,\,$\mathbb{C}$)-vector $E_m$ transforms as
\begin{equation}
\delta E_m = \tensor{\epsilon}{_m^n} E_n,
\label{eq:7A}
\end{equation}
and a second rank tensor $H_{mn}$ as
\begin{equation}
\delta H_{mn} = \tensor{\epsilon}{_m^p} H_{pn} + \tensor{\epsilon}{_n^p} H_{mp}.
\label{eq:7B}
\end{equation}
In particular, $\delta_{mn}$ or $\delta^{mn}$ is an invariant tensor
\begin{equation}
\delta(\delta_{mn}) = \tensor{\epsilon}{_m^p} \delta_{pn} + \tensor{\epsilon}{_n^p} \delta_{mp} = \epsilon_{mn} + \epsilon_{nm} = 0.
\label{eq:7C}
\end{equation}
Another invariant is the totally antisymmetric tensor $\epsilon_{mnpq}$. 

%Absatz
In a Minkowski formulation with
\begin{align}
\begin{split}
H^\mathrm{(M)}_{00} &= -H_{00},\quad H^\mathrm{(M)}_{k0} = iH_{k0}, \\ 
H^\mathrm{(M)}_{0k} &= iH_{0k},\quad H^\mathrm{(M)}_{kl} = H_{kl},
\end{split}
\label{eq:7D}
\end{align}
one has
\begin{equation}
\delta H^\mathrm{(M)}_{mn} = \epsilon^{\mathrm{(M)}\,p}_{~~m} H^{\mathrm{(M)}}_{~~pn} + \epsilon^{\mathrm{(M)}\,p}_{~~n} H^{\mathrm{(M)}}_{~~mp} 
\label{eq:7E}
\end{equation}
The Minkowski formulation is not a different theory, but rather a simple rewriting by choosing other conventions for real and imaginary components of $e^m_\mu$ and $\epsilon_{mn}$.

\paragraph*{Complex gauge bosons}

%Absatz
The gauge bosons of the SO(4,\,$\mathbb{C}$)-symmetry group are described by complex fields $A_{\mu mn}=-A_{\mu nm}$. This differs from the usual description of the compact gauge group SO(4) in terms of real gauge fields. Complex gauge fields are not per se something new. For example, the charged gauge bosons of the weak-interaction gauge symmetry SU(2) $\hat{=}$ SO(3) are denoted by complex fields $W_\mu^{\pm}$ in an appropriate basis.

%Absatz
The real parts of the complex gauge bosons are the gauge bosons of the subgroup SO(4) of SO(4,\,$\mathbb{C}$). The SO(4)-gauge transformations with real gauge parameters $\tensor{\epsilon}{_m^n}$ remain within the subspace of real gauge bosons. We can again define an equivalent Minkowski formulation for the gauge bosons by multiplying fields with one lower Lorentz index by a factor $i$,
\begin{equation}
A_{\mu ok}^{(M)}=iA_{\mu ok}, \quad A_{\mu ko}^{(M)}=iA_{\mu ko}, \quad A_{\mu kl}^{(M)}=A_{\mu kl}.
\label{eq:15CA} 
\end{equation} 
The real parts of $A_{\mu mn}^{(M)}$ are the gauge fields of the subgroup SO(1,\,3). Correspondingly, the SO(1,\,3) - gauge fields with one zero Lorentz index are purely imaginary in the euclidean formulation. This property can be understood most easily by noting that real gauge parameters $\tensor{\epsilon}{^{(M)}_m^n}$ in the Minkowski formulation generate transformations that leave the space of real tensors $H^{(M)}_{mn}$ or gauge fields $A^{(M)}_{\mu mn}$ in the Minkowski formulation invariant. Real $A^{(M)}_{\mu mn}$ are transformed among themselves by the real $\tensor{\epsilon}{^{(M)}_m^n}$ which span the SO(1,\,3) - subgroup.

\paragraph*{Analytic continuation}

%Absatz
For the purpose of analytic continuation from euclidean geometry to Minkowski geometry we define a family of complex vierbeins
\begin{equation}
\tensor{e}{_\mu^0} = e^{i\varphi}\; e^{\mathrm{(E)}\,0}_{~~\mu},\quad \tensor{e}{_\mu^k} = e^{\mathrm{(E)}\,k}_{~~\mu},
\label{eq:7}
\end{equation}
with real $e^{\mathrm{(E)}\,m}_{~~\mu}$. We can view analytic continuation as a map in the space of vierbeins
\begin{equation}
\tensor{e}{_\mu^m} \to \tensor{e}{^\prime_\mu^m}, \quad \tensor{e}{^\prime_\mu^0}=\tensor{e}{^{i\varphi}}\tensor{e}{_\mu^0}, \quad \tensor{e}{^\prime_\mu^k}=\tensor{e}{_\mu^k}.
\label{eq:15CC} 
\end{equation}
Changing $\varphi$ from zero to $\pi/2$ changes $\tensor{e}{_\mu^0} = e^{\mathrm{(E)}\,0}_{~~\mu}$ to $\tensor{e}{^{\prime}_\mu^0} = ie^{\mathrm{(E)}\,0}_{~~\mu}$, such that the Minkowski form of the transformed vierbein equals the original vierbein
\begin{equation}
\tensor{e}{^{\prime (M)}_\mu^m}=\tensor{e}{_\mu^m}.
\label{eq:15CB} 
\end{equation}
This can be used for analytic continuation from euclidean to Minkowski space \cite{CWES}. 
We emphasize that this version of analytic continuation keeps the coordinates $x^\mu$ fixed. It is purely achieved by a change of field values of the vierbein, and the associated metric, which will be a composite of vierbeins. Analytic continuation in momentum space from flat euclidean to Minkowski space is realized for $q^2=g^{\mu \nu} q_\mu q_\nu =q^\mu q_\mu, \quad q^\mu=g^{\mu \nu} q_\nu, \quad q_\mu=-i\partial_\mu=-i\frac{\partial}{\partial x^\mu}$,  by a change of the value of the composite inverse metric $g^{\mu \nu}$. For flat space one has $g^{\mu \nu}=\delta^{\mu \nu}$ or $g^{\mu \nu}=\eta^{\mu \nu}$ for euclidean and Minkowski signature, respectively.

%Absatz
Analytic continuation of the inverse vierbein $\tensor{e}{_m^\mu}$ maps $\tensor{e}{_0^\mu} \to e^{-i\varphi}\tensor{e}{_0^\mu}$. As a consequence, one has for $\varphi=\pi/2$
\begin{equation}
\tensor{e}{_0^{(E)\mu}} \to  \tensor{e}{_0^{\prime \mu}}=-i\tensor{e}{_0^{(E)\mu}}.
\label{eq:AC1} 
\end{equation} 
The Minkowski formulation of the analytically continued euclidean vierbein therefore again equals the original vierbein in the euclidean form,
\begin{equation}
\tensor{e}{_0^{\prime (M)\mu}} = i\tensor{e}{_0^{\prime \mu}}=\tensor{e}{_0^{(E)\mu}}.
\label{eq:AC2} 
\end{equation}

%Absatz
Analytic continuation also concerns the gauge fields. Similar to the inverse vierbein it multiplies the components of $A_{\mu mn}$ with one Lorentz index taking the value zero by a factor $e^{-i\varphi}$. For $\varphi=\pi/2$ this results in the map
\begin{equation}
A_{\mu k0} \to A'_{\mu k0}=-iA_{\mu k0}^{(E)}, \quad A_{\mu kl} \to A'_{\mu kl}=A_{\mu kl}^{(E)}.
\label{eq:AC3} 
\end{equation}
Again the Minkowski version $\tensor{A}{^{\prime (M)}_{\mu mn}}$ of the transformed gauge field equals the euclidean version of the original gauge field. As a consequence, the analytic continuation of the gauge bosons for the SO(4) subgroup are the gauge bosons of the subgroup SO(1,\,3). 

%Absatz
Analytic continuation only concerns the Lorentz indices. For contractions with vierbeins as $A_{\mu \nu \rho}$ the phase factors of analytic continuation drop out. The metric will transform non-trivially due to contraction with the tensor $\delta^{mn}$, such that the phase factors of $\tensor{e}{_\mu^m}\tensor{e}{_\nu^n}$ are not canceled in this case. 

%Absatz
\paragraph*{Invariant action}
The inverse vierbein $\tensor{e}{_m^\mu}$ is defined as the inverse complex matrix
\begin{equation}
\tensor{e}{_\mu^m}\;\tensor{e}{_m^\nu} = \delta_\mu^\nu,\quad \tensor{e}{_m^\mu}\;\tensor{e}{_\mu^n} = \delta_m^n.
\label{eq:8}
\end{equation}
With respect to diffeomorphisms it is a contravariant vector which transforms under SO(4,\,$\mathbb{C}$) as 
\begin{equation}
\delta \tensor{e}{_m^\mu} = \tensor{\epsilon}{_m^n}\;\tensor{e}{_n^\mu}.
\label{eq:9}
\end{equation}
The determinant of the complex matrix $\tensor{e}{_\mu^m}$ is invariant
\begin{equation}
e = \det( \tensor{e}{_\mu^m} ),\quad \delta e = 0.
\label{eq:10}
\end{equation}
With respect to diffeomorphisms $e$ transforms as a density (similar to $\sqrt{g}$ in a metric formulation). Therefore an action $S$,
\begin{equation}
S = \sum_n \int \d^4x\; e L_n,
\label{eq:11}
\end{equation}
is invariant under both diffeomorphisms and SO(4,\,$\mathbb{C}$) transformations if all $L_n$ are invariant under SO(4,\,$\mathbb{C}$) and transform as scalars under diffeomorphisms.

\paragraph*{Fermions and gauge fields}

%Absatz
A Dirac fermion transforms with respect to SO(4,\,$\mathbb{C}$) as
\begin{equation}
\delta\psi = -\frac{1}{2} \epsilon_{mn} \Sigma^{mn} \psi,
\label{eq:12}
\end{equation}
where the SO(4,\,$\mathbb{C}$) generators $\Sigma^{mn}$ and $i\Sigma^{mn}$ can be constructed from the commutator of Dirac matrices $\gamma^m$,
\begin{equation}
\{ \gamma^m, \gamma^n \} = \delta^{mn},\quad \Sigma^{mn} = -\frac{1}{4} [\gamma^m,\gamma^n].
\label{eq:13}
\end{equation}
We can again define Minkowski quantities by
\begin{align*}
\begin{split}
\gamma^{\mathrm{(M)}\,0} &= -i\gamma^0,\quad \gamma^{\mathrm{(M)}\,k} = \gamma^k, \\
\Sigma^{\mathrm{(M)}\,0k} &= -i\Sigma^{0k},\quad \Sigma^{\mathrm{(M)}\,kl} = \Sigma^{kl}, \\
\Sigma^{\mathrm{(M)}\,mn} &= -\frac{1}{4}[\gamma^{\mathrm{(M)}\,m}, \gamma^{\mathrm{(M)}\,n}], \\
\delta\psi &= -\frac{1}{2} \epsilon^\mathrm{(M)}_{mn} \Sigma^{\mathrm{(M)}\,mn} \psi.
\end{split}
\label{eq:14}
\end{align*}
Denoting
\begin{equation}
\gamma^\mu = \gamma^m \tensor{e}{_m^\mu} = \gamma^{\mathrm{(M)}\,m} e^{\mathrm{(M)}\,\mu}_m,
\label{eq:15}
\end{equation}
one finds for the euclidean vierbein $\tensor{e}{_m^\mu} = \delta_m^\mu$ that $\gamma^{\mu =0} = \gamma^0$, while for a Minkowski vierbein $e^{\mathrm{(M)}\,\mu}_m = \delta_m^\mu$ one has $\gamma^{\mu =0} = -i\gamma^0$. In the following we employ the euclidean notation.

%Absatz
We introduce the conjugate Dirac spinor $\bar{\psi}$ which transforms as
\begin{equation}
\delta \bar{\psi} = \frac{1}{2} \epsilon_{mn} \bar{\psi} \Sigma^{mn},
\label{eq:16}
\end{equation}
such that $\bar{\psi}\psi$ is invariant. With $\gamma^5 = -\gamma^0 \gamma^1 \gamma^2 \gamma^3$, $\{ \gamma^5, \gamma^m \}=0$, $[\gamma^5, \Sigma^{mn}]=0$, a Dirac spinor decomposes into two Weyl spinors
\begin{equation}
\psi_\pm = \frac{1}{2} (1\pm \gamma^5)\psi,\quad \bar{\psi}_\pm = \frac{1}{2} \bar{\psi}(1\mp \gamma^5).
\label{eq:17}
\end{equation}

%Absatz
We want to impose local SO(4,\,$\mathbb{C}$)-gauge symmetry. For the construction of a kinetic term for the fermions we need a covariant derivative involving gauge fields,
\begin{equation}
\D_\mu \psi = \partial_\mu \psi - \frac{1}{2} A_{\mu mn} \Sigma^{mn} \psi.
\label{eq:18}
\end{equation}
There are six complex gauge fields $A_{\mu mn} = -A_{\mu nm}$ corresponding to twelve real gauge fields. Their SO(4,\,$\mathbb{C}$) transformation involves an inhomogeneous part
\begin{equation}
\delta A_{\mu mn} = \tensor{\epsilon}{_m^p} A_{\mu pn} + \tensor{\epsilon}{_n^p} A_{\mu mp} - \partial_\mu \epsilon_{mn},
\label{eq:19}
\end{equation}
such that
\begin{equation}
\delta(\D_\mu \psi) = -\frac{1}{2} \epsilon_{mn} \Sigma^{mn} (\D_\mu \psi).
\label{eq:20}
\end{equation}
The bilinear
\begin{equation}
\tensor{E}{_\mu^m} = i \bar{\psi} \gamma^m \D_\mu \psi
\label{eq:21}
\end{equation}
has the same transformation property as the vierbein
\begin{equation}
\delta\tensor{E}{_\mu^m} = -\tensor{E}{_\mu^n} \tensor{\epsilon}{_n^m}.
\label{eq:22}
\end{equation}
We observe that our construction is compatible with the possibility that in a more fundamental theory as spinor gravity the vierbein is the expectation value of a fermion bilinear, $\tensor{e}{_\mu^m} = \braket{\tensor{E}{_\mu^m}}$.

%Absatz
Eq.\,\eqref{eq:22} allows us to construct a gauge invariant kinetic term for fermions
\begin{equation}
L_\psi = \tensor{E}{_\mu^m} \tensor{e}{_m^\mu} = i \bar{\psi}\gamma^m \tensor{e}{_m^\mu} \D_\mu \psi = i \bar{\psi} \gamma^\mu \D_\mu \psi.
\label{eq:23}
\end{equation}
With respect to general coordinate transformations $\psi$ and $\bar{\psi}$ transform as scalars, $A_{\mu mn}$ and $\D_\mu\psi$ as covariant vectors and $\tensor{e}{_m^\mu}$ as a contravariant vector. Thus $L_\psi$ is indeed a scalar. The Grassmann variables $\psi$ and $\bar{\psi}$ are independent. For given values of the vierbein one can combine them into a complex Grassmann variable. The associated complex structure depends on the signature \cite{CWES}.

\paragraph*{Field strength and kinetic term for gauge bosons}

%Absatz
For the construction of a gauge invariant kinetic term for the gauge fields we introduce the complex field strength
\begin{equation}
F_{\mu\nu mn} = \partial_\mu A_{\nu mn} - \partial_\nu A_{\mu mn} + \tensor{A}{_\mu_m^p} A_{\nu pn} - \tensor{A}{_\nu_m^p} A_{\mu pn}.
\label{eq:24}
\end{equation}
It is an antisymmetric tensor with respect to diffeomorphisms and with respect to SO(4,\,$\mathbb{C}$),
\begin{equation}
\delta F_{\mu\nu mn} = \tensor{\epsilon}{_m^p} F_{\mu\nu pn} + \tensor{\epsilon}{_n^p} F_{\mu\nu mp}.
\label{eq:25}
\end{equation}
An SO(4,\,$\mathbb{C}$) invariant tensor is obtained by multiplication with the vierbein
\begin{equation}
F_{\mu\nu\rho\sigma} = F_{\mu\nu mn} \tensor{e}{_\rho^m} \tensor{e}{_\sigma^n} = -F_{\nu\mu\rho\sigma} = -F_{\mu\nu\sigma\rho}.
\label{eq:26}
\end{equation}

%Absatz
Contracting Lorentz-indices with the vierbein is a rather general way to obtain SO(4,\,$\mathbb{C}$) invariants. We note, however, that $F^*_{\mu\nu mn}$ transforms differently as compared to $F_{\mu\nu mn}$ since the transformation coefficients $\tensor{\epsilon}{_m^p}$ are complex. Constructing an invariant tensor $F^*_{\mu\nu \rho \sigma}$ involves contractions with the complex conjugate vierbein $e_\rho^{*m} e_\sigma^{*n}$.

%Absatz
Having at our disposal SO(4,\,$\mathbb{C}$)-invariant terms, we need to construct scalars with respect to diffeomorphisms. For this purpose we define the complex pseudo-metric as a bilinear of the complex vierbeins,
\begin{align}
\begin{split}
G_{\mu\nu} = \tensor{e}{_\mu^m} \tensor{e}{_\nu^n} \delta_{mn}&,\quad G^{\mu\nu} = \tensor{e}{_m^\mu} \tensor{e}{_n^\nu} \delta^{mn}, \quad G_{\mu\nu} G^{\nu\rho} = \tensor{\delta}{_\mu^\rho}, \\
G_{\mu\nu} \tensor{e}{_\nu^m} = \delta^{m n} \tensor{e}{_n^\mu} &= \tensor{e}{^{m\mu}} ,\quad \tensor{e}{_m^\mu} G_{\mu\nu}=\tensor{e}{_\nu^n}\delta_{n m}=\tensor{e}{_{\nu m}}.
\end{split}
\label{eq:27}
\end{align}
(Care is needed for the index position of objects carrying both world and Lorentz-indices.)
The pseudo-metric is a symmetric complex tensor which is invariant under SO(4,\,$\mathbb{C}$)-transformations. It can be used to lower and raise space-time indices
\begin{equation}
\tensor{F}{_\mu^\nu_\rho^\sigma} = G^{\nu\alpha} G^{\sigma\beta} F_{\mu\alpha\rho\beta}.
\label{eq:28}
\end{equation}

%Absatz
A kinetic invariant for the gauge fields can be constructed as
\begin{align}
\begin{split}
L_F &= L_F^* = \frac{Z_F}{8} F_\mu^*{}\indices{^\nu_\rho^\sigma} \tensor{F}{_\nu^\mu_\sigma^\rho} \\
  &= \frac{Z_F}{8} F^*_{\mu\nu\rho\sigma} F_{\alpha\beta\gamma\delta} G^{*\mu\alpha} G^{\nu\beta} G^{*\rho\gamma} G^{\sigma\delta}.
\end{split}
\label{eq:29}
\end{align}
For this particular structure with the complex conjugate field strength the kinetic terms for all gauge bosons are positive for euclidean space. This demonstrates in a simple way that non-compact gauge groups do not necessarily lead to ghosts. The positivity would not hold if one would use a real metric $g^{\mu\nu}$ for index contractions.

%Absatz
The real positive parameter $Z_F$ plays the role of an inverse squared gauge coupling, as in usual Yang-Mills theories. The kinetic term can be brought to a standard normalization by a rescaling $A_{\mu mn} = \sqrt{Z_F} A^R_{\mu mn}$. The renormalized field $A^R_\mu$ with standard normalization enters the covariant derivative of the fermions, as well as all other covariant derivatives, in the form $\partial_\mu - Z_F^{-1/2} A^R_\mu = \partial_\mu - g A^R_\mu$, with $g$ the gauge coupling. For large $Z_F$ the gauge coupling is small. The theory is weakly interacting and perturbative methods may, at least partially, apply. The invariant $L_F$ is not the only possible invariant constructed with up to two derivatives of the gauge fields. Other invariants can be found by different index contractions and different combinations of $F^*F$, $FF$, $F^*F^*$, and will be discussed later. Possible invariants also include an invariant which is linear in $F$.

\paragraph*{Kinetic term for vierbein}

%Absatz
We next need a kinetic term for the complex vierbein field $\tensor{e}{_\mu^m}$. For this purpose we employ the SO(4,\,$\mathbb{C}$)-invariant tensor
\begin{align}
\label{eq:29A}
& U_{\mu\nu\rho} = -U_{\mu\rho\nu} = \frac{1}{2} \left\{ e_{\mu m} (\partial_\rho \tensor{e}{_\nu^m} - \partial_\nu \tensor{e}{_\rho^m} ) \right. \\
\nonumber
&+ e_{\nu m} \left.( \partial_\rho \tensor{e}{_\mu^m} - \partial_\mu \tensor{e}{_\rho^m} ) - e_{\rho m} (\partial_\nu \tensor{e}{_\mu^m} - \partial_\mu \tensor{e}{_\nu^m} )  \right\} - A_{\mu\nu\rho}
\end{align}
where
\begin{equation}
A_{\mu\nu\rho} = A_{\mu mn} \tensor{e}{_\nu^m} \tensor{e}{_\rho^n}.
\label{eq:29B}
\end{equation}
It can be related to the covariant derivative of the vierbein \cite{CWGG},
\begin{equation}
\tensor{U}{_\mu_\nu^m} = \D_\mu \tensor{e}{_\nu^m} = \partial_\mu \tensor{e}{_\nu^m} - \tensor{\Gamma}{_\mu_\nu^\sigma} \tensor{e}{_\sigma^m} + \tensor{A}{_\mu^m_n} \tensor{e}{_\nu^n}.
\label{eq:30}
\end{equation}
Here $\tensor{\Gamma}{_\mu_\nu^\sigma}$ is constructed from $\tensor{e}{_\mu^m}$ via the pseudo-metric $G_{\mu\nu}$,
\begin{equation}
\tensor{\Gamma}{_\mu_\nu^\sigma} = \frac{1}{2} G^{\sigma\rho} (\partial_\mu G_{\nu\rho} + \partial_\nu G_{\mu\rho} - \partial_\rho G_{\mu\nu}).
\label{eq:31}
\end{equation}
It generalizes the concept of a geometric connection to a complex composite pseudo-metric. A SO(4,\,$\mathbb{C}$) invariant tensor obtains as
\begin{align}
\begin{split}
U_{\mu\nu\rho} &= e_{\rho m} \tensor{U}{_\mu_\nu^m} =  \tensor{U}{_\mu_\nu^m} \delta_{mn} \tensor{e}{_\rho^n}\\
&= e_{\rho m} \partial_\mu \tensor{e}{_\nu^m} - \Gamma_{\mu\nu\rho} + A_{\mu\rho\nu},
\end{split}
\label{eq:32}
\end{align}
where
\begin{equation}
\Gamma_{\mu\nu\rho} = \frac{1}{2} ( \partial_\mu G_{\nu\rho} + \partial_\nu G_{\mu\rho} - \partial_\rho G_{\mu\nu} ).
\label{eq:32A}
\end{equation}
Inserting $G_{\mu\nu} = e_{\mu m}\tensor{e}{_\nu^m}$ one finds eq.\,\eqref{eq:29A}. 

%Absatz
We define a kinetic invariant for the vierbein by
\begin{align}
\label{eq:32C}
L_U &= \frac{m^2}{8} ( U^*_\mu{}\indices{^\nu_\rho} \tensor{U}{^\mu_\nu^\rho} + U^*{}\indices{^\mu_\nu^\rho} \tensor{U}{_\mu^\nu_\rho} ) \\
\nonumber
&= \frac{m^2}{8} U^*_{\mu\nu\rho} U_{\alpha\beta\gamma} ( G^{\mu\alpha} G^{*\nu\beta} G^{\rho\gamma} + G^{*\mu\alpha} G^{\nu\beta} G^{*\rho\gamma} ).
\end{align}
Again, this is not the unique possible invariant and will be supplemented by other invariants later.

%Absatz
At this point we may briefly summarize our geometric setting and clarify the relation to other formulations of metric affine gravity \cite{HCM,HEI,RP,PS}. Consider first the case of real vierbeins $\tensor{e}{_\mu^m}$, real gauge fields $A_{\mu m n}$, and therefore also a real metric $g_{\mu \nu}$ instead of $G_{\mu \nu}$. We employ two different connections, the metric connection $\tensor{\Gamma}{_{\mu \nu}^\rho}$, which is the Levi-Civita connection expressed in terms or the vierbein, and a gauge connection, given by the gauge fields  $A_{\mu m n}$. In principle, the corresponding fibre bundles need not to be related. This is the usual setting for Yang-Mills theories coupled to gravity. The geometric connection shows neither torsion nor non-metricity.

%Absatz
We could introduce a different connection
\begin{equation}
\tensor{B}{_{\mu \nu}^\rho}=\tensor{\Gamma}{_{\mu \nu}^\rho} +\tensor{U}{_{\mu \nu}^\rho}.
\label{eq:AA1}
%\tag*{AA1}
\end{equation}
With this new geometric connection the vierbein is covariantly conserved,
\begin{equation}
\tilde{D}_\mu \tensor{e}{_\nu^m}=\partial_\mu  \tensor{e}{_\nu^m}-\tensor{B}{_{\mu \nu}^\rho} \tensor{e}{_\rho^m} +\tensor{A}{_\mu^m_n} \tensor{e}{_\nu^n}=0.
\label{eq:AA2}
%\tag*{AA2}
\end{equation}
The connection $\tensor{B}{_{\mu \nu }^\rho}$ leads to torsion
\begin{equation}
\tensor{B}{_{\mu \nu }^\rho}- \tensor{B}{_{\nu \mu }^\rho}=\tensor{U}{_{\mu \nu }^\rho}- \tensor{U}{_{\nu \mu }^\rho},
\label{eq:AA3}
%\tag*{(AA3)}
\end{equation}
while non-metricity is still absent, according to 
\begin{align}
\begin{split}
 \tilde{D}_\mu g_{\nu\rho}&=D_\mu g_{\nu \rho}- (\tensor{U}{_{\mu \nu \rho}}+\tensor{U}{_{\mu \rho \nu}})=-(\tensor{U}{_{\mu \nu \rho}}+\tensor{U}{_{\mu \rho \nu}})=0, \\
 D_\mu g_{\nu\rho}&=(D_\mu \tensor{e}{_\nu^m})\tensor{e}{_{\rho m}}+(D_\mu \tensor{e}{_{\rho m}})\tensor{e}{_\nu^m}=\tensor{U}{_{\mu \nu \rho}}+\tensor{U}{_{\mu \rho \nu}}=0.
\label{eq:AA4}
\end{split}
%\tag*{(AA4)}
\end{align}
We find it more convenient to keep the usual language of Yang-Mills-theories without torsion and $D_\mu \tensor{e}{_\nu^m}\ne0$. This setting is extended to complex vierbeins, gauge fields, $G_{\mu \nu}$  and $\tensor{\Gamma}{_{\mu \nu}^\rho}$. The usual product rules for covariant derivatives apply, where $\tensor{\Gamma}{_{\mu \nu}^\rho}$ appears for each world index of a generalized tensor, and $\tensor{A}{_\mu^m_p}$ for each Lorentz-index. An example is 
\begin{align}
\begin{split}
D_\mu \tensor{U}{_{\nu \rho \sigma}}&=D_\mu(\tensor{U}{_{\nu \rho}^m} \tensor{e}{_{\sigma m}})\\&=(D_\mu \tensor{U}{_{\nu \rho}^m})\tensor{e}{_{\sigma m}}+\tensor{U}{_{\nu \rho}^m}D_\mu(\tensor{e}{_{\sigma m}}),
\label{eq:AA5}
\end{split}
%\tag*{(AA5)}
\end{align} 
where the left hand side only involves the geometric connection. We observe that index conversion between latin and greek indices and covariant differentiation do not commute,
\begin{equation}
\tensor{e}{_{\sigma m}} D_\mu \tensor{U}{_{\nu \rho}^m} = D_\mu \tensor{U}{_{\nu \rho \sigma}}-\tensor{U}{_{\nu \rho}^m} \tensor{U}{_{\mu \sigma m}}.
\label{eq:AA6}
%\tag*{(AA6)}
\end{equation}
This is a direct consequence of the non-vanishing covariant derivative of the generalized vierbein.

\section{Solutions of field equations and mode expansion\label{sec:4}}

%Absatz
In this section we address the question if a classical action based on $L_{F}+L_U+L_\psi$ can be used for a consistent definition of a functional integral. We want to avoid that already the quadratic approximation in the vicinity of flat euclidean space with vanishing gauge fields needs to an action that diverges to minus infinity as the field values grow. We will find that this stability property is indeed realized. Beyond the positive eigenvalues of the second functional derivative of the action we also observe zero eigenvalues. Those include, of course, the gauge fluctuations which can be stabilized by a suitable gauge fixing. For the action based on $L_F+L_U$ there are also some physical fluctuation modes that do not have a kinetic term. In the next section we add an invariant that cures this disease.

%Absatz
The field equations derived from an action based on $L_F+L_U+L_\psi$ admit flat space as a solution, both with euclidean and Minkowski signature. The gauge fields vanish for this solution. In order to investigate the stability of the high-momentum fluctuations of our model we can employ a mode expansion around these flat space solutions. We base this analysis first on $L_F+L_U+L_\psi$ and add further invariants subsequently. 

\paragraph*{Field equations and solutions}

%Absatz
The field equations obtain from the action
\begin{equation}
S = \int \d^4x\; e(L_\psi + L_F + L_U)
\label{eq:33}
\end{equation}
by variation with respect to $\bar{\psi}$, $\tensor{e}{_\mu^m}$ and $A_{\mu mn}$. For the fermions one finds the Dirac equation
\begin{equation}
\gamma^\mu \D_\mu \psi = 0.
\label{eq:34}
\end{equation}
Since fermions have no expectation values we omit $L_\psi$ in the following. (Incoherent fermion fluctuations can provide for source terms to the field equations for the bosonic field. We will omit this here.)

%Absatz
The field equation for the gauge fields reads
\begin{equation}
e(\D_\mu \tensor{F}{^\nu_{\mu mn}} - J_{\mu mn} ) = 0,
\label{eq:35}
\end{equation}
where
\begin{align}
\begin{split}
\D_\nu \tensor{F}{^\nu_{\mu mn}} &=\partial_\nu \tensor{F}{^\nu_{\mu m n}} - \tensor{\Gamma}{_{\nu \mu}^\rho} \tensor{F}{^\nu_{\rho m n}}\\+\tensor{\Gamma}{_{\nu \rho}^\nu}  &\tensor{F}{^\rho_{\mu m n}}+\tensor{A}{_\nu_m^p} \tensor{F}{^\nu _{\mu p n}}+\tensor{A}{_{\nu n}^p} \tensor{F}{^\nu_{\mu m p}}.
\end{split}
\label{eq:36}
\end{align}
For the vierbein one obtains
\begin{equation}
e(\D_\nu \tensor{U}{^\nu_\mu^m} - \tensor{K}{_\mu^m} ) = 0.
\label{eq:37}
\end{equation}
The source terms $J_{\mu mn}$ and $K_{\mu m}$ vanish for $F_{\mu\nu mn}=0$ and $\tensor{U}{_\mu_\nu^m}=0$. This is realized for $A_{\mu mn}=0$ and $\tensor{e}{_\mu^m}$ independent of $x$. In this case one also has $\D_\nu\tensor{F}{^\nu_{\mu mn}}=0$, $\D_\nu \tensor{U}{^\nu_\mu^m}=0$. As a result we find a family of solutions with constant vierbein and vanishing gauge fields
\begin{equation}
\partial_\nu \tensor{e}{_\mu^m}=0,\quad A_{\mu mn}=0.
\label{eq:38}
\end{equation}
In particular, this includes flat euclidean space $\tensor{e}{_\mu^m}=\delta_\mu^m$, as well as flat Minkowski space $e_{~~\mu}^\mathrm{(M)}{}\indices{^m} = \delta_\mu^m$. 

%Absatz
For any given choice of a non-vanishing constant vierbein the gauge group SO(4,\,$\mathbb{C}$) is spontaneously broken. Also diffeomorphism symmetry is spontaneously broken. For flat euclidean space the vierbein $\tensor{e}{_\mu^m}=\delta_\mu^m$ is left invariant by global SO(4)-transformations which combine particular SO(4,\,$\mathbb{C}$)-transformations with suitable general coordinate transformations. In case of flat Minkowski space this global symmetry group is replaced by the Lorentz group SO(1,\,3). The field equations derived from the classical action $L_{F}+L_{U}$ provide an example for which the difference between space and time is a result of spontaneous symmetry breaking \cite{CWTS}. We observe that $\tensor{e}{_\mu^m}=0$, $G_{\mu\nu}=0$, $A_{\mu mn}=0$ is also a solution of the field equations.

\paragraph*{Mode expansion for flat space}

%Absatz
For an investigation of stability we expand the action \eqref{eq:33} around a flat space solution with constant $\tensor{\bar{e}}{_\mu^m}$ and $\bar{A}_{\mu mn}=0$. Choosing the parametrization 
\begin{equation}
\tensor{e}{_\mu^m} = \tensor{\bar{e}}{_\mu^m} + \frac{1}{2}\tensor{H}{_\mu^\nu} \tensor{\bar{e}}{_\nu^m},
\label{eq:E1}
\end{equation}
we expand in second order in $H$ and $A$. Since $F_{\mu\nu mn}$ is at least linear in $A$ we can employ in $eL_F$ the lowest order $H=0$ for the vierbein and replace covariant derivatives by partial derivatives
\begin{align}
\begin{split}
eL_F = \frac{Z_F \bar{e}}{8} & \bar{G}^{*\mu\alpha}\bar{G}^{\nu\beta} ( \partial_\mu A^*_\nu{}\indices{^\rho_\sigma} - \partial_\nu A^*_\mu{}\indices{^\rho_\sigma} ) \\
&\times( \partial_\alpha \tensor{A}{_\beta_\rho^\sigma} - \partial_\beta \tensor{A}{_\alpha_\rho^\sigma} ).
\end{split}
\label{eq:E2}
\end{align}
Here the indices are changed with $\tensor{\bar{e}}{_\mu^m}$ and $\bar{G}_{\mu\nu}$. For euclidean flat space one has $\bar{G}^{\mu\alpha} = \delta^{\mu\alpha}$ and $\bar{e}=1$. 

%Absatz
For constant real $G_{\mu \nu} $ the expression 
\begin{align}
\begin{split}
eL_F^\mathrm{(E)} = \frac{Z_F \bar{e}}{8} 
	( \partial_\mu A_{\nu\rho\sigma} - \partial_\nu A_{\mu\rho\sigma} )^* 
	( \partial^\mu A^{\nu\rho\sigma} - \partial^\nu A^{\mu\rho\sigma} ).
\end{split}
\label{eq:E3}
\end{align}
yields in momentum space, with $q^2 = q^\mu q_\mu=q_\mu q_\nu G^{\mu \nu}$, a kinetic term both for the ``real'' and ``imaginary'' gauge bosons
\begin{equation}
\int_x eL_F = \frac{Z_F \bar{e}}{4} \int_q A^*_{\mu\rho\sigma} ( q^2 \bar{G}^{\mu\nu} - q^\mu q^\nu ) \tensor{A}{_\nu^{\rho\sigma}}.
\label{eq:E4}
\end{equation}
For flat euclidean space, $G^{\mu \nu}=\delta^{\mu \nu}$ this term is positive semidefinite. 

%Absatz
Indeed, decomposing $A$ into its real and imaginary parts $A_{\mu\rho\sigma} = A_{\mathrm{R},\mu\rho\sigma} + iA_{\mathrm{I},\mu\rho\sigma}$ and using a multi-index $z=(\chi,\rho,\sigma)$, $\chi=(\mathrm{R,I})$, $z=1...12$, this yields for euclidean space, $\tensor{\bar{e}}{_\mu^m}=\delta_\mu^m$, standard kinetic terms for all twelve gauge bosons $A_\mu^z$
\begin{equation}
\int_x eL_F = \frac{Z_F}{2} \sum_z \int_q A_\mu^z (q^2\delta^{\mu\nu} - q^\mu q^\nu ) A_\nu^z.
\label{eq:E5}
\end{equation}
The sector of gauge fields alone is stable as in usual gauge theories. In the quadratic approximation it describes twelve massless free photons. In flat Minkowski space one has $\bar{G}^{\mu\nu} = \eta^{\mu\nu}$ and $\bar{e}=i$. The expressions \eqref{eq:E3}, \eqref{eq:E4} remain the same, with indices raised and lowered with $\eta$. 

%Absatz
For $L_U$ we need $U_{\mu\nu\rho}$ in linear order in $H$ and $A$
\begin{equation}
U_{\mu\nu\rho} = \frac{1}{2} \left\{ \partial_\mu H_{\nu\rho}^\mathrm{(A)} + \partial_\rho H_{\mu\nu}^\mathrm{(S)} - \partial_\nu H_{\mu\rho}^\mathrm{(S)} \right\} - A_{\mu\nu\rho},
\label{eq:E6}
\end{equation}
where we have decomposed $H_{\mu\nu} = \tensor{H}{_\mu^\sigma} \bar{G}_{\sigma\nu}$ into its symmetric and antisymmetric parts
\begin{equation}
H_{\mu\nu}^\mathrm{(S)} = \frac{1}{2} (H_{\mu\nu} + H_{\nu\mu}),\quad H_{\mu\nu}^\mathrm{(A)} = \frac{1}{2}( H_{\mu\nu} - H_{\nu\mu} ).
\label{eq:E7}
\end{equation}
The quadratic approximation to $L_U$ contains three terms, $L_U = L_U^{(1)} + L_U^{(2)} + L_U^{(3)}$. The first term is a kinetic term for $H$
\begin{align}
\nonumber
L_U^{(1)} = &\frac{m^2}{32} \left\{ 
	( \partial_\mu \tensor{H}{^{\mathrm{(A)}\nu}_\rho} )^* 
	\partial^\mu H^\mathrm{(A)}_{~~\nu}{}\indices{^\rho} \right. \\ 
\nonumber
	&+ ( \partial_\rho H^\mathrm{(S)}_{~~\mu}{}\indices{^\nu} 
			- \partial^\nu H^\mathrm{(S)}_{\mu\rho} )^* 
		( \partial^\rho \tensor{H}{^{\mathrm{(S)}\mu}_\nu} 
			- \partial_\nu H^{\mathrm{(S)}\mu\rho} ) \\ 
%\nonumber\label{eq:E8}
	&+ ( \partial_\mu \tensor{H}{^{\mathrm{(A)}\nu}_\rho} )^*
		( \partial^\rho \tensor{H}{^{\mathrm{(S)}\mu}_\nu} 
			- \partial_\nu H^{\mathrm{(S)}\mu\rho} ) \\ 
\nonumber
	&+ \left.( \partial^\mu H^\mathrm{(A)}_{~~\nu}{}\indices{^\rho} )^*
		( \partial_\rho H^\mathrm{(S)}_{~~\mu}{}\indices{^\nu} 
			- \partial^\nu H^\mathrm{(S)}_{\mu\rho} ) + \mathrm{c.\,c.} \right\} 
\end{align}
For real $\bar{G}_{\mu\nu}$ and using partial integration this simplifies to
\begin{align}
\begin{split}
L_U^\mathrm{(1)} &= -\frac{m^2}{16} \Bigl\{  H^{\mathrm{(A)}*}_{\mu\nu} \partial^2 H^{\mathrm{(A)}\mu\nu}  \\ 
 &  + 2H^{\mathrm{(S)}*}_{\mu\nu} ( \partial^2 \delta_\rho^\nu - \partial^\nu  \partial_\rho ) H^{\mathrm{(S)}\rho\mu} \\ 
 &+  [ 2H^{\mathrm{(A)}*}_{\mu\nu} \partial^\nu \partial_\rho H^{\mathrm{(S)}\rho\mu} + \mathrm{c.\,c.} ] \Bigr\}, 
\end{split}
\label{eq:E9}
\end{align}
where $\partial^2=\bar{G}_{\mu\nu}\partial^\mu \partial^\nu$.

%Absatz
The second term is a mass term for the gauge bosons
\begin{equation}
L_U^{(2)} = \frac{m^2}{8} ( A^*_\mu{}\indices{^\nu_\rho} \tensor{A}{^\mu_\nu^\rho} + \mathrm{c.\,c.} ).
\label{eq:E10}
\end{equation}
For real $\bar{G}_{\mu\nu}$ this provides a mass for all twelve gauge bosons
\begin{equation}
L_U^{(2)} = \frac{m^2}{4} A^*_{\mu\nu\rho} A^{\mu\nu\rho} = \frac{m^2}{2} \sum_z A^z_\mu A^z_\nu \bar{G}^{\mu\nu},
\label{eq:E11}
\end{equation}
where the last identity holds for euclidean space. For $\bar{e}\neq 0$ the pseudo-metric $\bar{G}_{\mu\nu}$ has no zero eigenvalues. All gauge bosons acquire a mass. For euclidean flat space all twelve gauge bosons have the same mass $m$. The massive gauge bosons are a direct consequence of the spontaneous breaking of the SO(4,\,$\mathbb{C}$)-symmetry by a constant vierbein $\tensor{\bar{e}}{_\mu^m} \neq 0$, $\bar{e} \neq 0$. The local symmetry is broken completely, with no local unbroken subgroup left. Therefore all gauge bosons become massive. The mechanism is the analogue to the Higgs mechanism, with the difference that $\tensor{e}{_\mu^m}$ are covariant vectors and not scalars.

%Absatz
The third part is a source term for the gauge bosons
\begin{align}
\begin{split}   
L_U^{(3)} = -\frac{m^2}{8} \Bigl\{ & A^*_\mu{}\indices{^\nu_\rho} 
( \partial^\mu H^\mathrm{(A)}_{~~\nu}{}\indices{^\rho} 
+ \partial^\rho H^{\mathrm{(S)}\mu}{}\indices{_\nu} \\ 
& - \partial_\nu H^{\mathrm{(S)}\mu\rho} ) + \mathrm{c.\,c.}  \Bigr\} \\
= -\frac{1}{2} ( & A^{*\mu  mn}  J_{\mu mn} + \mathrm{c.\,c.} ).
\end{split}
\label{eq:E12}
\end{align}
The source,
\begin{align}
J_{\mu mn} = \frac{m^2}{4} \bar{G}^*_{\mu\alpha} \bar{e}^*_m{}\indices{^\nu} \bar{e}^*_{\rho n} 
	( \partial^\alpha H^\mathrm{(A)}_{~~\nu}{}\indices{^\rho} 
		+ \partial^\rho \tensor{H}{^{\mathrm{(S)}\alpha}_\nu} 
		- \partial_\nu H^{\mathrm{(S)}\alpha\rho} ),
\label{eq:E13}
\end{align}
is the linear approximation to eq.\,\eqref{eq:35}.

\paragraph*{High momentum limit}

%Absatz
The short-distance limit corresponds to momenta going to infinity. For smooth ``background'' vierbeins and pseudo-metrics $\bar{e}\indices{_\mu^m}(x)$, $\bar{G}_{\mu\nu}(x)$ one can neglect in this limit the dependence on $x$ and approximate them by constant $\bar{e}\indices{_\mu^m}$ and $\bar{G}_{\mu\nu}$, as discussed previously. In the high momentum limit we can also neglect the source term and the mass for the gauge bosons. For the euclidean theory we find in this limit a consistent theory for twelve photons. The high momentum limit may be associated to $q^2/m^2 \gg 1$.

%Absatz
The kinetic term $L_U^{(1)}$ for the complex vierbein fluctuation $H_{\mu\nu}$ needs a more detailed discussion. We consider real $\bar{G}_{\mu\nu}$ and decompose $H^\mathrm{(A)}_{\mu\nu}$ and $H^\mathrm{(S)}_{\mu\nu}$ as \cite{CWMF}
\begin{align}
\begin{split}
H^\mathrm{(S)}_{\mu\nu} &= T_{\mu\nu} + \partial_\mu K_\nu + \partial_\nu K_\mu \\
& + \frac{1}{3} \left( \bar{G}_{\mu\nu} - \frac{\partial_\mu \partial_\nu}{\partial^2} \right) S + \frac{\partial_\mu \partial_\nu}{\partial^2} U, \\
H^\mathrm{(A)}_{\mu\nu} &= B_{\mu\nu} + \partial_\mu C_\nu - \partial_\nu C_\mu,
\end{split}
\label{eq:L1}
\end{align}
with
\begin{align}
\nonumber
\partial^\nu T_{\mu\nu} = \partial^\mu T_{\mu\nu} = 0&, \;\; T_{\mu\nu} \bar{G}^{\mu\nu}=0, \;\; \partial^\nu B_{\mu\nu} = \partial^\mu B_{\mu\nu}=0, \\
\label{eq:L2}
\partial^\mu K_\mu = 0&, \;\; \partial^\mu C_\mu = 0,
\end{align}
and
\begin{align}
\nonumber
\partial^\mu H^\mathrm{(S)}_{\mu\nu} &= \partial^2 K_\nu + \partial_\nu U, \quad \partial^\mu H^\mathrm{(A)}_{\mu\nu} = \partial^2 C_\nu, \\
\label{eq:L3}
( \partial^2 \delta_\mu^\rho &- \partial_\mu \partial^\rho ) H^\mathrm{(S)}_{\rho\nu} 
	= \partial^2 T_{\mu\nu} + \partial^2 \partial_\nu K_\mu \\
\nonumber	
	&\qquad\qquad\qquad+ \frac{1}{3} (  \bar{G}_{\mu\nu}\partial^2 - \partial_\mu \partial_\nu ) S.
\end{align}
Inserting these relations yields in momentum space yields
\begin{align}
\begin{split}
L_U^{(1)} = &\int_q \frac{m^2\bar{e}}{16} \left\{  \vphantom{\dfrac{0}{0}}
2T^*_{\mu\nu} q^2 T^{\mu\nu} + B^*_{\mu\nu} q^2 B^{\mu\nu} \right. \\
&\quad + \left. 2( K^*_\mu - C^*_\mu ) q^4 ( K^\mu - C^\mu ) + \frac{2}{3} S^* q^2 S \right\}.
\end{split}
\label{eq:L4}
\end{align}
In euclidean space all terms are positive, diverge for $q^2 \to \infty$, and vanish for $q^2 \to 0$. These are standard kinetic terms for both the real and imaginary parts of $T_{\mu\nu}$, $B_{\mu\nu}$, $K_\mu - C_\mu$ and $S$. (The increase $\sim q^4$ for $K_\mu - C_\mu$ is due to the normalization convention. By a different normalization of $K_\mu$ and $C_\mu$ the kinetic term increases $\sim q^2$, as expected for an action involving two derivatives.) We observe that $L_U^{(1)}$ does not involve the combinations $K_\mu + C_\mu$ and $U$.

%Absatz
We could add further kinetic invariants formed by different contractions of indices. For example, the vector
\begin{align}
\begin{split}
\tilde{U}_\rho &= G^{\mu\nu} U_{\mu\nu\rho} \\
	&= \frac{1}{2} ( \partial^\mu H^\mathrm{(A)}_{\mu\rho} + \partial_\rho \bar{G}^{\mu\nu} H^\mathrm{(S)}_{\mu\nu} - \partial^\mu H^\mathrm{(S)}_{\mu\rho} ) - \tensor{A}{^\mu_{\mu\rho}} \\
	&= \frac{1}{2} \left[ \partial_\rho S - \partial^2 ( K_\rho - C_\rho ) \right] - \tensor{A}{^\mu_{\mu\rho}}
\end{split}
\label{eq:L5}
\end{align}
involves the scalar $S$ and the vectors $K_\rho - C_\rho$ and $\tensor{A}{^\mu_{\mu\rho}}$. An invariant term $\sim \tilde{U}^*_\rho \tilde{U}^\rho + \tilde{U}_\rho \tilde{U}^{*\rho}$ modifies the kinetic term, mass term and source term for these fields. We observe that $\tilde{U}_\rho$ does not involve the fields $K_\mu + C_\mu$ or $U$. The same holds for the antisymmetric combination
\begin{align}
\nonumber
U_{\mu\nu\rho} &- U_{\nu\mu\rho} = ( \partial_\mu \tensor{e}{_\nu^m} - \partial_\nu \tensor{e}{_\mu^m} ) e_{\rho m} - A_{\mu\nu\rho} + A_{\nu\mu\rho} \\
\nonumber
	&= \frac{1}{2} ( \partial_\mu H_{\nu\rho} - \partial_\nu H_{\mu\rho} ) - ( A_{\mu\nu\rho} - A_{\nu\mu\rho} ) \\
\label{eq:L6}
	&= \frac{1}{2} \{ \partial_\mu T_{\nu\rho} - \partial_\nu T_{\mu\rho} + \partial_\mu B_{\nu\rho} - \partial_\nu B_{\mu\rho}  \\
\nonumber
	&\qquad + \left. \partial_\rho [ \partial_\mu ( K_\nu - C_\nu ) - \partial_\nu ( K_\mu - C_\mu ) ] \right\}  \\
\nonumber
	&\qquad + \frac{1}{6} ( \bar{G}_{\nu\rho} \partial_\mu S - \bar{G}_{\mu\rho} \partial_\nu S ) - ( A_{\mu\nu\rho} - A_{\nu\mu\rho} ).
\end{align}
In fact, with the linear expansion
\begin{align}
\begin{split}
U_{\mu\nu\rho} = \frac{1}{2} &\left\{ \vphantom{\frac{0}{0}}
	\partial_\rho T_{\mu\nu} - \partial_\nu T_{\mu\rho} + \partial_\mu B_{\nu\rho} \right. \\
	&\quad - \partial_\mu [ \partial_\nu ( K_\rho - C_\rho ) - \partial_\rho ( K_\nu - C_\nu ) ] \\
	&\quad \left. +\frac{1}{3} ( \bar{G}_{\mu\nu} \partial_\rho S - \bar{G}_{\mu\rho} \partial_\nu S ) \right\} - A_{\mu\nu\rho},
\end{split}
\label{eq:L7}
\end{align}
it is clear that the fields $K_\mu + C_\mu$ and $U$ can never appear in quadratic order in any invariant constructed from $U_{\mu\nu\rho}$. The detailed mode decomposition of the gauge fields $A_{\mu \nu \rho}$ can be found in the appendix \ref{sec: Ap.A}.

\paragraph*{Physical and gauge modes}

%Absatz
At this stage several fields do not have a kinetic term and therefore no valid propagator: 
\begin{enumerate*}[label=(\roman*)]
%\begin{inparaenum}[i)]
\item The longitudinal components of the gauge bosons. These are the gauge degrees of freedom of the local SO(4,\,$\mathbb{C}$)-gauge group for a pure Yang-Mills theory without additional matter fields.
\item The real part of the transversal vectors $K_\mu + C_\mu$ and the real part of $U$. These four degrees of freedom are the gauge degrees of freedom of diffeomorphisms.  The absence of a kinetic term for the gauge degrees of freedom is a direct consequence of gauge symmetry and not a problem. A valid continuum formulation can be implemented by the usual gauge fixing procedure.
\item Beyond the gauge degrees of freedom another four degrees of freedom without a kinetic term are the imaginary parts of $K_\mu + C_\mu$ and $U$. 
\end{enumerate*}

%Absatz
We will see in the next section that these fields actually acquire a kinetic term if we include additional invariants.
%\end{inparaenum}
In particular, we can add an invariant $L_{W}$, cf. eq.~\eqref{eq:M23A}, such that the combination $L_{F}+L_{U}+L_{ W}$ leads for euclidean flat space to positive kinetic terms that increase $\sim q^{2}$ for all modes. This combination is a candidate for a classical action that defines a well defined functional integral.

\section{Invariants for the composite pseudo-metric\label{sec:5}}

%Absatz
A metric is a real symmetric second rank tensor that is invariant under generalized Lorentz transformations, in our case invariant under SO(4,\,$\mathbb{C}$). It can be constructed as a bilinear of the complex vierbein. In our complex formulation the SO(4,\,$\mathbb{C}$) invariant vierbein-bilinears form a complex pseudo-metric. The imaginary part of the pseudo-metric constitutes an additional tensor which permits the construction of further kinetic terms. The additional kinetic terms provide for a well behaved propagator for all physical excitations. 

\paragraph*{Complex pseudo-metric}

%Absatz
The composite pseudo metric $G_{\mu\nu}$ is a bilinear in the complex vierbein 
\begin{equation}
\label{84A}
G_{\mu\nu}=\tensor{e}{_\mu^m} \tensor{e}{_\nu^n}\delta_{mn}\ .
\end{equation}
It is invariant under SO(4,\,$\mathbb{C}$ transformations.
We define the metric $g_{\mu\nu}$ as the real part of the pseudo-metric $G_{\mu\nu}$
\begin{align}
g_{\mu\nu} = \frac{1}{2} (G_{\mu\nu} + G^*_{\mu\nu}) = \frac{1}{2} ( \tensor{e}{_\mu^m} \tensor{e}{_\nu^n} + e^*_\mu{}\indices{^m} e^*_\nu{}\indices{^n} ) \delta_{mn}.
\label{eq:M1}
\end{align}
A second SO(4,\,$\mathbb{C}$)-invariant tensor is given by the imaginary part of the pseudo-metric,
\begin{equation}
w_{\mu\nu} = -\frac{i}{2} ( G_{\mu\nu} - G^*_{\mu\nu} ),
\label{eq:M2}
\end{equation}
such that
\begin{equation}
G_{\mu\nu} = g_{\mu\nu} + iw_{\mu\nu}.
\label{eq:M3}
\end{equation}
We may call the tensor $w_{\mu\nu}$ the ``cometric".

%Absatz
The inverse metric $g_{\mu\nu}$ obeys, as usual,
\begin{equation}
g^{\mu\nu} g_{\nu\rho} = \delta_\rho^\mu.
\label{eq:M4}
\end{equation}
Using the relation
\begin{equation}
G^{\mu\nu} = g^{\mu\nu} - ig^{\mu\rho} w_{\rho\sigma} G^{\sigma\nu} = g^{\mu\nu} - iG^{\mu\rho} w_{\rho\sigma} g^{\sigma\nu},
\label{eq:M5}
\end{equation}
we can convert the raising of indices with $G_{\mu\nu}$ into the more usual raising with $g^{\mu\nu}$, and similarly for the lowering of indices by use of eq.\,\eqref{eq:M3}. With
\begin{equation}
G^{\mu\nu} = g^{\mu\nu} - iG^{\mu\rho} w_{\rho\sigma} G^{\sigma\nu} + u^{\mu\nu} = g^{\mu\nu} - iw^{\mu\nu} + u^{\mu\nu}
\label{eq:M6}
\end{equation}
one has
\begin{equation}
u^{\mu\nu} = (w^{\mu\rho} + iu^{\mu\rho} ) \tensor{w}{_\rho^\nu}.
\label{eq:M7}
\end{equation}
For small $w_{\mu\nu}$ the tensor $u^{\mu\nu}$ vanishes in linear order in $w$, $u^{\mu\nu} = w^{\mu\rho} \tensor{w}{_\rho^\nu} + \mathcal{O}(w^3)$, and we can often neglect it.

\paragraph*{Further kinetic invariants}

%Absatz
The tensor $w_{\mu\nu}$ allows for the construction of further kinetic invariants. They will provide the missing kinetic term for the imaginary part of $K_\mu + C_\mu$ and $U$. The covariant derivative
\begin{equation}
Y_{\mu\nu\rho} = \D_\mu w_{\nu\rho} = \partial_\mu w_{\nu\rho} - \tensor{\Gamma}{_{\mu\nu}^\sigma} w_{\sigma\rho} - \tensor{\Gamma}{_{\mu\rho}^\sigma} w_{\nu\sigma}
\label{eq:M23}
\end{equation}
transforms as a tensor. Here $\tensor{\Gamma}{_{\mu\nu}^\sigma}$ is constructed from $G_{\mu\nu}$ or $\tensor{e}{_\mu^m}$ according to eq.\,\eqref{eq:31}. We can add to the action an invariant $eL_W$ with ($\kappa = \kappa^*$)
\begin{align}
\begin{split}
L_W &= \frac{\kappa^2}{2} ( Y^*_\mu{}\indices{^\nu_\rho} Y\indices{^\mu_\nu^\rho} + Y\indices{_\mu^\nu_\rho} Y^*{}\indices{^\mu_\nu^\rho} ) \\
&= \frac{\kappa^2}{2} Y^*_{\mu\nu\rho} Y_{\alpha\beta\gamma} G^{\nu\beta} G^{*\rho\gamma} ( G^{\mu\alpha} + G^{*\mu\alpha} ).
\end{split}
\label{eq:M23A}
\end{align}

%Absatz
In a quadratic expansion around a constant real $\bar{G}_{\mu\nu}$ we can omit $\tensor{\Gamma}{_\mu_\nu^\sigma}$, such that the term linear in $H$ reads
\begin{align}
\begin{split}
Y_{\mu\nu\rho} &= \partial_\mu w_{\nu\rho} = -\frac{i}{2} \partial_\mu ( G_{\nu\rho} - G^*_{\nu\rho} ) = Y^*_{\mu\nu\rho} \qquad \qquad \\
&= -\frac{i}{2} \partial_\mu ( H^\mathrm{(S)}_{\nu\rho} - H^\mathrm{(S)*}_{\nu\rho} ) = \partial_\mu \Im [ H^\mathrm{(S)}_{\nu\rho} ] \\
&= \partial_\mu \Im \left[ \vphantom{\frac{0}{0}} T_{\nu\rho} + \partial_\nu K_\rho + \partial_\rho K_\nu \right. \\
&\qquad\qquad + \left.\frac{1}{3} \left( \bar{G}_{\nu\rho} - \frac{\partial_\nu \partial_\rho}{\partial^2}  \right) S + \frac{\partial_\nu \partial_\rho}{\partial^2} U \right].
\end{split}
\label{eq:M24}
\end{align}
In contrast to $\tensor{U}{_{\mu\nu\rho}}$ in eq. \eqref{eq:L7} , this expression contains the imaginary part of $K_\mu$ separately, as well as the imaginary part of $U$. 

\paragraph*{Mode expansion of complex vierbein}

%Absatz
We express the complex field fluctuations~\eqref{eq:L1} by real fields, corresponding to the real and imaginary parts
\begin{align}
\begin{split}
T_{\mu\nu} &= t_{\mu\nu} + is_{\mu\nu},\quad K_\mu = \kappa_\mu + i\lambda_\mu, \\
S &= \sigma + i\tau,\quad U = u + iv, \\
B_{\mu\nu} &= b_{\mu\nu} + id_{\mu\nu},\quad C_\mu = \gamma_\mu + i\delta_\mu.
\end{split}
\label{eq:M25}
\end{align}
In terms of these fields $L_W$ reads in momentum space
\begin{align}
\nonumber
\int \d^4x\; eL_W = \int_q &\bar{e} \kappa^2 \{ s^{\mu\nu}(-q) q^2 s_{\mu\nu}(q) + 2\lambda^\mu(-q) q^4 \lambda_\mu(q) \\
\label{eq:M26}
&+ \frac{1}{3} \tau(-q) q^2 \tau(q) + v(-q) q^2 v(q) \}.
\end{align}
For the fields $s_{\mu\nu}$ and $\tau$ this term simply adds a contribution $\sim \kappa^2$ to the term $\sim m^2$ from eq.\,\eqref{eq:L4}, such that the inverse propagator for $s_{\mu\nu}$ is $\sim ( m^2/8 + \kappa^2)q^2$, and for $\tau$ it is $\sim(m^2/24 + \kappa^2/3)q^2$. The scalar field $v$ acquires now a standard inverse propagator $\sim \kappa^2 q^2$, curing the deficiency of an action based only on $L_U$. For the vector fields $\lambda_\mu$ and $\delta_\mu$ we observe a mixing, with an inverse propagator given by a matrix
\begin{equation}
P_V = q^2
\begin{pmatrix}
\frac{m^2}{16} + \kappa^2  &  -\frac{m^2}{16}  \\
-\frac{m^2}{16}  &  \frac{m^2}{16}
\end{pmatrix},
\label{eq:M27}
\end{equation}
where the first line and row correspond to $\lambda_\mu$, and the second to $\delta_\mu$, and we have absorbed a factor $2q^2$ in a standard normalization of the vector fields. For $\kappa \neq 0$ both eigenvalues $\lambda_\pm^{(V)}$ of $P_V/q^2$ are positive, $\lambda_\pm^{(V)} > 0$, such that stability is assured for all vector fields. The absence of a kinetic term for the combination $\lambda_\mu + \delta_\mu$ in an action based solely on $L_U$ is cured for $\kappa \neq 0$. The kinetic terms in the sector $(t_{\mu\nu}, b_{\mu\nu}, \kappa_\mu, \gamma_\mu, \sigma, u )$ are not affected by $L_W$. In particular, the gauge degrees of freedom $\kappa_\mu + \gamma_\mu$ and $u$ do not appear in the quadratic action.

%Absatz
We conclude that the bosonic part of the classical action based on $L_F+L_U+L_W$ leads to well behaved kinetic terms and therefore well behaved propagators for all physical modes. Such a classical action seems to be a good candidate for the formulation of a functional integral for pregeometry. There seems to be no obvious obstructions from this side for the formulation of a Yang-Mills theory for a non-compact gauge group. Replacing the real metric by a complex pseudo metric is crucial for well behaved propagators of all gauge fields. This section concludes our discussion of the classical action.

\section{Time-space symmetry breaking\label{sec:6}}

%Absatz
This section addresses the central question of this paper about the origin of the asymmetry between time and space. For this purpose we need the discussion of solutions of the field equations that may either be flat Minkowski or euclidean space. The relevant field equations are derived from the quantum effective action $\Gamma$. We therefore start in this section to include additional terms in a derivative expansion of the effective action that are invariant with respect to diffeomorphisms and local SO(4,\,$\mathbb{C}$) symmetry. These invariants need not to be present in the classical action.

\paragraph*{Potential for the cometric}

%Absatz
We can construct real scalar invariants $W_i$ from $w_{\mu\nu}$, as
\begin{align}
\begin{split}
W_1 &= w_{\mu\nu} w_{\rho\sigma} G^{\mu\rho} G^{*\nu\sigma} = w^*_\mu{}\indices{^\nu} \tensor{w}{_\nu^\mu}, \\
W_2 &= w^*_\mu{}\indices{^\nu} \tensor{w}{_\nu^\rho} w^*_\rho{}\indices{^\sigma} \tensor{w}{_\sigma^\mu}.
\end{split}
\label{eq:M8}
\end{align}
This allows us to supplement in the effective action $\Gamma$ a potential term for the cometric
\begin{equation}
\Gamma_V = \int \d^4x \; e V(W_1,W_2),
\label{eq:M9}
\end{equation}
with
\begin{equation}
V = \gamma_1 W_1 + \gamma_2 W_2 + \frac{\lambda_1}{2} W_1^2 + \frac{\lambda_2}{2} W_2^2 + ...
\label{eq:M10}
\end{equation}
We will discuss other terms, as a constant $V_0$ or a term linear in $\tensor{w}{_\mu^\mu}$, later. For $\gamma_1 = \gamma_2 = 0$, $\lambda_1 > 0$, $\lambda_2 > 0$, the potential is positive, $V > 0$, whenever $W_1 \neq 0$ or $W_2 \neq 0$. Its minimum occurs in this case for $W_1 = W_2 = 0$. 

%Absatz
We can consider $\tensor{w}{_\mu^\nu}$ as the elements of a complex matrix $\hat{W}$, with
\begin{equation}
W_1 = \tr(\hat{W}^*\hat{W}),\quad W_2 = \tr(\hat{W}^*\hat{W}\hat{W}^*\hat{W}).
\end{equation}
One can evaluate the traces in a basis where $\hat{W}' = D\hat{W}D^{-1}$ is diagonal. For the particular case where both $\hat{W}'$ and $(\hat{W}^*)'$ are diagonal,
\begin{align}
\begin{split}
\hat{W}' = \diag(\lambda_i)&,\quad \hat{W}^{\prime *} = \diag(\lambda^*_i), \\
\hat{W}^{\prime *} \hat{W}' &= \diag(|\lambda_i|^2),
\end{split}
\label{eq:M12}
\end{align}
one has $W_1 \geq 0$, $W_2 \geq 0$. More generally, in the subspace of configurations $G_{\mu\nu}$ for which $W_1 \geq 0$, $W_2 \geq 0$ the potential has its minimum at $W_1=W_2=0$ for all $\gamma_1 \geq 0$, $\gamma_2 \geq 0$.

%Absatz
Expanding in small $w_{\mu\nu}$ one has $W_1 \sim w^2$, $W_2 \sim w^4$. In the vicinity of a real euclidean ``background metric'' $\bar{G}_{\mu\nu} = \delta_{\mu\nu}$ one finds $W_1 >0$,
\begin{equation}
W_1 = \sum_{\mu,\nu} (w_{\mu\nu})^2.
\label{eq:M13}
\end{equation}
This differs from an expansion around Minkowski space, $\bar{G}_{\mu\nu} = \eta_{\mu\nu}$, where $W_1$ can take negative values $(k,l=1...3)$
\begin{equation}
W_1 = \sum_{k,l} w^2_{kl} + w^2_{00} - 2\sum_k w^2_{0k}.
\label{eq:M14}
\end{equation}
In this case the configuration $G_{\mu\nu} = \eta_{\mu\nu}$, $w_{\mu\nu}=0$, corresponds to a saddle point if $\gamma_1 \neq 0$.

\paragraph*{Spontaneous time-space asymmetry}

%Absatz
The presence of a potential modifies the field equations. Arbitrary constant $\bar{e}\indices{_\mu^m}$ are no longer solutions. Solutions of the field equations are found for constant vierbeins for which the pseudo-metric $G_{\mu\nu}$ is real,
\begin{equation}
\bar{G}_{\mu\nu} = \bar{G}^*_{\mu\nu},\quad w_{\mu\nu} =0.
\label{eq:M15}
\end{equation}
In this case, one has
\begin{equation}
\frac{\partial V}{\partial w_{\mu\nu}} = 0,\quad \frac{\partial V}{\partial G_{\mu\nu}} = 0,\quad \frac{\partial V}{\partial \tensor{e}{_\mu^m}} = 0.
\label{eq:M16}
\end{equation}
The condition \eqref{eq:M15} singles out the subspace of real pseudo-metrics. It still contains metrics $g_{\mu\nu}$ with arbitrary signature.

%Absatz
Two metrics with different signature are continuously connected in the space of all complex vierbeins. For example, the Minkowski metric $G_{\mu\nu}=\eta_{\mu\nu}$ can be obtained from the euclidean metric $G_{\mu\nu}=\delta_{\mu\nu}$ by a phase change of $\tensor{e}{_\mu^0}$ that corresponds to analytic continuation \eqref{eq:7}. This continuous change of phase does not remain, however, within the space of solutions of the field equations. Within the space of real $G_{\mu\nu}$ every continuous path from $\delta_{\mu\nu}$ to $\eta_{\mu\nu}$ has to pass through the point where $\det(G_{\mu\nu})=0$ or $e=0$. The difference between time and space is generated by spontaneous symmetry breaking \cite{CWTS}, with euclidean and Minkowski signature of the metric corresponding to different extrema, for which different subgroups of SO(4,\,$\mathbb{C}$) remain unbroken. For all constant complex vierbeins that lead to a real constant $G_{\mu\nu}=g_{\mu\nu}$ both the potential and the action vanish, $V=0$, $S=0$. This still provides for a continuous family of solutions corresponding to flat directions of the potential $V$. 

\paragraph*{Conditions for flat space solutions}

%Absatz
The condition for flat space to be a solution of the field equations is the absence of terms linear in the fields in an expansion around flat space. Since we are mainly interested in this case here, we have not included a cosmological constant, corresponding to an invariant piece in the action
\begin{equation}
\Gamma_{V_0}=V_0 \int\limits_{x}^{} e .	
\label{eq:M16A}
%\tag*{M16A}
\end{equation}
As for standard gravity, a flat space solution requires $V_0=0$, since $e$ involves a term linear in $H$. A term \eqref{eq:M16A} is irrelevant for momenta much larger than $V_0^{1/4}$. It plays no role for the stability issue at high momenta. For low momenta and cosmology the cosmological constant should be included. In order to keep the focus of this paper simple we set here $V_0=0$, however. 

%Absatz
Similarly, the particular role of real $G_{\mu\nu}$ would not hold if the potential $eV$ contains a term linear in $w_{\mu\nu}$. Potential candidates would be terms that are linear in the scalar
\begin{equation}
y = w_{\mu\nu} G^{\nu\mu} = \tensor{w}{_\mu^\mu} = \tr \hat{W}.
\label{eq:M17}
\end{equation}
Expanding around an arbitrary real metric $g_{\mu\nu}$, $G_{\mu\nu} = g_{\mu\nu} + iw_{\mu\nu}$, the real part of $y$ is indeed linear in $w$
\begin{equation}
\frac{1}{2} (y+y^*) = \frac{1}{2} w_{\mu\nu} ( G^{\nu\mu} + G^{*\nu\mu} ) = w_{\mu\nu} g^{\nu\mu} + \mathcal{O}(w^2).
\label{eq:M18}
\end{equation}
In contrast, the imaginary part is of the order $w^2$,
\begin{equation}
-\frac{i}{2} ( y - y^* ) = -\frac{i}{2} w_{\mu\nu} ( G^{\nu\mu} - G^{*\nu\mu} ) = -w_{\mu\nu} w^{\nu\mu} + \mathcal{O}(w^3),
\label{eq:M19}
\end{equation}
where we employ the relation \eqref{eq:M6}. If we impose on $S/e$ a discrete symmetry
\begin{equation}
\tensor{e}{_\mu^m} \leftrightarrow e_\mu^{*\,m}
\label{eq:M20}
\end{equation}
a term linear in $y+y^*$ cannot appear in the potential. The combination
\begin{equation}
\frac{1}{2} ( y + y^* ) = -\frac{i}{4} ( G_{\mu\nu} - G^*_{\mu\nu} ) ( G^{\nu\mu} + G^{*\nu\mu} )
\label{eq:M21}
\end{equation}
is odd under the discrete symmetry, since eq.\,\eqref{eq:M20} implies
\begin{equation}
G_{\mu\nu} \leftrightarrow G^*_{\mu\nu}.
\label{eq:M22}
\end{equation}
Supplementing eq.\,\eqref{eq:M20} with $A_{\mu mn} \leftrightarrow A^*_{\mu mn}$ the invariants $L_F$ and $L_U$ are invariant under this discrete transformation.

%Absatz
A constant term $V_0$ in the potential also induces in the expansion of $eV_0$ a term linear in $\tensor{w}{_\mu^\mu}$, since
\begin{equation}
e = \bar{e} \left[ 1+ \frac{1}{2} \tensor{H}{_\mu^\mu} -\frac{1}{8} ( \tensor{H}{_\mu^\nu} \tensor{H}{_\nu^\mu} - \tensor{H}{_\mu^\mu} \tensor{H}{_\nu^\nu} ) \right] + ...
\label{eq:M22A}
\end{equation}
and
\begin{align}
\nonumber
\tensor{w}{_\mu^\mu} &= -\frac{i}{2} G^{\nu\mu} ( G_{\mu\nu} - G^*_{\mu\nu} ) = -\frac{i}{2} G^{\nu\mu} ( H^\mathrm{(S)}_{\mu\nu} - H^\mathrm{(S)*}_{\mu\nu} ) \\
&= -\frac{i}{2} ( \tensor{H}{_\mu^\mu} - H^{*\;\mu}_\mu )
\label{eq:M22B}
\end{align}
implies
\begin{equation}
e = \bar{e} \left[ 1 + \frac{1}{4} ( \tensor{H}{_\mu^\mu} + H^{*\;\mu}_\mu ) + \frac{i}{2} \tensor{w}{_\mu^\mu} \right] + ...
\label{eq:M22C}
\end{equation}
In the presence of a term linear in $\tensor{w}{_\mu^\mu}$ in $eV$ the solutions of the field equation typically lead to a pseudo-metric
\begin{equation}
G_{\mu\nu} = e^{i\alpha} ( g_{\mu\nu} + i\tilde{w}_{\mu\nu} ),
\label{eq:M22D}
\end{equation}
with real $g_{\mu\nu}$ and $\tilde{w}_{\mu\nu}$. This modifies the relation \eqref{eq:M1} between the metric $g_{\mu\nu}$ and the pseudo-metric $G_{\mu\nu}$. We avoid here a discussion of this complication by choosing couplings for which the term linear in $\tensor{w}{_\mu^\mu}$ in an expansion of $eV$ vanishes.

%Absatz
We conclude that the invariants based on the imaginary part $w_{\mu\nu}$ of the pseudo-metric play an important double role for the quantum effective action. First, the combination of the terms $L_F + L_U + L_W$ renders for a euclidean metric a well behaved propagator for all physical fluctuation fields in the limit of large $q^2$. This holds provided $Z$, $m^2$ and $\kappa^2$ are positive, where these couplings need not to coincide with the ones in the classical action. All fields except for the gauge degrees of freedom have a standard kinetic term with positive coefficients. If interactions remain under control for large $q^2$, such an effective action is expected to have a well behaved short distance limit. Second, the potential $V$ selects a subsector with a real metric. The dependence of the effective action on the real part $g_{\mu\nu}$ and the imaginary part $w_{\mu\nu}$ of the pseudo-metric $G_{\mu\nu}$ is different.

\paragraph*{Real metric limit and real gauge field limit}

%Absatz
An interesting limit is the ``real metric limit'' achieved by taking $\kappa\to\infty$, $\gamma_i\to\infty$, $\lambda_i\to\infty$. Expanded around a euclidean metric $g_{\mu\nu}(x)$,
\begin{equation}
G_{\mu\nu}(x) = g_{\mu\nu}(x) + iw_{\mu\nu}(x),
\label{eq:M28}
\end{equation}
all field configurations that lead to a small $w_{\mu\nu} \neq 0$ have a divergent positive action if $e$ is real, positive, and different from zero. We may therefore define an effective theory that discards all fields for which $w_{\mu\nu}\neq 0$, setting them to zero in the effective action. This imposes a non-linear constraint in the space of complex vierbeins. All vierbeins that are related by SO(4,\,$\mathbb{C}$)-transformations lead to the same pseudo-metric. (The pseudo-metric $G_{\mu\nu}$ can be used to characterize orbits of SO(4,\,$\mathbb{C}$)-transformations.) Out of the 16 complex components of $\tensor{e}{_\mu^m}$ only ten can lead to different $G_{\mu\nu}$. The constraint $G^*_{\mu\nu} = G_{\mu\nu}$ eliminates ten real degrees of freedom, such that ten real degrees of freedom are left. They correspond to the ten degrees of freedom of $g_{\mu\nu}$. Only six of them are physical, since four degrees of freedom correspond to a diffeomorphism transformation of the ``physical metric''\cite{CWPM}.

%Absatz
The real metric limit can be supplemented by a ``real gauge field limit''. For this purpose we consider the invariant
\begin{equation}
L_\mathrm{I} = -\frac{Z_\mathrm{I}}{32} ( F^*_\mu{}\indices{^\nu_\rho^\sigma} - \tensor{F}{_\mu^\nu_\rho^\sigma} ) ( F^*_\nu{}\indices{^\mu_\sigma^\rho} - \tensor{F}{_\nu^\mu_\sigma^\rho} ). 
\label{eq:M29}
\end{equation}
It is similar to the invariant $L_F$ in eq.\,\eqref{eq:29}, but only involves the imaginary part of $\tensor{F}{_\mu^\nu_\rho^\sigma}$. For real $G_{\mu\nu}$ this invariant only involves the imaginary part of $A_{\mu\nu\rho}$, similar to eq.\,\eqref{eq:E2}. The kinetic term for gauge bosons leading to imaginary $A_{\mu\nu\rho}$ is multiplied by $Z_F + Z_\mathrm{I}$, while the one for the gauge bosons leading to real $A_{\mu\nu\rho}$ remains multiplied by $Z_F$. For $Z_\mathrm{I} \to\infty$ the imaginary part of $A_{\mu\nu\rho}$ decouples effectively from the other fields since the gauge coupling $(Z_F + Z_\mathrm{I})^{-1/2}$ goes to zero. Different gauge couplings for different gauge bosons of the group SO(4,\,$\mathbb{C}$) may be surprising at first sight. This effect arises since the kinetic term for the gauge bosons is directly affected by the value of the vierbein and the associated spontaneous symmetry breaking from any $\tensor{e}{_\mu^m} \neq 0$. (The kinetic term vanishes for $\tensor{e}{_\mu^m} \neq 0$ due to $e=0$.) A similar effect is known for grand unified gauge theories where scalar expectation values breaking the symmetry can give rise to different kinetic terms and therefore different gauge coupling for gauge bosons belonging to different subgroups \cite{HILL,SHAW}. We notice that the fields that decouple are the gauge bosons corresponding to imaginary $A_{\mu\nu\rho}$. For complex vierbeins this is not equivalent to imaginary $A_{\mu mn}$.

\paragraph*{Mass split for the gauge bosons}

%Absatz
Another invariant for imaginary fields is given similar to eq.\,\eqref{eq:32C} by
\begin{equation}
L_J = -\frac{m^2_J}{16} ( U^*_\mu{}\indices{^\nu_\rho} - \tensor{U}{_\mu^\nu_\rho} ) ( \tensor{U}{^{*\mu}_\nu^\rho} - \tensor{U}{^\mu_\nu^\rho} ).
\label{eq:M30}
\end{equation}
Besides a contribution to the kinetic terms for the imaginary parts in $H_S$ and $H_A$ similar to eq.\,\eqref{eq:E9} and correspondingly eq.\,\eqref{eq:L4}, and a contribution to the source term similar to eq.\,\eqref{eq:E12}, this term contributes a mass term for the gauge bosons with imaginary $A_{\mu\nu\rho}$ similar to eq.\,\eqref{eq:E10}. The squared mass for these bosons is given by $m^2 + m_J^2$. For $m^2_J \to\infty$ these bosons not only decouple as for $Z_\mathrm{I}\to\infty$, but also have a divergent mass.   They can be discarded from the effective theory. Combining the real metric limit with the real gauge field limit $Z_\mathrm{I}\to\infty$, $m_J^2\to\infty$, all configurations that do not lead to real $G_{\mu\nu}$ and real $A_{\mu\nu\rho}$ can be omitted from the effective theory.

%Absatz
In this limit our model is very similar to a model of euclidean gravity with real vierbeins and real gauge fields $A_{\mu mn}$ for an SO(4)-gauge group. Such a model shows the acceptable ultraviolet behavior of a Yang-Mills theory with massless vector-matter fields $\tensor{e}{_\mu^m}$ and fermions. Restricting our setting from the beginning to real $\tensor{e}{_\mu^m}$ and $A_{\mu mn}$, the inverse propagator for both the physical gauge and vector-matter fields increases $\sim q^2$ for large $q^2$. The action based on $L_F + L_U$ is positive, with a minimum at $S=0$ corresponding to a constant $\tensor{\bar{e}}{_\mu^m}$ and $\bar{A}_{\mu mn}=0$. One may therefore employ numerical Monte-Carlo simulations for its investigation. It will be interesting to see if this model is asymptotically free. The unusual features of this gauge theory are the appearance of the inverse vierbein in the action, and the additional invariance under diffeomorphisms. It remains to be seen if a quantum field theory of this type shows acceptable features for euclidean quantum gravity. If valid, one may obtain a model for quantum gravity with Minkowski signature by analytic continuation.

%Absatz
Despite many similarities with the euclidean gravity model above the real metric and real gauge field limit of our SO(4,\,$\mathbb{C}$)-invariant pregeometry shows also an important difference. The reason is that the constraint of real $G_{\mu\nu}$ and real $A_{\mu \nu \rho}$ does not enforce real $\tensor{e}{_\mu^m}$ and real $A_{\mu mn}$. A real metric with Minkowski signature and imaginary $\tensor{e}{_\mu^0}$ obeys these constraints as well. The real metric and gauge field limit describes simultaneously theories with a different signature of the metric. If we take very large but finite parameters $\kappa$, $\gamma_i$, $\lambda_i$, $Z_\mathrm{I}$ and $m_J^2$, small imaginary fluctuations around a euclidean metric lead first to a sharp increase of $L$, while the action develops an imaginary part due to the prefactor $e$. This holds, for example, if we follow the path of analytic continuation given by eq.\,\eqref{eq:7}. For $\varphi = \pi/2$, however, the terms $V$, $L_W$, $L_\mathrm{I}$, $L_J$ vanish again since $G_{\mu\nu}$ is again real. The effective theory describes simultaneously the euclidean signature and the Minkowski signature obtained by analytic continuation. Both ``branches'', as well as other branches with different signatures, are actually continuously connected at the configuration $\tensor{e}{_\mu^m} = 0$.

%Absatz
For a real metric $\bar{g}_{\mu\nu}$ we can consider standard real metric fluctuations
\begin{equation}
g_{\mu\nu} = \bar{g}_{\mu\nu} + h_{\mu\nu}.
\label{eq:M31}
\end{equation}
They are connected to the vierbein fluctuations by
\begin{equation}
h_{\mu\nu} = \Re [ H^\mathrm{(S)}_{\mu\nu} + \frac{1}{4} \tensor{H}{_\mu^\rho} H_{\rho\nu} ].
\label{eq:M32}
\end{equation}
This is the reason for the chosen normalization of $H_{\mu\nu}$.

%Absatz
At this level no obvious obstructions to the formulation of a consistent model of pregeometry based on a SO(4,\,$\mathbb{C}$) Yang-Mills theory are visible. What remains to be achieved is a realistic phenomenology. This requires the presence of further invariants in the quantum effective action that dominate for low enough momenta. These invariants could lead to new instabilities. This issue will be discussed in the remaining parts of this paper.

\section{Low momentum limit and the $\quad\quad$ emergence of general relativity\label{sec:7}}

%Absatz
Even in the real metric and the real gauge field limit our theory remains a pregeometry. At short distances the degrees of freedom are vierbeins and gauge fields. These constitute a larger number of physical degrees of freedom than the ones contained in the metric. For high momenta the composite metric field does not play any apparently crucial role. This changes in the long distance or low momentum limit, for which standard general relativity emerges. In this limit we can still use a vierbein $\tensor{e}{_\mu^m}$ and a gauge field $A_{\mu m n}$. The gauge field will equal the spin connection $\omega_{\mu m n}$. These objects will be directly related to the metric, however, and do not describe any longer independent fluctuations. General relativity emerges as the effective low energy theory. In the presence of a suitable invariant in pregeometry one obtains the Einstein-Hilbert action for the composite metric.

\paragraph*{Freezing of the gauge bosons}

%Absatz
For momenta much smaller than $m^2$ the mass term \eqref{eq:E10} for the gauge bosons dominates over the kinetic term \eqref{eq:E2}. The gauge bosons are ``frozen'' and cease to be relevant propagating degrees of freedom. This does not mean, however, that we can set $A_{\mu mn} =0$. The reason is the source term \eqref{eq:E12} which forbids solutions $A_{\mu mn}=0$ for inhomogeneous vierbeins, $\partial_\mu \tensor{e}{_\nu^m} \neq 0$.

%Absatz
If we neglect for a moment the kinetic invariants for the gauge fields, the field equation for the gauge fields reads
\begin{equation}
U_{\mu\nu\rho} =0.
\label{eq:LM1}
\end{equation}
From eq.\,\eqref{eq:29A} we infer that $A_{\mu\nu\rho}$ becomes a function of the vierbein, $A_{\mu\nu\rho} = \omega_{\mu\nu\rho}(e)$,
\begin{align}
\nonumber
\omega_{\mu\nu\rho}(e) = \frac{1}{2} \{ &e_{\mu m} ( \partial_\rho \tensor{e}{_\nu^m} - \partial_\nu \tensor{e}{_\rho^m} ) + e_{\nu m} ( \partial_\rho \tensor{e}{_\mu^m} - \partial_\mu \tensor{e}{_\rho^m} ) \\
\label{eq:LM2}
&- e_{\rho m} ( \partial_\nu \tensor{e}{_\mu^m} - \partial_\mu \tensor{e}{_\nu^m} ) \}.
\end{align}
For $A=\omega(e)$ the covariant derivative of the vierbein vanishes
\begin{equation}
\D_\mu \tensor{e}{_\nu^m} = 0.
\label{eq:LM3}
\end{equation}
For real $\tensor{e}{_\mu^m}$ and $\omega_{\mu mn}$ eq.\,\eqref{eq:LM2} is the expression of the spin connection in Cartan's geometry \cite{CAR}. In our context this generalizes to complex $\tensor{e}{_\mu^m}$ and $\omega_{\mu mn}$. With eq.\,\eqref{eq:LM2} the gauge fields are no longer independent degrees of freedom, only the vierbein remains as a relevant bosonic degree of freedom. In the real metric limit we may choose for a euclidean signature of $g_{\mu\nu}$ real vierbeins $\tensor{e}{_\mu^m}$. For a Minkowski signature we can choose real $e^{\mathrm{(M)}\,m}_{~~\mu}$, corresponding to imaginary $\tensor{e}{_\mu^0}$. Then $\omega^\mathrm{(M)}_{\mu mn}$ is real. 

\paragraph*{Low momentum effective theory}

%Absatz
We next include the kinetic invariants for the gauge fields. We can write the gauge fields $A_{\mu\nu\rho}$ as a difference between the (generalized) spin connection $\omega_{\mu\nu\rho}(e)$ and the SO(4,\,$\mathbb{C}$)-invariant tensor $U_{\mu\nu\rho}$,
\begin{equation}
A_{\mu\nu\rho} = \omega_{\mu\nu\rho} - U_{\mu\nu\rho}.
\label{eq:LM4}
\end{equation}
The inhomogeneous transformation part of $A_{\mu\nu\rho}$ cancels for the difference $A_{\mu\nu\rho} - \omega_{\mu\nu\rho}(e)$. Instead of the independent gauge fields $A_{\mu\nu\rho}$ we could choose the tensor $U_{\mu\nu\rho}$ as a field variable. In terms of $U_{\mu\nu\rho}$ the kinetic invariant $L_F$ reads 
\begin{equation}
L_F = \frac{Z_F}{8} ( R^*_\mu{}\indices{^\nu_\rho^\sigma} - V^*_\mu{}\indices{^\nu_\rho^\sigma} ) ( \tensor{R}{_\nu^\mu_\sigma^\rho} - \tensor{V}{_\nu^\mu_\sigma^\rho} ),
\label{eq:LM5}
\end{equation}
with
\begin{equation}
\tensor{V}{_{\mu\nu\rho}^\sigma} = \tensor{e}{_m^\sigma} ( \D_\mu \tensor{U}{_\nu_\rho^m} - \D_\nu \tensor{U}{_\mu_\rho^m} ).
\label{eq:LM6}
\end{equation}
Here the complex curvature tensor $R_{\mu\nu\rho\sigma}$ is constructed from the complex connection $\tensor{\Gamma}{_{\mu\nu}^\rho}$ in eq.\,\eqref{eq:32A} in the usual way. It only depends on the pseudo-metric $G_{\mu\nu}$. Eq.\,\eqref{eq:LM5} can be derived directly from the commutator of covariant derivatives of the vierbein \cite{CWGG}
\begin{equation}
( \D_\mu \D_\nu - \D_\nu \D_\mu ) \tensor{e}{_\rho^m} = \tensor{F}{_\mu_\nu^m_n} \tensor{e}{_\rho^n} - \tensor{R}{_\mu_\nu^\sigma_\rho} \tensor{e}{_\sigma^m} = \tensor{V}{_{\mu\nu\rho}^m},
\label{eq:LM7}
\end{equation}
inserting into eq.\,\eqref{eq:29} the relation
\begin{equation}
F_{\mu\nu\rho\sigma} = R_{\mu\nu\rho\sigma} - V_{\mu\nu\rho\sigma}.
\label{eq:LM8}
\end{equation}

%Absatz
The low momentum limit can be formulated in a more concise way by considering the field equation for $U_{\mu\nu\rho}$, evaluated for a given fixed $\tensor{e}{_\mu^m}(x)$. The term $L_F$ contains a kinetic term $(\D U)^2$ for $U$ as well as a mixed term $\sim RV$. (We omit here indices for simplicity.) Employing partial integration the mixed term turns to a source term $U\D R$ linear in $U$, with source containing covariant derivatives $\D R$ of the curvature tensor $R$. The relevant mass scale $m$ is set by $L_U \sim m^2U^2$. If the curvature tensor varies only smoothly on a length scale set by $m^{-1}$, the influence of the source on the solution for $U$ becomes small. To a good approximation the field equation for $U$ has the solution $U\sim \D R /m^2$, which vanishes for $m^2\to\infty$. Inserting this solution into $L_F$ yields
\begin{equation}
L_F = \frac{Z_F}{8} R^*_{\mu\nu\rho\sigma} R^{\mu\nu\rho\sigma} + \mathcal{O}\left( \frac{Z_F^2 R \D^2 R}{m^2} \right).
\label{eq:LM9}
\end{equation}
Also $L_U$ turns out of the order $Z_F^2 R \D^2 R /m^2$. The condition for the neglection of the second term in eq.\,\eqref{eq:LM4}, and therefore for the applicability of the low momentum limit, can be formulated as
\begin{equation}
\left| \frac{Z_F \D^2 R}{m^2 R} \right| \ll 1.
\label{eq:LM10}
\end{equation}

%Absatz
For an effective theory at low momenta the invariants with a small number of derivatives play a leading role. The term with no derivatives, the potential $V$ in eq.\,\eqref{eq:M10}, only affects the imaginary part of the pseudo-metric. The field equations have solutions $G_{\mu\nu} = G^*_{\mu\nu} = g_{\mu\nu}$, $w_{\mu\nu}=0$, cf.\ eq.\,\eqref{eq:M15}. For these solutions the pseudo-metric and the curvature tensor $R_{\mu\nu\rho\sigma}$ are real. The terms $L_U + L_F$ reduce in the low momentum limit to a metric theory with effective action ($e=\sqrt{g}$) involving the squared Riemann curvature tensor
\begin{equation}
\Gamma_F = \frac{Z_F}{8} \int_x \sqrt{g} R_{\mu\nu\rho\sigma} R^{\mu\nu\rho\sigma}.
\label{eq:LM11}
\end{equation}
With different contractions of $FF$-terms the effective action for pregeometry can also induce terms $\sim R_{\mu\nu}R^{\mu\nu}$ and $R^2$. This is four-derivative gravity. For suitable coefficients of the various terms no inconsistency of such an effective action is visible. It is, however, not a realistic theory for gravity since the Einstein-Hilbert term is missing.

\paragraph*{Emergence of Einstein-Hilbert action}

%Absatz
For a realistic effective action of pregeometry one needs for small enough momenta an effective term proportional to the curvature scalar $R$ which contains only two derivatives of the metric. Consistent with all symmetries pregeometry indeed admits an invariant with one derivative,
\begin{align}
\nonumber
L_R &= -\frac{M^2}{4} ( \tensor{e}{_m^\mu} \tensor{e}{_n^\nu} \tensor{F}{_{\mu\nu}^{mn}} + \mathrm{c.\,c.} ) = -\frac{M^2}{4} ( \tensor{F}{_{\mu\nu}^{\mu\nu}} + \mathrm{c.\,c.} ) \\
&= -\frac{M^2}{4} ( R - \tensor{V}{_{\mu\nu}^{\mu\nu}} + \mathrm{c.\,c.} ),
\label{eq:LM12}
\end{align}
with curvature scalar
\begin{equation}
R = \tensor{R}{_{\mu\nu}^{\mu\nu}}.
\label{eq:LM14}
\end{equation}
In the low momentum limit we can neglect
\begin{equation}
\tensor{V}{_{\mu\nu}^{\mu\nu}} = \tensor{e}{_m^\mu}(D_\mu \tensor{U}{_\nu^{\nu m}}-D_\nu \tensor{U}{_\mu^{\nu m}}).
\label{eq:LM15}
\end{equation}
For low momenta the leading term in $L_R$ therefore reduces to the Einstein-Hilbert action
\begin{equation}
\Gamma_R =-\frac{M^2}{2} \int_x \sqrt{g} R.
\label{eq:LM16}
\end{equation}
This term dominates over $\Gamma_F$ in eq.\,\eqref{eq:LM11} for $|R/M^2| \ll 1$. 

%Absatz
This low momentum limit of our SO(4,\,$\mathbb{C}$)-gauge theory is simply Einstein's general relativity. The parameter $M$ can be identified with the (reduced) Planck mass. We notice the presence of two mass scales $m$ and $M$. For $m \gg M$ the low momentum limit can be taken in two steps. In the first step the condition \eqref{eq:LM10}, together with real metric solutions, leads to an intermediate effective action which depends only in the metric and involves up to four derivatives
\begin{equation}
\Gamma = \int_x \sqrt{g} \left\{ -\frac{M^2}{2} R + \frac{\tilde{Z}}{8} R_{\mu\nu\sigma\lambda} R^{\mu\nu\sigma\lambda}-\frac{C}{2}R^{2} \right\}.
\label{eq:LM17}
\end{equation}
In leading order in $M^2/m^2$ the coefficient $\tilde{Z}$ equals $Z_F$, and we have omitted the Gauss-Bonnet topological invariant. In a second step for $R/M^2 \ll 1$ the second term can be neglected, resulting in Einstein's gravity.

\section{Stable propagators\label{sec:8}}

%Absatz
The effective action~\eqref{eq:LM17} is Stelle's gravity with up to four derivatives acting on the metric.
For $M^{2}=0$, $\tilde{Z}>0$, $C>0$ there seems to be no problem for an effective action of this type. For $M^2>0$, as needed for realistic gravity, one encounters again the tachyons or ghosts of four-derivative gravity. It may seem at first sight that our model of pregeometry brings no advantage in this respect. The effective action~\eqref{eq:LM17} is, however, a low momentum expansion which looses its validity in the momentum region where the problematic tachyon or ghost poles in the graviton propagator appear. We will demonstrate that for simple forms of the effective action for pregeometry the instabilities of eq.~\eqref{eq:LM17} are artifacts of the derivative expansion. The graviton propagator remains well behaved.

%Absatz
The invariants $L_U$ and $L_R$ are crucial for the emergence of the effective low energy theory. In their presence the propagating modes in flat space mix vierbein fluctuations and gauge fields. This mixing is a key element for the transition from a stable high-momentum behavior to a stable low-momentum behavior of the propagators. The inverse propagators for the physical particles are not polynomials in $q^2$. They rather interpolate smoothly between two different approximate polynomials for small and large $q^2$, thereby avoiding ghost or tachyonic instabilities as for the graviton propagator in eq. \eqref{eq:IN1}.

\paragraph*{Mode mixing and stability}

%Absatz
The intermediate effective action \eqref{eq:LM17} involves a term with four derivatives, while our pregeometry is formulated with terms involving only up to two derivatives. The origin of this behavior is the relation \eqref{eq:LM4}, \eqref{eq:LM2}, which expresses the gauge fields as derivatives of the vierbein. The action \eqref{eq:LM17} is known to be unstable, containing ghosts as characteristic for higher derivative theories. If we start with a stable pregeometry the instability of the effective theory is an artifact of the low momentum approximation. Since this issue is of wider interest for the relation between pregeometry and higher derivative gravity, we will sketch the origin of the apparent instability in more detail. The issue is unrelated to a complex vierbein and gauge fields -- it is the same for a theory with real vierbein and gauge fields. The essence is the mixing between the gauge fields and derivatives of the vierbein. We discuss this issue for the graviton and scalar degrees of freedom in detail in appendix \ref{sec: Ap.A}, focusing on euclidean gravity. This leads to the well-behaved graviton propagator in eq.~\eqref{eq:IN1}. In the present section we describe the general structure of the stability issue.

%Absatz
Omitting indices the inverse propagator matrix or the matrix of second functional derivatives of $S$ takes in momentum space the form
\begin{equation}
P(q^2) = \begin{pmatrix}
Zq^2 + m^2  &  -icm^2q  \\
icm^2q  &  bm^2q^2
\end{pmatrix}.
\label{eq:LM18}
\end{equation}
This type of inverse propagator matrix characterizes the different irreducible representations separately, with different effective constants $Z$, $m^2$ and $c$. The first line and row correspond in eq. \eqref{eq:LM18} to the gauge bosons, the second to the vierbein. The terms $\sim m^2$ arise from $L_U$, including the off-diagonal terms that reflect the source term $L_U^{(3)}$. The term $\sim Z$ reflects $L_F$. The term $L_R$ induces additional off-diagonal terms $\sim M^2q$. We include it in $c$, which becomes a function $c(M^{2}/m^{2})$. The issue of stability and the apparent instability of the low momentum approximation can be discussed in terms of the simple $2\times 2$-matrix $P$, with real $Z>0$, $m^2>0$, $b>0$ and $c$.

%Absatz
We first perform a general discussion of the propagator $P^{-1}$ and turn to our specific model later. Poles of the propagator in the complex plane for $q^2$ correspond to vanishing eigenvalues of $P$. In turn, those are given by a vanishing determinant
\begin{equation}
\det P = bm^2q^2 \left( Z q^2 + \frac{b-c^2}{b} m^2 \right).
\label{eq:LM19}
\end{equation}
In the absence of mixing, $c=0$, we recognize a massless particle with pole at $q^2=0$ and a massive particle with pole at $q^2 = -m^2/Z$. For $c\neq 0$ the massive particle pole is shifted to lower values. All poles are on the real axis. A stable particle obtains for 
\begin{equation}
c^2 < b,
\label{eq:123A}
%\tag*{(123A)}
\end{equation} 
while for $c^2 > b$ one observes a tachyon instability. For $c^2=b$ poles can only occur for $q^2=0$.

%Absatz
We absorb a factor of $m$ in the definition of the vierbein field, such that all entries in $P_R$ have the same dimension
\begin{equation}
P_R(q^2)=\begin{pmatrix}
Zq^2 + m^2  &  -icmq   \\   icmq  &  bq^2
\end{pmatrix}.
\label{eq:LM21}
\end{equation}\\
The eigenvalues of $P_R$ obey
\begin{equation}
\lambda_\pm = \frac{1}{2} \{ Zq^2 + m^2 + bq^2 \pm [(Zq^2+m^2-bq^2)^2 + 4c^2m^2q^2]^{1/2} \}.
\label{eq:LM20}
\end{equation}

%Absatz
The pole at $q^2=0$ corresponds to $\lambda_-$. Expanding $\lambda_-$ around $q^2=0$,
\begin{equation}
\lambda_- = (b-c^2)q^2 + \mathcal{O}(q^4),
\label{eq:LM22}
\end{equation}
yields a positive coefficient for $q^2$ if $b>c^2$. This corresponds to a stable massless particle. Similarly, for an expansion of $\lambda_+$ around the location of the propagator-pole at 
\begin{equation}
q^2_p = -\frac{m^2}{Z} \left( 1 - \frac{c^2}{b} \right),
\label{eq:LM23}
\end{equation}
one obtains
\begin{equation}
\lambda_+ = \alpha Z (q^2 - q^2_p),\quad \alpha = \left( 1-\frac{c^2}{b} \right)\left(1-\frac{(Z+b)c^2}{b^2} \right)^{-1}.
\label{eq:LM24}
\end{equation}
This holds for 
\begin{equation}
(Z+b)c^2 < b^2,
\label{eq:LM25}
\end{equation}
and the massive particles is stable in this case. Instabilities occur for $c^2>b$ or $(Z+b)c^2 > b^2$. For $(Z+b)c^2 > b^2$ the massive particle pole is described by a second zero of $\lambda_-$, while $\lambda_+$ has no zero. Eq.\,\eqref{eq:LM24} holds now for $\lambda_-$, and $\alpha<0$ indicates a ghost-instability. For the boundary case $(Z+b)c^2=b^2$ both $\lambda_+$ and $\lambda_-$ are proportional $\sqrt{q^2-q_p^2}$ for $q^2\to q_p^2$.

%Absatz
Finally, for the boundary of stability $c^2=b$ one has
\begin{equation}
\lambda_\pm = \frac{1}{2} \{ m^2 + (Z+b)q^2 \pm [ ( m^2 + ( Z+b )q^2)^2 - 4Zbq^4 ]^{1/2} \}.
\label{eq:LM26}
\end{equation}
For $Z=0$ this yields
\begin{equation}
\lambda_+ = m^2 + bq^2,\quad \lambda_- =0,
\label{eq:LM27}
\end{equation}
corresponding to a single stable massive particle with squared mass $m^2/b$, plus a flat direction. For $Z>0$ one finds for small $q^2$
\begin{equation}
\lambda_- = \frac{Zbq^4}{m^2}.
\label{eq:LM28}
\end{equation}
The eigenvalue $\lambda_+$ no longer has a zero.

%Absatz
We conclude that for positive $Z$ and $m^2$ the parameters have to obey two stability conditions \eqref{eq:123A}, \eqref{eq:LM25} in order to avoid ghosts or tachyons. If these conditions are met no instability occurs in the corresponding sector of excitations.

\paragraph*{Impact of the invariant $L_R$ and graviton propagator}

%Absatz
Taking in our model only the invariant $L_U$ corresponds to the case $c^2=b$, $Z=0$. The action depends in this case only on a particular combination of $A$ and $\D e$, while the direction in field space orthogonal to it is flat, cf.\ eq.\,\eqref{eq:LM27}. Adding $L_F$ introduces $Z>0$. The inverse propagator $\sim q^4$ corresponds to the four derivatives in eq.\,\eqref{eq:LM11}. Finally, adding the term $L_R$ induces additional non-diagonal elements such that
\begin{equation}
c = \sqrt{b} \left( 1 - f\frac{M^2}{m^2} \right),
\label{eq:LM29}
\end{equation}
or
\begin{equation}
b-c^2 = \left( 2f \frac{M^2}{m^2} - \frac{f^2 M^4}{m^4} \right) b.
\label{eq:LM30}
\end{equation}
Furthermore, the contribution of $L_R$ to the gauge boson mass term shifts $m^2$ in eq. \eqref{eq:LM21} to $m^2_g$. (For the general stability discussion of this section we omit the index $g$.)

%Absatz
There is a stable massless particle for 
\begin{equation}
0 < \frac{f M^2}{m^2} < 2,
\label{eq:LM31}
\end{equation}
corresponding to the graviton, cf.\ eq.\,\eqref{eq:LM22}. There is also a massive particle with squared mass $-q_p^2$ given by eq.\,\eqref{eq:LM23},
\begin{equation}
m_U^2 = -q_p^2 = \frac{2fM^2 - f^2 M^4/m^2}{Z}.
\label{eq:LM32}
\end{equation}
According to eq.\,\eqref{eq:LM25} the massive particle is stable provided
\begin{equation}
\frac{Z}{b} < \frac{1}{\left(1-\frac{fM^2}{m^2}\right)^2} -1.
\label{eq:LM33}
\end{equation}
For larger $Z$ the massive particle becomes a ghost. 

%Absatz
We conclude that there is a possible range of parameters for which both the massless graviton and a heavy particle are stable. No instability occurs in this sector. The inverse heavy particle propagator obeys near the pole
\begin{equation}
\lambda_+ = \frac{Z(2x-x^2)}{2x-x^2-\frac{Z}{b}(1-x)^2} (q^2 + m_U^2),\quad x = \frac{fM^2}{m^2}.
\label{eq:LM34}
\end{equation}
This corresponds to the heavy degree of freedom related to the tensor field $U_{\mu\nu\rho}$.

%Absatz
For the graviton we find in appendix \ref{sec: Ap.A} the values
\begin{align}
\begin{split}
& m^2_g=(1-y)m^2, \quad y=\frac{M^2}{m^2}, \\
& b=1, \quad c=\sqrt{1-y}, \quad f=\frac{(1-\sqrt{1-y})(1-y)}{y},\\
& x=\frac{f y}{1-y}=1-\sqrt{1-y}.
\end{split}
\label{eq:NB1} 
\end{align}
Insertion of these values into $\lambda_-(q^2)$ in eq. \eqref{eq:LM20} yields the inverse graviton propagator \eqref{eq:IN1}. The conditions of stability $0<M^2<m^2$ and $0<Z<Z_c$ in eq. \eqref{eq:IN3} correspond to the stability conditions \eqref{eq:LM31}, \eqref{eq:LM33}.

\paragraph*{Propagator in effective low momentum theory}

%Absatz
The effective low momentum theory discards the heavy degree of freedom related to $U_{\mu\nu\rho}$. It only keeps the massless field whose inverse propagator is described by $\lambda_-(q^2)$,
\begin{align}
\nonumber
\lambda_- &= \frac{1}{2} \left\{ m^2 + (Z+b)q^2 \vphantom{\sqrt{0^0}} \right. \\
\nonumber
  &\qquad\qquad \left. - \sqrt{(m^2 +(Z-b)q^2)^2 + 4bm^2(1-x)^2 q^2 }  \right\} \\
\label{eq:LM35}
  &= \frac{m^2 + (Z+b)q^2}{2} \\
\nonumber
  &\qquad \times\left\{ 1 - \epsilon(q^2) \sqrt{ 1- \frac{4bq^2(Zq^2 +m^2(2x-x^2))}{(m^2 + (Z+b)q^2)^2} } \right\}
\end{align}
with
\begin{equation}
\epsilon(q^2) = \sign \left( m^2 + (Z+b)q^2 \right).
\label{eq:LM36}
\end{equation}
 If the stability condition \eqref{eq:LM33} is respected, we know that the only zero of $\lambda_-$ in the complex $q^2$-plane occurs for $q^2 =0$. Indeed from eq.\,\eqref{eq:LM19} we infer that the only other possible zero could be at $q^2 = q_p^2$, as given by eq.\,\eqref{eq:LM32}. Evaluating
\begin{equation}
\lambda_- (q_p^2) = \frac{m^2 + (Z+b)q_p^2}{2} (1-\epsilon(q_p^2)),
\label{eq:LM37}
\end{equation}
we observe $\epsilon(q_p^2) =-1$ if the condition \eqref{eq:LM33} is obeyed. With $\lambda_-(q_p^2) <0$ the only possible zero is at $q^2 =0$.

\paragraph*{Expansion for small momentum and fake ghosts}

%Absatz
Expanding $\lambda_-(q^2)$ for small $q^2$,
\begin{equation}
\lambda_-(q^2) = b(2x-x^2)q^2 + b(1-x)^2(Z-(2x-x^2)b) \frac{q^4}{m^2} + ...,
\label{eq:LM38}
\end{equation}
we observe for small $x$ the structure of the low momentum effective theory \eqref{eq:LM17}
\begin{equation}
m^2 \lambda_-(q^2) \approx 2fbM^2q^2 + Zbq^4.
\label{eq:LM39}
\end{equation}

%Absatz
If one determines the propagator-poles in the low-momentum effective theory \eqref{eq:LM38} or \eqref{eq:LM39} one finds besides the stable graviton pole at $q^2=0$ a ghost pole at $q^2=-2fM^2/Z$. This is the usual ghost instability of higher derivative gravity. This pole is a pure artifact of the polynomial expansion of $\lambda_-(q^2)$. As we have seen, the full expression for $\lambda_-(q^2)$ does not contain further zeros and all excitations are stable if the condition \eqref{eq:LM33} is met. We observe that the fake pole occurs at $q_p^2$ as given by eq.\,\eqref{eq:LM32}. If the condition \eqref{eq:LM33} is violated, this zero is indeed present in $\lambda_-(q^2)$ and the model has a ghost instability. On the other hand, if eq.\,\eqref{eq:LM33} holds, the true pole corresponds to a zero of $\lambda_+(q^2)$ and describes a stable massive excitation.

\paragraph*{Propagator for high momentum}

%Absatz
For large values of $q^2$ we observe
\begin{align}
\lim_{q^2\to\infty} \lambda_-(q^2) = \begin{cases}
Zq^2,\quad &\textnormal{for}\quad Z<b \\
bq^2,\quad &\textnormal{for}\quad Z>b.
\end{cases}
\label{eq:LM40}
\end{align}
For both possibilities this momentum behavior is stable. The mode $\lambda_-$ either corresponds to the gauge bosons ($Z<b$) or the vierbein ($Z>b$). The limiting behavior \eqref{eq:LM40} clearly demonstrates that the term $\sim q^4$ corresponds to a transient behavior, switching from $\lambda_- \sim q^2$ for small $q^2$ to $\lambda_- \sim q^2$ for large $q^2$, though with different coefficients. Any Taylor expansion describing such a crossover will typically contain a term $\sim q^4$ in the small momentum expansion.

\paragraph*{General stability analysis}

%Absatz
While the simple inverse propagator matrix \eqref{eq:LM18} contains all essential features, the complete stability analysis of our model is more involved. First of all, for $\bar{G}_{\mu\nu}=\delta_{\mu\nu}$ two fields $H_{\mu\nu}$ and $\partial_\rho \tensor{A}{_\mu^\rho_\nu}$, $\partial_\nu \tensor{A}{^\rho_\rho_\mu}$ etc.\ can mix only if they belong to the same SO(4)-representation. The elements of the matrix \eqref{eq:LM18} may depend on the SO(4)-representation and one would like to require that for all mixings they lead to a stable situation. Part of the stability analysis can be found in appendix \ref{sec: Ap.A}.

%Absatz
Furthermore, our discussion reveals that a more accurate treatment of the low-momentum effective theory should omit the modes corresponding to $\lambda_+(q^2)$ after proper diagonalization of the inverse propagator matrix, rather than simply setting $U_{\mu\nu\rho}=0$. The invariants $L_F$ and $L_R$ contain terms linear in $U_{\mu\nu\rho}$ and one should insert the solution of the field equations for $\tensor{U}{_{\mu \nu \rho}}$ in terms of $G_{\mu\nu}$. This will add higher derivative terms to the effective action \eqref{eq:LM17}, corresponding to the momentum dependence of $\lambda_-(q^2)$. At the present stage at least the excitations in the graviton sector (traceless transversal tensor fluctuations of the metric) should be stable for a suitable choice of parameters. For Einstein gravity the physical scalar mode in the metric is not bounded from below in euclidean flat space. Nevertheless, this theory is stable due to the positive energy condition in Minkowski space. The situation may be similar in our model. It will be an interesting task to find for the various couplings multiplying invariants the range for which the excitations in flat space are stable.

\paragraph*{Emergence of general four-derivative gravity}

%Absatz
Finally, in addition to $L_F$ we can also construct invariants involving different index contractions by forming the tensor $F_{\mu\nu} = \tensor{F}{_{\mu\rho\nu}^\rho}$ and the scalar $F=\tensor{F}{_\mu^\mu}$. Invariants based on $F^{*\,\nu}_\mu \tensor{F}{_\nu^\mu}$ and $F^*F$ yield for the effective low-momentum actions terms $\sim R_{\mu\nu}R^{\mu\nu}$ and $\sim R^2$. These additional terms modify the stability analysis quantitatively. 
Furthermore, invariants with two derivatives can be constructed by index contractions differing from eq.~\eqref{eq:29}, keeping in mind $F_{\nu\mu}\neq F_{\mu\nu}$, $F_{\mu\nu\rho\sigma}\neq F_{\rho\sigma\mu\nu}$. In addition, we can construct invariants employing the totally antisymmetric tensor $\varepsilon_{\mu\nu\rho\sigma}$. The coefficients of all these invariants may differ for the short-distance theory $q^{2}\gg m^{2}$ and the low momentum theory $q^{2}\ll m^{2}$ due to the renormalization flow of couplings.
One expects a renormalization flow from the couplings in the classical action for the short distance limit to the couplings in the quantum effective action in the long distance limit. For the short distance theory one wants to establish a set of couplings for which the functional integral is well defined. For the couplings in the effective action the set of couplings should guarantee stability of Minkowski space.

\section{Discussion\label{sec:9}}

%Absatz
We have proposed a model of pregeometry based on the local gauge symmetry SO(4,\,$\mathbb{C}$) and diffeomorphism invariance. If this is a consistent quantum field theory, general relativity emerges naturally in the low-momentum limit. Our theory would then be a valid model for quantum gravity.

%Absatz
Four issues need a more profound clarification: 
\begin{enumerate*}[label=(\roman*)]
\item renormalizability, 
\item unitarity,
\item stability,
\item consistency of Yang-Mills theories with non-compact gauge group, together with the associated questions of time-space asymmetry and analytic continuation.
\end{enumerate*}
We address these open points in a short discussion. 

%Absatz
In a quantum field theory the exact field equations, propagators and interaction vertices are all derived from the quantum effective action $\Gamma$. This includes all effects of quantum fluctuations. Questions of stability and causality concern the solutions of the field equations derived from the quantum effective action. In the second part of this paper our discussion of field equations, solutions and stability should be interpreted as an approximation to the quantum effective action which includes the corresponding invariants. 

%Absatz
In a quantum field theory the couplings multiplying these invariants are ``running'' or scale-dependent couplings. Instead of explicitly discussing the dependence of propagators and vertices on the momenta involved, it is convenient to introduce an infrared cutoff $k$ such that only quantum fluctuations with momenta $q^2 > k^2$ are effectively included in the effective average action $\Gamma_k$. This allows one to investigate the running of couplings in an approximation with a finite number of derivatives, instead of an explicit discussion of the non-localities related to running couplings in dependence on external momenta. Since external momenta do not always act as universal infrared cutoffs the effective average action implements the stepwise inclusion of fluctuations more directly. There is, in general, no direct translation between the running of couplings with $k$ and the dependence of vertices on external momenta.

%Absatz
We can consider the couplings $Z(k)$, $m^2(k)$, $M^2(k)$ etc.\ as $k$-dependent couplings. In principle, their running with $k$ can be derived from suitable approximations to an exact flow equation \cite{CWFE} for the $k$-dependence of $\Gamma_k$. For $k\to 0$ the effective average action $\Gamma_k$ equals the quantum effective action $\Gamma$ since all fluctuations are included. For $k\to\infty$ no fluctuations are included and $\Gamma_k$ equals the microscopic or ``bare'' action $S$. For an assessment if the proposed SO(4,\,$\mathbb{C}$)-gauge theory constitutes a renormalizable quantum field theory one needs to understand the running of couplings. For example, it is possible that $Z(k\to\infty) \to \infty$, in which case the gauge interactions are asymptotically free. On the other hand, it seems hard to turn off the interactions of the vierbein in the ultraviolet (UV) limit $k\to\infty$. In this case renormalizability can be achieved by asymptotic safety if there exists an UV-fixed point for all dimensionless couplings as $\tilde{m}^2(k) = m^2(k)/k^2$ etc. The issue of renormalizability involves then the establishment of the existence of an UV-fixed point. Many features of asymptotic safety for gravity, as predictions for parameters (renormalizable couplings) in the standard model of particle physics, will be similar for our model.

%Absatz
Unitarity concerns the microscopic formulation of the theory or the classical action. It is a statement that the microscopic Hamiltonian that describes the evolution from a given time-hypersurface to the infinitesimally neighboring one should be hermitian. For formulating this problem in a consistent way the functional integral defining the quantum field theory with Minkowski signature should be well defined. This issue involves to a large extent the properties of the microscopic or classical propagators. It is believed that a model has high chances to be unitary if all poles of propagators correspond to physical particles rather than tachyons or ghosts. No poles in the complex plane should hinder analytic continuation. The propagators should fall off sufficiently fast for $|q^2|\to\infty$ and show no essential singularities. If the high-momentum behavior of the action is well approximated by the invariants $L_F + L_U + L_W$ we have shown that these properties are obeyed in the short distance limit. The propagators of all physical excitations are proportional $q^{-2}$ or $(q^2+m^2)^{-1}$. The question arises if these properties hold for all momenta, including the low-momentum limit. 

%Absatz
In perturbation theory the issue of unitarity is closely related to the issue of stability of small fluctuations around a given ground state or cosmological solution. Beyond perturbation theory the connection gets more loose since stability concerns the quantum effective action, while unitarity involves properties of the classical action. More in detail, stability concerns the behavior of solutions of the field equations and therefore involves the properties of the quantum effective action from which they are derived. In a non-perturbative context the relevant couplings in the effective action may differ substantially from the classical action. More generally, even the degrees of freedom may differ. We follow here a simple assumption about the form of the effective action by considering a set of invariants with up to two derivatives. The couplings multiplying these invariants may differ from those for the classical action. Many couplings, as for example the terms linear in derivatives~$\sim F$, may actually be absent in the classical action.

%Absatz
A first discussion of stability is rather encouraging. For euclidean flat space we have found for many modes that also for low momenta the diagonalized propagators, corresponding to $\lambda_\pm^{-1}(q^2)$, do not lead to tachyonic or ghost poles if the parameters are in a suitable range. In this case the ghosts appearing in a four-derivative approximation to gravity are fake -- they are artifacts of the approximation \cite{PLCW}. The important question is therefore if the running couplings remain within this ``range of stability''. The issue of metastability of euclidean flat space in the scalar sector seems similar to Einstein-gravity where no instability occurs due to the positive energy theorem. 

%Absatz
The issues of renormalizability, unitarity and stability can first be investigated in a simpler theory for euclidean gravity. For this purpose the vierbein $\tensor{e}{_\mu^m}$ and the gauge fields $A_{\mu mn}$ can be taken real, and the gauge group is the compact group SO(4). An extension to pregeometry with gauge group SO(4,\,$\mathbb{C}$) presumably relies strongly on the properties of analytic continuation. We emphasize, however, that the theory with complex vierbeins and gauge fields has additional degrees of freedom which will influence the running of couplings and the properties of a possible UV-fixed point.

%Absatz
Our first discussion of consistency of a gauge theory based on the non-compact gauge group SO(4,\,$\mathbb{C}$) is also encouraging. For a suitable form of the effective action all physical degrees of freedom have acceptable propagators for large $q^2$ when expanded around euclidean flat space. 
This is not trivial due to the non-compact character of the gauge group SO(4,\,$\mathbb{C}$). For example, the invariant $\eta_{mn}$ for the SO(1,\,3)-subgroup has one negative eigenvalue, such that a standard kinetic term constructed only with the real metric $\delta_{\mu\nu}$ of euclidean space would lead to the presence of a ghost-mode with a negative sign of the kinetic term. The term $L_{F}$ in eq.~\eqref{eq:29} demonstrates how such ghost modes are avoided by use of the complex conjugate $F_{\mu\nu\rho\sigma}^{*}\,$. The construction of the SO(4,\,$\mathbb{C}$) invariant field strength $F_{\mu\nu\rho\sigma}$ requires the presence of the complex vierbein and would not be possible if only real vierbeins are admitted. 

%Absatz
If stability holds for euclidean space, the propagators in Minkowski space can be obtained by analytic continuation. As well known from gauge fields, the second functional derivative of the effective action can change sign when continued from euclidean to Minkowski space. This occurs for all components of fields with an odd number of zero indices, due to contractions with $\eta_{\mu\nu}$ instead of $\delta_{\mu\nu}$. The issue is well understood for gauge fields $A_\mu^z$ with a single world index and poses no problem. Better understanding may be needed for objects with several indices as $H_{\mu\nu}$ or $A_{\mu\nu\rho}$. Our description of analytic continuation as continuation in the value of the complex vierbein field gives an unambiguous prescription for analytic continuation, including chiral fermion fields \cite{CWES}. This provides for a solid starting point for a discussion of stability in ground states with a metric signature different from euclidean signature.

%Absatz
For the effective action at low momenta one wants to end with a geometric description involving a real metric $g_{\mu\nu}\,$. We have found a simple scenario how the properties of the ground state can single out a real metric $g_{\mu\nu}$ as a composite field from complex vierbeins. This definition is valid for a vanishing ``cosmological constant'' $V_0$, or more generally the vanishing of a term linear in the trace of the complex pseudo-metric. Adaptation to $V_0 \neq 0$ needs additional work.

%Absatz
Even tough our proposal for pregeometry still has important open issues, it nevertheless also seems to lead to important simplifications as compared to other scenarios of quantum gravity based only on the metric, or associated geometric quantities in discrete settings. A main advantage is the simple form of the inverse propagator in the short distance limit. It can be well described by an action containing only two derivatives, avoiding issues of causality, locality and lack of unitarity. Furthermore, the difference between time and space needs not to be postulated a priori. Time-space asymmetry can emerge from spontaneous symmetry breaking as a property of the ground state or cosmological state.

\section*{Acknowledgement}

%Absatz
The author would like to thank M. Yamada and V. Rubakov for discussions.

\renewcommand*\appendixpagename{\Large Appendices}
\appendix
\appendixpage

\section{Mode expansion and propagators for euclidean pregeometry \label{sec: Ap.A}}

%Absatz
In this appendix we discuss a simpler version of pregeometry, based on the compact gauge group SO(4). The vierbein and the gauge fields are here real fields, in contrast to the complex fields in the main text. The emergent low-momentum effective theory will be euclidean gravity. By analytic continuation we can reach an SO(1,\,3)-gauge theory. The emergent gravity is then general relativity with Minkowski signature. The SO(4)-gauge theory discussed in this appendix is a genuine euclidean quantum field theory with a well defined functional integral, once regularized, for example, on a lattice.  As we have discussed in section \ref{sec:4}, it can also be taken as a partial description of the "real metric and real gauge field limit" of the SO(4,\,$\mathbb{C}$)-gauge theory. 

%Absatz
Many results of the main text can be taken over by simply omitting the imaginary parts of the complex vierbein and gauge fields. For this case  the action is real and bounded from below. This will allow us to discuss further aspects of our proposal of pregeometry based on Yang-Mills theories. 
The number of degrees of freedom being smaller than for the SO(4,\,$\mathbb{C}$) - gauge theory, we can also perform a simpler discussion of the mode expansion and stability. We recall, however, that the full discussion of stability for the SO(4,\,$\mathbb{C}$) - gauge theory needs to include the imaginary parts of the vierbein and gauge fields. Also the renormalization group running will be different for the SO(4,\,$\mathbb{C}$)~- and SO(4) - gauge theories.
This appendix partly overlaps with material in ref.~\cite{Wetterich:2021ywr}. This should enhance the self-consistency of the present paper and give the reader direct access to examples illustrating the general discussion in sect.~\ref{sec:7}.

\subsection{Mode decomposition\label{subsec: Mode decomposition}}
\medskip

%Absatz
We decompose the fluctuations of the vierbein and the gauge fields around euclidean flat space and zero gauge fields. The decomposition involves different representations of the "euclidean Lorentz group" SO(4) that cannot mix in quadratic order. Mixing is observed, however, between modes belonging to the same representation. This renders the discussion of stability rather complex. We discuss the general structure of the mixing and focus subsequently on the graviton propagator and the scalar propagator. We find a stable graviton propagator without ghosts or tachyons. The mode decomposition of the vierbein can be taken over from sections \ref{sec:3}, \ref{sec:4}. We supplement this here by a decomposition of the gauge fields.

%Absatz 
We decompose the linear gauge fields $A_{\mu\nu\rho}$ in flat space into transversal modes $B_{\mu\nu\rho}$ and longitudinal modes $L_{\nu\rho}$,
\begin{equation}\label{eq: DC01}
A_{\mu\nu\rho}=B_{\mu\nu\rho}+\partial_{\mu}L_{\nu\rho}\,,\quad \partial^{\mu}B_{\mu\nu\rho}=0\;.
\end{equation}
With
\begin{equation}\label{eq: DC02}
P_{\mu}^{\nu}=\delta_{\mu}^{\nu}-\dfrac{\partial_{\mu}\partial^{\nu}}{\partial^{2}}\,,\quad\partial^{\mu}P_{\mu}^{\nu}=0\,,\quad P_{\mu}^{\nu}\partial_{\nu}=0\;,
\end{equation}
we can write
\begin{equation}\label{eq: DC03}
B_{\mu\nu\rho}=P_{\mu}^{\sigma}A_{\sigma\nu\rho}=P_{\mu}^{\sigma}B_{\sigma\nu\rho}\;.
\end{equation}
The transversal fluctuations can be decomposed as
\begin{align}\label{eq: DC04}
B_{\mu\nu\rho}=\dfrac{1}{4}\varepsilon_{\nu\rho}^{\quad\sigma\tau}(P_{\mu\sigma}v_{\tau}-P_{\mu\tau}v_{\sigma})\quad\quad \nn \\
+\dfrac{1}{3}(P_{\mu\nu}w_{\rho}-P_{\mu\rho}w_{\nu})+D_{\mu\nu\rho}\;,\nn\\
D_{\mu\nu\rho}=\dfrac{1}{2}(\partial_{\nu}E_{\mu\rho}-\partial_{\rho}E_{\mu\nu})+C_{\mu\nu\rho}\;,
\end{align}
with transversal traceless symmetric tensor $E_{\mu\nu}$
\begin{equation}\label{eq: DC05}
\partial^{\mu}E_{\mu\nu}=0\,,\quad\delta^{\mu\nu}E_{\mu\nu}=0\,,\quad E_{\mu\nu}=E_{\nu\mu}\;,
\end{equation}
and $C_{\mu\nu\rho}$ obeying
\begin{align}\label{eq: DC06}
C_{\mu\nu\rho}=-C_{\mu\rho\nu}\,,\quad \partial^{\mu}C_{\mu\nu\rho}=0\,,\quad\varepsilon^{\mu\nu\rho\sigma}C_{\mu\nu\rho}=0\,,\nn\\
\delta^{\mu\nu}C_{\mu\nu\rho}=\delta^{\mu\rho}C_{\mu\nu\rho}=0\,,\quad \tilde{P}_{\sigma\tau}^{\quad\nu\rho}C_{\mu\nu\rho}=0\;.
\end{align}
Here the projector
\begin{equation}\label{eq: DC07}
\tilde{P}_{\sigma\tau}^{\quad\nu\rho}=\dfrac{1}{2\partial^{2}}(\partial_{\sigma}\partial^{\nu}\delta_{\tau}^{\rho}-\partial_{\tau}\partial^{\nu}\delta_{\sigma}^{\rho}-\partial_{\sigma}\partial^{\rho}\delta_{\tau}^{\nu}+\partial_{\tau}\partial^{\rho}\delta_{\sigma}^{\nu})\;,
\end{equation}
obeys
\begin{align}\label{eq: DC08}
\tilde{P}_{\sigma\tau}^{\quad\alpha\beta}\tilde{P}_{\alpha\beta}^{\quad\nu\rho}=\tilde{P}_{\sigma\tau}^{\quad\nu\rho}\,,\quad \tilde{P}_{\nu\rho}^{\quad\nu\rho}=3\,,\nn\\ \quad \partial^{\tau}\tilde{P}_{\sigma\tau}^{\quad\nu\rho} =\dfrac{1}{2}(\partial^{\rho}\delta_{\sigma}^{\nu}-\partial^{\nu}\delta_{\sigma}^{\rho})\;,
\end{align}
and
\begin{align}\label{eq: DC09}
\tilde{P}_{\sigma\tau}^{\quad\nu\rho} D_{\mu\nu\rho}=(\partial_{\sigma}E_{\mu\tau}-\partial_{\tau}E_{\mu\sigma})\,,
\nn\\
\tilde{P}_{\sigma\tau}^{\quad\nu\rho} (\partial_{\nu}E_{\mu\rho}-\partial_{\rho}E_{\mu\nu})=(\partial_{\sigma}E_{\mu\tau}-\partial_{\tau}E_{\mu\sigma})\,,\nn\\
C_{\mu\nu\rho}=D_{\mu\nu\rho}-\tilde{P}_{\nu\rho}^{\quad\sigma\tau}D_{\mu\sigma\tau}\;.
\end{align}

%Absatz
Out of the 24 components $A_{\mu\nu\rho}$ six degrees of freedom $L_{\nu\rho}=-L_{\rho\nu}$ are longitudinal degrees of freedom. Thus there are 18 transversal modes $B_{\mu\nu\rho}$. The four modes $v_{\mu}$ correspond to the totally antisymmetric part $A_{[\mu\nu\rho]}$, while the four modes $w_{\rho}$ account for the trace $A_{\mu\nu\rho}\delta^{\mu\nu}$. There remain 10 modes for $D_{\mu\nu\rho}$. The projector $\tilde{P}$ eliminates half of them, and the traceless transversal symmetric tensor $E_{\mu\nu}$ accounts indeed for five modes. The other five modes correspond to $C_{\mu\nu\rho}=-C_{\mu\rho\nu}$, which indeed is subject to 19 constraints. The vectors $v_{\mu}$ and $w_{\mu}$ can be decomposed into transversal vectors and scalars
\begin{align}\label{eq: DC10}
&v_{\mu}=v_{\mu}^{(t)}+\partial_{\mu}\tilde{v}\,,\quad w_{\mu}=w_{\mu}^{(t)}+\partial_{\mu}\tilde{w}\,,\nn\\ &\partial^{\mu}v_{\mu}^{(t)}=0\,,\quad \partial^{\mu}w_{\mu}^{(t)}=0\;.
\end{align}
The irreducible representations of the transversal modes $B_{\mu\nu\rho}$ are $2\times 5 + 2\times 3+2\times 1$. In particular, the scalar part of $B_{\mu\nu\rho}$ reads
\be\label{eq: DC11}
B_{\mu\nu\rho}^{(s)}=\dfrac{1}{2}\varepsilon_{\mu\nu\rho}^{\quad\;\,\tau}\partial_{\tau}\tilde{v}+\dfrac{1}{3}(\delta_{\mu\nu}\partial_{\rho}\tilde{w}-\delta_{\mu\rho}\partial_{\nu}\tilde{w})\;.
\ee
Finally, the six longitudinal modes are two triplets
\be\label{eq: DC12}
L_{\nu\rho}=M_{\nu\rho}+\partial_{\nu}l_{\rho}-\partial_{\rho}l_{\nu}\;,
\ee
with
\be\label{eq: DC13}
\partial^{\nu}M_{\nu\rho}=\partial^{\rho}M_{\nu\rho}=0\,,\quad\partial^{\nu}l_{\nu}=0\;.
\ee

%Absatz
This decomposition can be performed equally for complex fields, or separately for their real and imaginary parts. It can be the basis for a stability discussion of the full SO(4,\,$\mathbb{C}$) - gauge theory. Since all fields carry only word-indices they are SO(4,\,$\mathbb{C}$) - invariant.

\subsection{High momentum limit for real vierbein and gauge fields.\label{subsec: High momentum limit}}

%Absatz
In the following we focus the discussion by considering the euclidean gauge group SO(4) with real gauge fields $A_{\mu mn}$ and real vierbeins $e_{\mu}^{\quad\!\!\!m}$. We expand around flat space, $\overline{e}_{\mu}^{\quad \!\!\!m}=\delta_{\mu}^{m}$.  The source term $L_{U}^{(3)}$ mixes the gauge fields $A_{\mu\nu\rho}$ and the vierbein fluctuations $H_{\mu\nu}$. In quadratic order one finds from eq.
\be\label{eq: DC14}
\begin{split}
	\int_{x}\!\overline{e}L_{U}^{(3)}\!=-\dfrac{m^{2}}{4}\!\!\!\int_{x}\!\!\!A^{\mu\nu\rho}\Bigl(\partial_{\mu}H_{\nu\rho}^{(A)}\!+\!\partial_{\rho}H_{\mu\nu}^{(S)}\!-\!\partial_{\nu}H_{\mu\rho}^{(S)}\Bigr)\\
	=\dfrac{m^{2}}{4}\int_{x}\biggl{\lbrace}L^{\nu\rho}\Bigl(\partial^{2}H_{\nu\rho}^{(A)}+\partial_{\rho}\partial^{\mu}H_{\mu\nu}^{(S)}-\partial_{\nu}\partial^{\mu}H_{\mu\rho}^{(S)}\Bigr) \\
	+2\partial_{\rho}B^{\mu\nu\rho}H_{\mu\nu}^{(S)}\biggr{\rbrace}=\dfrac{m^{2}}{4}\int_{x}(Y_{l}+Y_{t})\;.
\end{split}
\ee
For the longitudinal part $Y_{l}$ we employ eqs. \eqref{eq:L1}-\eqref{eq:L3}, \eqref{eq:M25}
\be\label{eq: DC15}
Y_{l}=M^{\nu\rho}\partial^{2}b_{\nu\rho}+2l^{\mu}\partial^{4}(\kappa_{\mu}-\gamma_{\mu})\;.
\ee

%Absatz
For the transversal modes we decompose the tensor~$\partial_{\rho}B^{\mu\nu\rho}$ into its traceless symmetric, antisymmetric, and (modified) trace parts
\be\label{eq: DC16}
\partial_{\rho}B^{\mu\nu\rho}=\tilde{B}^{(S)\mu\nu}+\tilde{B}^{(A)\mu\nu}+\dfrac{1}{3}P^{\mu\nu}\tilde{b}\;,
\ee
with
\begin{align}\label{eq: DC17}
\!\!\!\!\tilde{B}^{(S)\mu\nu}\!\!=\!\tilde{B}^{(S)\nu\mu}&,\quad\!\! \tilde{B}^{(S)\mu\nu}\delta_{\mu\nu}\!\!=\!0,\quad\!\!\tilde{B}^{(A)\mu\nu}\!=\!-\tilde{B}^{(A)\nu\mu}\,,\!\!\!\!\nn\\
\partial_{\mu}&\tilde{B}^{(S)\mu\nu}+\partial_{\mu}\tilde{B}^{(A)\mu\nu}=0\;.\!\!\!\!
\end{align}
Comparing with the expansion~\eqref{eq: DC04} we identify
\ba\label{eq: DC18}
&\tilde{B}^{(S)\mu\nu}=-\dfrac{1}{2}\partial^{2}E^{\mu\nu}\,,\quad \tilde{B}^{(A)\mu\nu}=\dfrac{1}{2}\varepsilon^{\mu\nu\rho\tau}\partial_{\rho}v_{\tau}^{(t)}\,,\nn\\ 
&\quad\quad\tilde{b}=\partial^{2}\tilde{w}\;.
\end{eqnarray}
The components $C_{\mu\nu\rho}$ do not appear due to the identity
\be\label{eq: DC19}
\partial^{\rho}C_{\mu\nu\rho}=0\;,
\ee
which follows from eqs.\eqref{eq: DC08},\eqref{eq: DC09}. Inserting also the expansion \eqref{eq:L1},\eqref{eq:M25} for $H_{\mu\nu}^{(S)}$ the transversal part of eq. \eqref{eq: DC14} reads
\be\label{eq: DC20}
\!\!\!\! Y_{t}\!=\!\Bigl(\!2\tilde{B}^{(S)\mu\nu}\!+\dfrac{2}{3}P^{\mu\nu}\tilde{b}\Bigr)H_{\mu\nu}^{(S)}=\!-E^{\mu\nu}\partial^{2}t_{\mu\nu}\!+\dfrac{2}{3}\tilde{w}\partial^{2}\sigma\,.\!\!\!
\ee
We observe that the twelve transversal gauge field fluctuations $v^{(t)}$, $w^{(t)}$, $\tilde{v}$, $C_{\mu\nu\rho}$ do not appear in $L_{ U}^{(3)}$.
Also the four gauge modes $\kappa_{\mu}+\gamma_{\mu}$ and $u$ are absent.

%Absatz
Similarly, we can decompose the gauge boson mass term~$L_{U}^{(2)}$,
\be\label{eq: DC21}  
\begin{split}
	L_{U}^{(2)}=\dfrac{m^{2}}{4}A^{\mu\nu\rho}A_{\mu\nu\rho}=\dfrac{m^{2}}{4}(B^{\mu\nu\rho}B_{\mu\nu\rho}-L^{\nu\rho}\partial^{2}L_{\nu\rho})\\
	=\dfrac{m^{2}}{4}\Bigl{\lbrace}-\dfrac{1}{2}E^{\nu\rho}\partial^{2}E_{\nu\rho}+C^{\mu\nu\rho}C_{\mu\nu\rho}+v^{(t)\mu}v_{\mu}^{(t)}-\dfrac{3}{2}\tilde{v}\partial^{2}\tilde{v}\\
	+\dfrac{4}{9}w^{(t)\mu}w_{\mu}^{(t)}-\dfrac{2}{3}\tilde{w}\partial^{2}\tilde{w}-M^{\nu\rho}\partial^{2}M_{\nu\rho}+2l^{\mu}\partial^{4}l_{\mu}\Bigr{\rbrace}\;,
\end{split}
\ee
and employ eq.\eqref{eq:L4} for $L_{U}^{(1)}$,
\be\label{eq: DC22}  
\begin{split}
	L_{U}^{(1)}=\dfrac{m^{2}}{4}\Bigl{\lbrace}-\dfrac{1}{2}t^{\mu\nu}\partial^{2}t_{\mu\nu}-\dfrac{1}{4}b^{\mu\nu}\partial^{2}b_{\mu\nu}\\+\dfrac{1}{2}(\kappa^{\mu}-\gamma^{\mu})\partial^{4}(\kappa_{\mu}-\gamma_{\mu}) 
	-\dfrac{1}{6}\sigma\partial^{2}\sigma\Bigr{\rbrace}\;.
\end{split}
\ee
Taking things together, $L_{U}$ involves a couple of independent pieces
\be\label{eq: DC23}
L_{U}=\dfrac{m^{2}}{4}\bigl{\lbrace} L_{tE}+L_{\sigma\tilde{w}}+L_{bM}+L_{\kappa\gamma l}+L_{m}'\bigr{\rbrace}\;.
\ee

%Absatz
For the transversal traceless symmetric tensors $t$ and $E$ one has
\be\label{eq: DC24}
L_{tE}=-\dfrac{1}{2}(t^{\mu\nu}+E^{\mu\nu})\partial^{2}(t_{\mu\nu}+E_{\mu\nu})\;,
\ee
while in the scalar sector $\sigma$ and $\tilde{w}$ are connected
\be\label{eq: DC25}
L_{\sigma\tilde{w}}=-\dfrac{1}{6}(\sigma-2\tilde{w})\partial^{2}(\sigma-2\tilde{w})\;.
\ee
The sector for the longitudinal gauge bosons involves
\be\label{eq: DC26}
L_{b M}=-\dfrac{1}{4}(b^{\mu\nu}-2M^{\mu\nu})\partial^{2}(b_{\mu\nu}-2M_{\mu\nu})\;,
\ee
and
\be\label{eq: DC27}
L_{k\gamma l}=\dfrac{1}{2}(\kappa^{\mu}-\gamma^{\mu}+2l^{\mu})\partial^{4}(\kappa_{\mu}-\gamma_{\mu}+2l_{\mu})\;.
\ee
Up to absorption of momentum-dependent normalization factors the mixing takes the form of the terms $m^{2}$ in eq. \eqref{eq:LM18}.
The remaining part $L_{m}'$ involves only mass terms for the gauge bosons and no mixing
\be\label{eq: DC28}
L_{m}'=C^{\mu\nu\rho}C_{\mu\nu\rho}+v^{(t)\mu}v_{\mu}^{(t)}-\dfrac{3}{2}\tilde{v}\partial^{2}\tilde{v}+\dfrac{4}{9}w^{(t)\mu}w_{\mu}^{(t)}\;.
\ee

%Absatz
The kinetic term for the gauge bosons \eqref{eq:E4} involves only the transversal gauge bosons,
\be\label{eq: DC29}
\!\! L_{F}=-\dfrac{Z_{F}}{4}A^{\mu\nu\rho}\partial^{2}P_{\mu}^{\;\,\sigma} A_{\sigma\nu\rho}=-\dfrac{Z_{F}}{4}B^{\mu\nu\rho}\partial^{2}B_{\mu\nu\rho}\;.
\ee
The decomposition is similar to the transversal part of eq.~\eqref{eq: DC21}, replacing $m^{2}$ by $-Z_{F}\partial^{2}$,
\be\label{eq: DC30}
\begin{split}
	L_{F}=\dfrac{Z_{F}}{4}\biggl{\lbrace}\dfrac{1}{2}E^{\nu\rho}\partial^{4}E_{\nu\rho}-C^{\mu\nu\rho}\partial^{2}C_{\mu\nu\rho}-v^{(t)\mu}\partial^{2}v_{\mu}^{(t)} \\
	+\dfrac{3}{2}\tilde{v}\partial^{4}\tilde{v}-\dfrac{4}{9}w^{(t)\mu}\partial^{2}w_{\mu}^{(t)}+\dfrac{2}{3}\tilde{w}\partial^{4}\tilde{w}\biggr{\rbrace}\;.
\end{split}
\ee

\subsection{Unitarity\label{subsec Ap.A: Unitarity}}

%Absatz
Let us consider a microscopic or classical action based on $L_{F}+L_{U}$.
With an appropriate normalization, the inverse propagator for the gauge boson fluctuations $C$, $v^{(t)}$, $w^{(t)}$ and $\tilde{v}$ is the standard one for massive particles in momentum space
\be\label{eq: DC31}
P=G^{-1}=q^{2}+\dfrac{m^{2}}{Z_{F}}\;.
\ee
In the $t-E$ - sector the inverse propagator matrix takes the form
\be\label{eq: DC32}
P=G^{-1}=\dfrac{q^{2}}{4}\begin{pmatrix}
	Z_{F}q^{2}+m^{2}&, \,m^{2}\\
	m^{2}&,\,m^{2}
\end{pmatrix}\;.
\ee
The only zero eigenvalue of $P$ occurs for $q^{2}=0$, corresponding to eq.\eqref{eq:LM19} for $c^{2}=b$. The eigenvalues of $P$ are
\be\label{eq: DC34}
\lambda_{\pm}=\dfrac{q^{2}}{8}\Big(Z_{F}q^{2}+2m^{2}\pm\sqrt{Z_{F}^{2}q^{4}+4m^{4}}\,\Bigr)\;.
\ee
Up to an overall normalization factor $q^{2}/4$ this corresponds to eq. \eqref{eq:LM20} with $b=m^{2}/q^{2}$, $c^{2}=b$. The different normalizations factors can be absorbed in a momentum dependent renormalization of the fields, $E_{R\mu\nu}=\dfrac{q}{2}E_{\mu\nu}$, $q=\sqrt{|q^{2}|}$, $t_{R\mu\nu}=\dfrac{m}{2}t_{\mu\nu}$, resulting in 
\be\label{eq: DC35}
P_{R}=
\begin{pmatrix}
	Z_{F}q^{2}+m^{2}&,\;mq\\
	mq&,\;q^{2}
\end{pmatrix}
\ee
This equals eq. \eqref{eq:LM21} for $b=c=1$, up to factors of $i$ that do not affect the eigenvalues and have been used in eq. \eqref{eq:LM21} for a formal correspondence $\partial_{\mu}=iq_{\mu}$. The $q$- dependence in the renormalization of $E$ corresponds to the standard normalization of gauge fields $A$, noting $A\sim qE$, while the normalization of $t$ provides a canonical mass dimension to this field. For the renormalized fields the eigenvalues of~$P_{R}$ are $(Z=Z_{F})$
\be\label{eq: DC36}
\begin{split}
	&\lambda_{\pm}=\dfrac{1}{2}\Bigl{\lbrace}(Z+1)q^{2}+m^{2}\\ 
	&\quad\quad\quad\quad\pm\sqrt{(Z-1)^{2}q^{4}+2(Z+1)q^{2}m^{2}+m^{4} }\Bigr{\rbrace}\;.
\end{split}
\ee
For the high momentum behavior, $q^{2}\gg m^{2}$, this yields
\begin{align}\label{eq: DC37}
\lambda_{+}=
\begin{cases}
Z\Bigl(q^{2}+\dfrac{m^{2}}{Z-1}\Bigr) \phantom{\Bigg(}&\text{for $Z>1$}\\
q^{2}+\dfrac{m^{2}}{1-Z}&\text{for $Z<1$} \;,
\end{cases}
\nn\\ \nn \\
\lambda_{-}=
\begin{cases}
q^{2}-\dfrac{m^{2}}{Z-1} \phantom{\Bigg(}&\text{for $Z>1$}\\
Z\Bigl(q^{2}-\dfrac{m^{2}}{1-Z}\Bigr)&\text{for $Z<1$} \;.
\end{cases}
\end{align}
Near the pole at $q^{2}=0$ one finds
\be\label{eq: DC38}
\lambda_{-}=\dfrac{Zq^{4}}{m^{2}}\quad ,\quad\lambda_{+}=m^{2}+(Z+1)q^{2}\;,
\ee
in accordance with eq. \eqref{eq:LM28}. 

%Absatz
In the $\tilde{w}-\sigma$ - sector the inverse propagator matrix
\be\label{eq: DC39}
P=\dfrac{q^{2}}{3}\begin{pmatrix}
	Zq^{2}+m^{2}&,\; -\dfrac{m^{2}}{2}\\
	-\dfrac{m^{2}}{2}&\!\! ,\; \phantom{\Bigg|}\phantom{-}\dfrac{m^{2}}{4}\;
\end{pmatrix}
\ee
has the same structure as eq.~\eqref{eq: DC32}, up to an overall~factor~$4/3$ and a rescaling of $\sigma$ by a factor $(-2)$.

%Absatz
For the longitudinal gauge bosons there is no contribution from $L_{F}$. This corresponds to the case $Z=0$, $c^{2}=b$ in eq. \eqref{eq:LM26}. The inverse propagator matrix in the $M-b$ - sector obtains from $L_{bM}$ in eq.~\eqref{eq: DC26}
\be\label{eq: DC40}
P=\dfrac{q^{2}m^{2}}{2}\begin{pmatrix}
	\phantom{\Bigg|}\;1&-\dfrac{1}{2}\\-\dfrac{1}{2}&\phantom{-}\dfrac{1}{4}\;\;
\end{pmatrix}\;.
\ee
After proper renormalization of the fields this becomes
\be\label{eq: DC41}
P_{R}=\begin{pmatrix}
	\phantom{\Big|}m^{2}&-mq\\-mq&q^{2}
\end{pmatrix}\;.
\ee
The eigenvalue $\lambda_{-}$ is zero for all $q^{2}$, corresponding to $\det P=0$ or $\det P_{R}=0$. This reflects the gauge mode of local SO(4)-gauge transformations. Indeed, these gauge transformations, applied to a "vacuum state" $A_{\mu mn}=0$, $e_{\mu}^{m}=\delta_{\mu}^{m}$, do not only shift the longitudinal components of $A_{\mu mn}$, but also rotate $e_{\mu}^{m}$, contributing infinitesimally to~$H_{\mu\nu}^{(A)}$.
For this reason the gauge modes are a linear combination of $L_{\mu\nu}$ and $b_{\mu\nu}$. The combination corresponding to the gauge mode is the one that does not appear in the SO(4)-invariant tensor $U_{\mu\nu\rho}$. The other linear combination is a physical mode, corresponding to the eigenvalue $\lambda_{+}$ of $P_{R}$
\be\label{eq: DC42}
\lambda_{+}=q^{2}+m^{2}\;.
\ee
This corresponds to the standard propagator for a massive particle.

%Absatz
In summary, all euclidean propagators are well behaved, without tachyons and ghosts for $Z_{F}>0$, $m^{2}>0$. The only touchy point is the double pole in the transversal traceless sector, corresponding to $\lambda_{-}\sim q^{4}$ in eq.~\eqref{eq: DC38}. It is not obvious why this behavior should obstruct the consistent definition of a functional integral.

\subsection{Stability and invariant linear in field strength\label{subsec: Stability and invariant linear}}

%Absatz
For a discussion of stability one has to investigate the quantum effective action. As compared to the classical action this contains additional terms. For example, the invariant $L_{R}$ is crucial for a realistic low-momentum theory for gravity. The new invariants modify the propagators and could induce new tachyonic or ghost instabilities. We investigate this issue by adding to $L_{F}+L_{U}$ the invariant $L_{R}$, cf eq. \eqref{eq:LM12}. In contrast to $L_{F}$ and $L_{U}$ the invariant $L_{R}$ is not a square and can therefore take negative values. Since $L_{R}$ is linear in $q$, it cannot modify the leading behavior $\sim{q}^{2}$ for large $q^{2}$. In the infrared limit of small $q^{2}$ it can play an important role, however. We well see that it dominates the behavior near massless propagator poles at $q^{2}=0$.

%Absatz
We need to expand
\be\label{eq: DC43}
eL_{R}=-\dfrac{M^{2}}{2}ee^{m\mu}e^{n\nu}F_{\mu\nu m n}
\ee
in quadratic order in $H$ and $A$ around $e_{\mu}^{m}=\delta_{\mu}^{m}\,,\, A_{\mu mn}=0$.
Since $F_{\mu\nu mn}$ is a least linear in $A_{\mu mn}$ according to eq. \eqref{eq:24}, we need the vierbein in linear order in $H$,
\ba\label{eq: DC44}
& ee^{m\mu}e^{n\nu}=\delta^{m\mu}\delta^{n\nu} \quad\quad\quad\quad\quad\quad\quad\quad\quad\quad\quad\quad\quad\quad\quad\quad \\
& \quad\quad\quad\quad +\dfrac{1}{2}H_{\rho\sigma}\bigl(\delta^{m\mu}\delta^{n\nu}\delta^{\rho\sigma}-\delta^{m\mu}\delta^{n\rho}\delta^{\nu\sigma}-\delta^{m\rho}\delta^{n\nu}\delta^{\mu\sigma}\bigr)\;. \nn
\end{eqnarray}
The linear term $\partial_{\mu}A_{\nu}^{\quad\mu\nu}-\partial_{\nu}A_{\mu}^{\quad\mu\nu}$ is a total derivative and therefore vanishes. In quadratic order one finds two contributions, $L_{R}=L_{R}^{(1)}+L_{R}^{(2)}$, 
\ba\label{eq: DC45}
&eL_{R}^{(1)}=-\dfrac{M^{2}}{4}H_{\rho\sigma}\Bigl{\lbrace}\bigl(\partial_{\mu}A_{\nu}^{\quad\mu\nu}-\partial_{\nu}A_{\mu}^{\!\!\quad\mu\nu}\bigr)\delta^{\rho\sigma}\\
&\quad\quad -\bigl(\partial_{\mu}A_{\nu}^{\!\!\quad\mu\rho}\!\!-\partial_{\nu}A_{\mu}^{\!\!\quad\mu\rho}\bigr)\delta^{\nu\sigma}\!\!-\!\bigl(\partial_{\mu}A_{\nu}^{\!\!\quad\rho\nu}-\partial_{\nu}A_{\mu}^{\!\!\quad\rho\nu}\bigr)\delta^{\mu\sigma}\!\Bigr{\rbrace}\;,\nn
\end{eqnarray}
and
\be\label{eq: DC46}
eL_{R}^{(2)}=-\dfrac{M^{2}}{2}\Bigl{\lbrace}A_{\mu}^{\quad\mu\rho}A_{\nu\rho}^{\quad\nu}-A_{\nu}^{\quad\mu\rho}A_{\mu\rho}^{\quad\nu}\Bigr{\rbrace}\;.
\ee

%Absatz
For the first term we observe that only the transversal gauge bosons contribute to $F_{\mu\nu mn}$ in linear order,
\be\label{eq: DC47}
\begin{split}
	\!\!\!\!\!\!\!\!eL_{R}^{(1)}\!=\!\dfrac{M^{2}}{2}H_{\rho\sigma}\Bigl{\lbrace}\delta^{\rho\sigma}\partial_{\mu}B_{\nu}^{\!\!\quad\nu\mu}-\partial_{\mu}B^{\sigma\rho\mu}\!-\partial^{\sigma}B_{\mu}^{\!\!\quad\mu\rho}\Big{\rbrace}\phantom{\biggr{\rbrace}}\!\!\!\!\!\!\!\!\\
	\!\!\!\!=\dfrac{M^{2}}{2}H_{\rho\sigma}\biggl{\lbrace}\Bigl(\dfrac{2}{3}\delta^{\rho\sigma}+\dfrac{1}{3}\dfrac{\partial^{\rho}\partial^{\sigma}}{\partial^{2}}\Bigr)\tilde{b}-\tilde{B}^{(S)\sigma\rho}-\tilde{B}^{(A)\sigma\rho}\\-\dfrac{2}{3}\partial^{\sigma}w^{(t)\rho}-\partial^{\sigma}\partial^{\rho}\tilde{w}\biggr{\rbrace}\;.\!\!\!\!\!\!
\end{split}
\ee
The decomposition yields
\be\label{eq: DC48}
\begin{split}
	eL_{R}^{(1)}=\dfrac{M^{2}}{2}\biggl{\lbrace}\dfrac{2}{3}\tilde{w}\partial^{2}\sigma+\dfrac{1}{2}E^{\rho\sigma}\partial^{2}t_{\sigma\rho}\\+\dfrac{1}{2}\varepsilon^{\mu\nu\rho\sigma}v_{\mu}^{(t)}\partial_{\nu}b_{\rho\sigma}
	+\dfrac{2}{3}w^{(t)\mu}\partial^{2}(\kappa_{\mu}-\gamma_{\mu})\biggr{\rbrace}\;.
\end{split}
\ee
This contributes to the mixing between transversal gauge bosons and vierbein fluctuations, similar to eq.~\eqref{eq: DC20}. As it should be, the gauge modes $\kappa_{\mu}+\gamma_{\mu}$ and $u$ do not appear. 
The second term decomposes as
\ba\label{eq: DC49}
& eL_{R}^{(2)}=\dfrac{M^{2}}{2}\biggl{\lbrace}\dfrac{1}{4}E^{\mu\nu}\partial^{2}E_{\mu\nu}+C_{\mu\nu\rho}C^{\nu\rho\mu}  \\
&\quad\quad\quad\quad+\dfrac{2}{9}w^{(t)\mu}w_{\mu}^{(t)}-\dfrac{2}{3}\tilde{w}\partial^{2}\tilde{w}
+\dfrac{1}{2}v^{(t)\mu}v_{\mu}^{(t)}-\dfrac{3}{2}\tilde{v}\partial^{2}\tilde{v}\nn  \\
& +\dfrac{4}{3}w^{(t)\mu}\partial^{2}l_{\mu}-\varepsilon^{\mu\nu\rho\sigma}v_{\mu}^{(t)}\partial_{\nu}M_{\rho\sigma}\biggr{\rbrace}\;. \nn
\end{eqnarray}

%Absatz
The different contributions of $eL_{R}$ can be listed as follows
\be\label{eq: DC50}
eL_{R}=\dfrac{M^{2}}{4}\bigl(\Delta L_{tE}+\Delta L_{\sigma\tilde{w}}+L_{vM}+L_{wL}+L_{C}+L_{\tilde{v}}\bigr)\;.
\ee
Here
\be\label{eq: DC51}
\Delta L_{tE}=E^{\mu\nu}\partial^{2}t_{\mu\nu}+\dfrac{1}{2}E^{\mu\nu}\partial^{2}E_{\mu\nu}\;,
\ee
and
\be\label{eq: DC52}
\Delta L_{\sigma\tilde{w}}=\dfrac{4}{3}(\tilde{w}\partial^{2}\sigma-\tilde{w}\partial^{2}\tilde{w})\;,
\ee
add to eqs.\eqref{eq: DC24}\eqref{eq: DC25} further contributions, while
\be\label{eq: DC53}
L_{vM}=\varepsilon^{\mu\nu\rho\sigma}v_{\mu}^{(t)}\partial_{\nu}(b_{\rho\sigma}-2M_{\rho\sigma})+v^{(t)\mu}v_{\mu}^{(t)}
\ee
can be combined with $L_{bM}$ in eq.~\eqref{eq: DC26}, involving the same physical combination $b_{\mu\nu}-2M_{\mu\nu}$.
Similarly
\be\label{eq: DC54}
L_{wl}=\dfrac{4}{3}w^{(t)\mu}\partial^{2}(\kappa_{\mu}-\gamma_{\mu}+2l_{\mu})+\dfrac{4}{9}w^{(t)\mu}w_{\mu}^{(t)\mu}
\ee
combines with $L_{k\gamma l}$ in eq.~\eqref{eq: DC27}. The remaining parts,
\be\label{eq: DC55}
L_{C}=2C_{\mu\nu\rho}C^{\nu\rho\mu}\;,\quad L_{\tilde{v}}=-3\tilde{v}\partial^{2}\tilde{v}\;,
\ee
add mass terms to the gauge boson fluctuations $C$ and $\tilde{v}$.

\subsection{Graviton propagator\label{subsec: Graviton Propagator}}

%Absatz
In the $t-E$ - sector the renormalized inverse propagator matrix~\eqref{eq: DC35} is replaced by
\be\label{eq: DC56}
P_{R}^{(tE)}=\begin{pmatrix}
	Z_{F}q^{2}+m^{2}-M^{2}&,&\;\dfrac{m^{2}-M^{2}}{m}q\phantom{\Bigg|}\\
	\dfrac{m^{2}-M^{2}}{m}q & , & q^{2}\phantom{\Bigg|}
\end{pmatrix}\;.
\ee
The poles of the propagators or zero eigenvalues of~$P_{R}$ occur for
\be\label{eq: DC57}
q^{2}=0\;,\quad q^{2}=-\mu^{2}\;,
\ee
with
\be\label{eq: DC58}
\mu^{2}=\dfrac{m^{2}}{Z_{F}}y(1-y)\;,\quad y=\dfrac{M^{2}}{m^{2}}\;.
\ee
A stable particle requires $\mu^{2} \geqslant 0$, or
\be\label{eq: DC59}
0\leqslant y \leqslant 1\;,\quad 0\leqslant M^{2}\leqslant m^{2}\;.
\ee
Otherwise one encounters a tachyonic instability. The inverse particle propagators are given by the eigenvalues of~$P_{R}$,
\ba\label{eq: DC60}
&\lambda_{\pm} = \dfrac{1}{2}\Biggl{\lbrace} (Z+1)q^{2}+m^{2}-M^{2} \\
&\quad\quad\;\pm\sqrt{\bigl[ (Z-1)q^{2}+m^{2}-M^{2}\bigr]^{2}+4\dfrac{q^{2}}{m^{2}}(m^{2}-M^{2})^{2}}\;\Biggr{\rbrace}\;.\nn
\end{eqnarray}

For $q^{2}\rightarrow 0$ one finds the massless graviton, corresponding to 
\be\label{eq: DC61}
\lambda_{-}=\dfrac{M^{2}q^{2}}{m^{2}}+\bigl(Z-\dfrac{M^{2}}{m^{2}}\bigr)\dfrac{q^{4}}{m^{2}}\;.
\ee
This reflects the low-momentum effective action \eqref{eq:LM17}.

%Absatz
Near the second pole at $q_{p}^{2}=-\mu^{2}$ one finds for
\be\label{eq: DC62}
-(Z+1)\mu^{2}+m^{2}-M^{2}<0
\ee
or
\be\label{eq: DC63}
Z<Z_{c}\;,\quad Z_{c}=\dfrac{y}{1-y}\;,
\ee
a negative $\lambda_{-} $ for $q^{2}$ near $-\mu^{2}$,
\be\label{eq: DC64}
\lambda_{-}=-m_{t}^{2}-\dfrac{1}{2}\Bigl(Z+1-\dfrac{A_{t}m^{2}}{m_{t}^{2}}\Bigr)(q^{2}+\mu^{2})
\ee
with positive $m_{t}^{2}$ for $Z<Z_{c}$,
\be\label{eq: DC65}
m_{t}^{2}=(Z+1)\mu^{2}-m^{2}+M^{2}=m^{2} y(1-y)\Bigl(\dfrac{1}{Z}-\dfrac{1}{Z_{c}}\Bigr)\;,
\ee
and
\be\label{eq: DC66}
A_{t}=(1-y)\biggl(Z+1-\Bigl(Z+\dfrac{1}{Z}\Bigr) y\biggr)\;.
\ee
Here we have always assumed the range $0\leqslant y\leqslant 1$ required by stability according to eq.~\eqref{eq: DC59}.
For large $|q^{2}|$ one obtains
\begin{align}\label{eq: DC67}
\lim_{q^{2}\rightarrow \infty}\lambda_{-}=
\begin{cases}
q^{2}&\text{for $Z>1$} \\
Zq^{2}&\text{for $Z<1$}\;,
\end{cases} \nn\\
\lim_{q^{2}\rightarrow -\infty}\lambda_{-}=
\begin{cases}
Zq^{2} &\text{for $Z>1$}\\
q^{2} &\text{for $Z<1$}\;.
\end{cases}
\end{align}
For $Z<Z_{c}$ the function $\lambda_{-}(q^{2}) $ remains negative and real for the range of real $q^{2}$ with $-\mu^{2}< q^{2}<0$ .

%Absatz
For $Z>Z_{c}$ the graviton propagator $\lambda_{-}^{-1}(q^{2})$ has a second pole at $q^{2}=-\mu^{2}$, with a behavior for $q^{2}$ near $-\mu^{2}$ given by
\be\label{eq: DC68}
\lambda_{-}(q^{2})=B_{t}(q^{2}+\mu^{2})\;,
\ee
where
\be\label{eq: DC69}
B_{t}=\dfrac{1}{2}\Bigl(Z+1+\dfrac{A_{t}m^{2}}{m_{t}^{2}}\Bigr)=\dfrac{Zy}{y-Z(1-y)}\;.
\ee
For $Z>Z_{c}$ one has $m_{t}^{2}<0$ and $B_{t}<0$. The negative prefactor of $q^{2}+\mu^{2}$ corresponds to a negative residuum at the pole at $q^{2}=-\mu^{2}$. This indicates a ghost instability. A stable theory with a well behaved graviton propagator for  all momenta therefore requires the "stability condition"~\eqref{eq: DC63}, $Z<Z_{c}$.

%Absatz
With this condition the graviton propagator has only a single pole in the complex $q^{2}$- plane at $q^{2}=0$. In the standard normalization with dimensionless metric or vierbein fields the graviton propagator is given by
\be\label{eq: DC69A}
G_\textup{grav}(q^{2})=\dfrac{4}{m^{2}\lambda_{-}(q^{2})}\;.
\ee
This is the graviton propagator \eqref{eq:IN1} mentioned in the introduction.
For $q^{2}$ near zero it reads, cf.~eq.~\eqref{eq: DC61},
\be\label{eq: DC70}
G_\textup{grav}(q^{2})=\dfrac{4}{M^{2}q^{2}}\Bigl(1+(Z-y)\dfrac{q^{2}}{M^{2}}\Bigr)^{-1}\;.
\ee
While the truncated propagator has a ghost pole at $q^{2}=-M^{2}/(Z-y)$, this pole is not present in the propagator~\eqref{eq: DC69A}.
It is therefore an artifact of the truncation \cite{PLCW}.

%Absatz
In the stable range for $Z<Z_{c}$ one obtains for $\lambda_{+}(q^{2})$ for $q^{2}$ in the vicinity of the pole at $q^{2}=-\mu^{2}$,
\be\label{eq: DC71}
\lambda_{+}=B_{t}(q^{2}+\mu^{2})\;.
\ee
With $m_{t}^{2}>0$ this yields $B_{t}>0$. The eigenmode $\lambda_{+}(q^{2})$ corresponds to a stable massive spin-two particle, with mass $\mu$ and residuum at the pole $B_{t}^{-1}>0$. In the stable range both eigenvalues of the inverse propagator matrix~\eqref{eq: DC56} correspond to stable modes. There is no instability in this sector. At $q^{2}=0$ one finds a positive value of $\lambda_{+}$, 
\be\label{eq: DC72}
\lambda_{+}(q^{2}=0)=m^{2}-M^{2}\;.
\ee
For the boundary case $M^{2}=m^{2}$ one observes a second stable massless particle, with
\be\label{eq: DC73}
\lambda_{+}=Zq^{2}\;,\quad \lambda_{-}=q^{2}\;.
\ee

%Absatz
For an overall picture of the $q^{2}$-dependence of the two eigenvalues $\lambda_{\pm}(q^{2})$ we further note that at a critical value $q_{c}^{2}<0$ both eigenvalue can coincide
\be\label{eq: DC74}
\lambda_{+}(q_{c}^{2})=\lambda_{-}(q_{c}^{2})\;.
\ee
This occurs when the square root in eq.~\eqref{eq: DC60} vanishes, determining
\be\label{eq: DC75}
\dfrac{q_{c\pm}^{2}}{m^{2}}=-\dfrac{1-y}{(1-Z)^{2}}\Bigl{\lbrace}Z+1-2y\mp 2 \sqrt{1-y}\;\sqrt{Z-y}\Bigr{\rbrace}\;.
\ee
The condition for the existence of real intersection points is $Z>y$. For $Z=y$, which belongs to the stable region, $\lambda_{+}(q^{2})$ and  $\lambda_{-}(q^{2})$ touch each other at $q_{c}^{2}=-m^{2}$. For $Z>y$ one has a finite region $q_{c_{-}}^{2}<q^{2}<q_{c_{+}}^{2}$ for which the argument in the square root of eq.~\eqref{eq: DC60} becomes negative. In this region $\lambda_{+}(q^{2})$ and $\lambda_{-}(q^{2})$ have a non-vanishing imaginary part for real $q^{2}$. (For $Z\rightarrow 1$ one has $q_{c_{-}}^{2}\rightarrow -\infty\;, \; q_{c_{+}}^{2}\rightarrow -m^{2}/4$, while for $Z\neq 1$, $Z>y$ both $q_{c_{-}}^{2}$ and $q_{c_{+}}^{2}$ are finite.) The region with an imaginary part of $\lambda_{\pm}(q^{2})$ occurs always beyond the location of the second pole of the propagator, $q_{c_{\pm}}<\mu^{2}$. This is visible from
\be\label{eq: DC76}
\dfrac{q_{c}^{2}}{m^{2}}=-\dfrac{\mu^{2}}{m^{2}}-x_{t}\;,
\ee
with
\begin{align}\label{eq: DC77}
\!\!\!\!\!\!\!\!x_{\pm}\!\!=&\dfrac{1}{(Z-1)^{2}}\Bigl(b\mp \sqrt{b^{2}-(Z-1)^{2}v^{2}}\Bigr)\,,\quad\!\! v=\dfrac{m_{t}^{2}}{m^{2}}\,,\!\!\!\!\!\!\!\!\nn\\
b&=2(1-y)^{2}+(Z-1)\Bigl(1-y-\dfrac{\mu^{2}}{m^{2}}\Bigr)\nn\\
&=(1-y)\Bigl[(Z+1)(1-y)+\dfrac{y}{Z}(Z-1)^{2}\Bigr]\;.
\end{align}
The argument of the square root being smaller than $b^{2}$, both $x_{+}$ and $x_{-}$ are positive since $b>0$.

%Absatz
For $y<Z<y/(1+y)$ the graviton propagator has a cut in the complex $q^{2}$- plane for real negative $q^{2}$, extending from $q_{c_{-}}^{2}$ to $q_{c_{+}}^{2}$. Except for the pole at $q^{2}=0$ and this cut, it is analytic, decaying~$\sim |q|^{-2}$ for large $|q|$. Nothing obstructs analytic continuation from euclidean space to Minkowski space. We can therefore extend our analysis to field configurations for which the metric is the Minkowski metric, $g_{\mu\nu}=\eta_{\mu\nu}$. In Minkowski space one has $q^{2}=-q_{0}^{2}+\vec{q}^{\,2}$. In the complex $q_{0}$ - plane the graviton propagator has two poles at $q_{0}=\pm\sqrt{\vec{q}^{\,2}}$. It inherits the cuts on the real $q_{0}$ - axis, extending from  $q_{0}^{2}=|q_{c_{-}}^{2}|+\vec{q}^{\,2}$ to $|q_{c_{+}}^{2}|+\vec{q}^{\,2}$. The prescription for the usual infinitesimal $i\varepsilon$ -  terms is dictated by analytic continuation. This graviton propagator is well behaved in Minkowski space, without any instability. For $Z<y$ the branch points in the complex $q^{2}$- plane acquire an imaginary part and analytic continuation has to be discussed more carefully.

%Absatz
We can map these results to the general discussion in sect. \ref{sec:5}. Comparing the inverse propagator matrix~\eqref{eq: DC56} with eq.\eqref{eq:LM21}, and denoting the gauge boson mass $m$ in eq.\eqref{eq:LM21} by $m_{g}$ in order to distinguish it from $m$ as used in this section, we can identify
\be\label{eq: DC78}
m_{g}^{2}=m^{2}-M^{2}=m^{2}(1-y)\;,\quad b=1 \;,\quad c=\sqrt{1-y}\;.
\ee
Here we disregard the factor $\pm i$ in the off-diagonal elements of eq. \eqref{eq:LM21} since they do not influence the eigenvalues $\lambda_{\pm}(q^{2})$. The relation
\be\label{eq: DC79}
c^{2}=b-y
\ee
identifies in eqs.\eqref{eq:LM30},\eqref{eq:LM34}
\be\label{eq: DC80}
\!\!\!\!x\!=\!\dfrac{f\!M^{2}}{m_g^{2}}\!=\!\dfrac{fy}{1\!-\!y}\,,\quad\!\!\! \dfrac{b\!-\!c^{2}}{b}\!=\!2x\!-\!x^{2}\!\!=\!y\,,\quad\!\! (1\!-\!x)^{\!2}\!\!=\!1\!-\!y\,.\!\!\!
\ee
The stability condition \eqref{eq:LM31} translates to~\eqref{eq: DC59}, $0<y<1$, and the tachyon mass in eqs. \eqref{eq:LM32} and~\eqref{eq: DC58} coincides,
\be\label{eq: DC81}
m_{U}^{2}=\dfrac{2x-x^{2}}{Z}m_{g}^{2}=\mu^{2}\;.
\ee
The stability condition for the absence of ghosts \eqref{eq:LM33} is equivalent to eq.~\eqref{eq: DC63}. For the inverse propagator of the stable massive particle near the pole eqs.\eqref{eq:LM34} and \eqref{eq: DC71} are identical. The Taylor expansion for $\lambda_{-}(q^{2})$ for small $q^{2}$ is identically given by eqs. \eqref{eq:LM39} or~\eqref{eq: DC61}.

%%%%%%%%%%%%%%

%%% Dieser Befehl führt dazu auch alle nicht zitierten Referenzen im Dokument aufzuzeigen!
%\nocite{*}

%\vspace{2.0cm}\noindent

%%%\bibliography{BIBLIOGRAPHY}
%%%\printbibliography

\bibliography{refs}

\end{document}